# Unifying equivalences across unsupervised learning, network science, and imaging/network neuroscience

Mika Rubinov

Departments of Biomedical Engineering, Computer Science, and Psychology, Vanderbilt University; Janelia Research Campus, Howard Hughes Medical Institute. Email: mika.rubinov@vanderbilt.edu

## Abstract

Modern scientific fields face the challenge of integrating a wealth of data, analyses, and results. We recently showed that a neglect of this integration can lead to circular analyses and redundant explanations. Here, we help advance scientific integration by describing equivalences that unify diverse analyses of datasets and networks. We describe equivalences across analyses of clustering and dimensionality reduction, network centrality and dynamics, and popular models in imaging and network neuroscience. First, we equate foundational objectives across unsupervised learning and network science (from $k$-means to modularity to UMAP), fuse classic algorithms for optimizing these objectives, and extend these objectives to simplify interpretations of popular dimensionality reduction methods. Second, we equate basic measures of connectional magnitude and dispersion with six measures of communication, control, and diversity in network science and network neuroscience. Third, we describe three semi-analytical vignettes that clarify and simplify the interpretation of structural and dynamical analyses in imaging and network neuroscience. We illustrate our results on example brain-imaging data and provide *abct*, an open multi-language toolbox that implements our analyses. Together, our study unifies diverse analyses across unsupervised learning, network science, imaging neuroscience, and network neuroscience.

## Introduction

The scientific literature is now growing by several million papers every year (Hanson et al., 2024). The integration of analyses and results from this literature is no longer merely an aspirational goal, but increasingly a methodological necessity. We recently showed that a neglect of this integration leads to circular analyses of knowledge, redundant explanations, and scientific stagnation (Rubinov, 2023). We described specific instances of this problem in neuroscience and estimated that the problem has affected more than three thousand studies in network neuroscience alone over the last decade. In this way, we stressed the necessity of scientific integration for real scientific progress (Gholipour, 2023).

Our study described a statistical framework to resolve this problem, that centers on:

1. The integration of existing knowledge into explanatory benchmark models.
2. The development of statistical tests of proposed discoveries against these models.
3. The formal revision of existing knowledge on the basis of these statistical tests.

This framework is general and principled. It is also daunting, however, insofar as it calls us to take stock of sprawling literatures, to formalize knowledge into complex statistical models, and to develop advanced numerical tests of these models. The inability to do much of this at present can leave scientific fields at an uneasy impasse (Yarkoni, 2022).

Here, we adopt a distinct integrative approach that allows us to sidestep some of the more daunting aspects of this general framework. Our approach centers on the showing of analytical equivalences that unify diverse analyses of datasets and networks. These equivalences go beyond strong correlations because they establish mathematical guarantees — necessary and sufficient conditions — that link analyses to each other. The guarantees often rest on assumptions that we describe in each case, and that we test on example brain-imaging data.

We use this approach to advance scientific integration in three ways. First, we equate variants of outwardly distinct analyses in unsupervised learning and network science. Second, we equate complicated and simple analyses in unsupervised learning and imaging neuroscience. Third, we equate popular, but often speculative, dynamical interpretations with basic statistics in network science and network neuroscience. All these equivalences ultimately help integration by reducing superfluous, overly complicated, or redundant explanations of the same aspects of the data.

The analytical approach we adopt does not replace our more general framework. Nonetheless, here we used this approach to unify diverse analyses of datasets and networks. The following list summarizes our main contributions (see **Table 1** for clarifications of terms):

1. We equated three variants of global residualization — first-component removal, degree correction, and global-signal regression — across unsupervised learning, network science, and imaging neuroscience.
2. We equated the $k$-means objective in residual data or networks with a normalized modularity that we term the $k$-modularity.
3. We fused Lloyd and Louvain methods — two classic algorithms for data clustering and module detection — and showed that our unified Loyvain method does $k$-means and spectral clustering better than standard methods on example brain-imaging data.
4. We combined these results with previous work to implement binary versions of popular unsupervised learning methods. As an example, we extended the Loyvain method to bipartite networks and used it to do binary canonical covariance analysis.
5. We described m-umap, a generalized modularity that unifies UMAP with traditional spring layouts.
6. We optimized m-umap via modularity maximization and showed that m-umap embeds brain-imaging data better than UMAP.
7. We equated components or modules of co-neighbor networks with standard or binary variants of diffusion-map embeddings, or co-activity gradients in imaging neuroscience.
8. We described the second network degree, a measure of connectional dispersion that complements the (first) network degree, a measure of connectional magnitude. We equated the degree and the second degree, separately and in combination, with six measures of communication, control, and diversity in network science and network neuroscience.
9. We described three semi-analytical vignettes that clarify and simplify analyses of the primary co-activity gradient in imaging neuroscience, as well as brain-network growth and dynamic node-module affinity in network neuroscience.

In what follows, we describe our results in three increasingly technical ways. First, the text and figures of the Results section show our main results informally and visually. Second, the boxes in the Results section summarize the mathematical underpinnings of these results. Third, the text and figures in the Methods section give more exhaustive mathematical details. We illustrate all our results on example brain-imaging data and provide *abct* (github.com/mikarubi/abct/), an open multi-language toolbox that implements our analyses.

**Table 1. Clarification of terms.**

| Term | Specific use in this study |
| --- | --- |
| **Exact and approximate equivalences** | We consider analyses to be exactly equivalent when results of these analyses are guaranteed to have a maximal linear (Pearson) correlation coefficient of $\pm 1$. Correspondingly, we consider analyses to be approximately equivalent when results of these analyses are guaranteed to have a strong, but typically not maximal, linear correlation coefficient. |
| **Semi-analytical equivalences** | Semi-analytical equivalences are equivalences that rest in part on numerical results. These equivalences generally have weaker guarantees but allow us to study more intricate relationships across datasets and representations. |
| **Unifications** | Unifications are showings that outwardly distinct analyses form closely related special cases of a single and more general analysis. |
| **Unsupervised learning and network science** | Unsupervised learning is a field that uses dimensionality reduction, clustering, and other methods to find (broadly defined) patterns in data. Network science is a field that uses network analysis methods to find patterns in networks. |
| **Imaging neuroscience and network neuroscience** | Imaging neuroscience is a field that often uses unsupervised learning methods to find patterns in brain-imaging data. Network neuroscience is a field that uses network-science methods to find patterns in brain networks. |
| **Structural and correlation brain networks** | Structural brain networks represent anatomical connections between brain regions, while correlation networks represent activity correlations (or co-activities) between brain regions. |
| **$\kappa$-nearest-neighbor and co-neighbor networks** | Symmetric $\kappa$-nearest-neighbor networks are binary networks that connect pairs of nodes if one of the nodes is a top-$\kappa$ nearest, or strongest, neighbor of the other node (in a structural, correlation, or another network). Co-neighbor networks are symmetric integer networks that connect pairs of nodes by the number of their shared $\kappa$-nearest neighbors. |
| **Components and Cauchy components** | Components are special vectors (eigenvectors) that capture structural patterns of maximal linear alignment with a network matrix. Each component has a weight (eigenvalue) that reflects the dominance of its corresponding pattern. In parallel, Cauchy components are special vectors that capture structural patterns of maximal nonlinear alignment with a network matrix (measured using a so-called Cauchy similarity). |
| **Clusters or modules** | Clusters are groups of data points that are more similar to each other than to other data points. Correspondingly, modules are groups of network nodes that are more strongly connected to each other than to other network nodes. |
| **Residual and shrunken networks** | Residual networks are networks after global residualization, the removal of the dominant structural pattern of variation (first component, degree, or global signal, depending on the analysis). Shrunken networks are networks after shrinkage, the weakening (but generally not complete removal) of the structural patterns of the first several components. |

# Results

This section consists of three parts. The first part focuses on clustering and dimensionality reduction, the second on network centrality and dynamics, and the third on semi-analytical vignettes in imaging and network neuroscience. **Appendix 1** summarizes the main results in this section. All but two of these results are new to the best of our knowledge (although the vastness of the literature makes it difficult to make this claim with absolute certainty). All results assume symmetric and non-negatively weighted networks. In some cases we can relax these conditions, while in other cases we further restrict them, as we discuss below.

Our main example data satisfy these conditions and come from the Human Connectome Project, a large brain-imaging resource (Van Essen et al., 2013). The data include estimates of anatomical connectivity and resting-state (or baseline) brain activity from 100 people (Rosen and Halgren, 2021; Smith et al., 2013). We defined the networks of each person by groups of weighted connections and nodes. Connection weights represent either estimates of anatomical connectivity, or correlated activity rescaled to lie between 0 and 1 (Methods). For all but one analysis, we defined nodes to be 360 areas of the cerebral cortex, the outer part of the human brain (Glasser et al., 2016). Together, these 360 centimeter-scale nodes comprise 59,412 millimeter-scale cortical vertices. In the analysis of UMAP and m-umap, we defined nodes at the finer resolution of these vertices.

**Figure 1** shows three variants of structural and correlation networks, averaged over all the data. Our later figures show relationships computed on these average networks. By contrast, our results report median (5–95%) values estimated from 100 bootstrap samples, each formed by averaging 100 networks sampled with replacement.

## Part 1. Clustering and dimensionality reduction

This part of the study describes equivalences between variants of several popular clustering and dimensionality reduction methods across unsupervised learning, network science, and imaging neuroscience. **Table 2** summarizes many of these equivalences and describes them as special cases of a single general objective.

**First-component removal ≅ degree correction ≅ global-signal regression**

We begin by noting that many datasets and networks contain dominant global patterns that sometimes represent artifactual, trivial, or irrelevant structure. Correspondingly, analyses often seek to remove these patterns to uncover or accentuate interesting underlying structure. Here, we use the term global residualization to describe this transformation of the data. We describe three approximately equivalent variants of this transformation across unsupervised learning, network science, and imaging neuroscience. In the next section, we build on these equivalences to unify two popular but outwardly distinct algorithms for clustering and module detection.

Variants of global residualization are common across diverse analyses, although they often have distinct names and motivations. These variants include removal of the trivial or constant eigenvector in unsupervised spectral learning (Von Luxburg, 2007) and correction for node degree (the sum of connection weights to all other nodes) in network science (Karrer and Newman, 2011). They also include field-specific removal of presumed artifactual or irrelevant patterns, such as

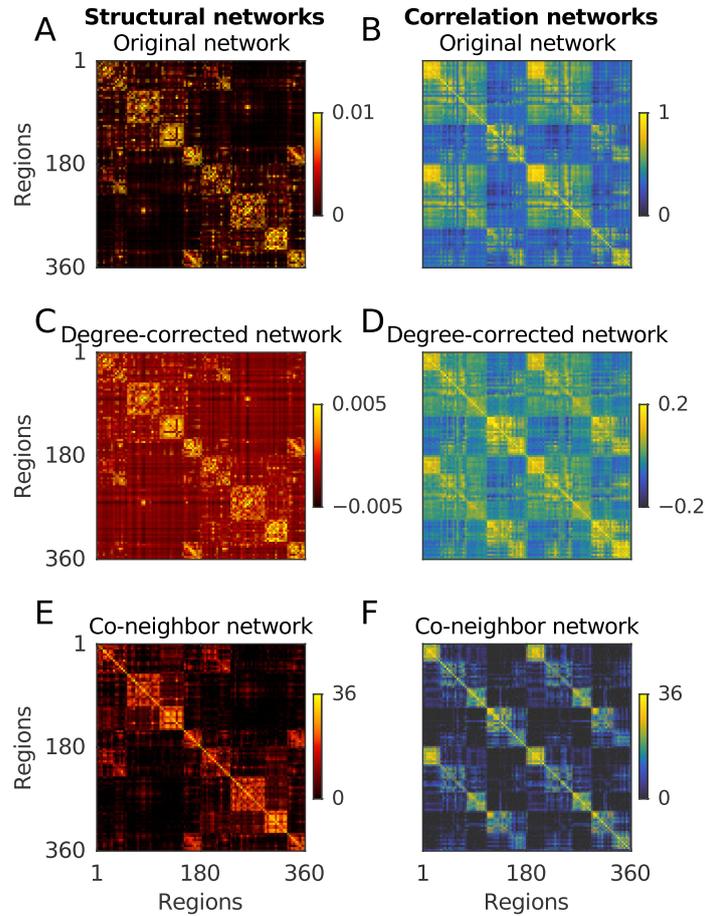

**Figure 1. Three variants of structural and correlation networks.**

**Column 1.** Matrices of structural (anatomical connectivity) networks averaged over all the data. Each network contains 360 nodes (rows and columns) that denote cortical (brain) regions. Nodes 1–180 denote regions in the left hemisphere, and nodes 181–360 denote corresponding regions in the right hemisphere. Individual elements $(i, j)$ represent connection weights between nodes $i$ and $j$. The node order emphasizes anatomical modules and is matched within hemispheres. The node order is the same for each matrix in Column 1.

**Column 2.** Matrices of correlation (correlated activity) networks averaged over all the data. Each network contains the same 360 nodes as the structural networks. The node order emphasizes correlated-activity modules and is matched for nodes in each hemisphere. The node order is the same for each matrix in Column 2.

**A–B.** Original networks.

**C–D.** Degree-corrected networks. Each connection in these networks denotes the difference between the original connection weight and the normalized product of the degrees of the corresponding pair of nodes. We use the degree to denote the sum of node connection weights to all other network nodes.

**E–F.** Co-neighbor networks. Each connection in these networks denotes the number of shared $\kappa$-nearest neighbors between the corresponding pair of nodes. We adopt a standard imaging-neuroscience choice of $\kappa = (10\% \text{ of } 360) = 36$, which implies that two nodes in these networks can have at most 36 co-neighbors.

**Table 2. Summary of main analyses in Part 1.**

Columns show types of components. Rows show types of networks. Residual networks are networks after global residualization (either first-component removal or degree correction). Adjacent rows and columns are related by type of network or component. In this way, adjacent cells show related variants of individual clustering and dimensionality reduction methods.

All objectives in this table form special cases of the following general objective:

$$\text{sum}(\text{transform}(\text{original network}) \odot \text{similarity}(\text{low-dimensional representation}))$$

where $\odot$ denotes elementwise matrix multiplication and the sum is over all matrix elements.

|  | **Binary components** | **Components** |
| --- | --- | --- |
| **Correlation network** | $k$-means clusters | Components |
| **Residual correlation network** | $k$-modularity modules | Components excluding the first component ($\cong$ degree) |
| **Residual co-neighbor network** | Binary co-activity gradients | Co-activity gradients |
|  | **Binary Cauchy components** | **Cauchy components** |
| **Residual symmetric $\kappa$-nearest-neighbor network** | m-umap modules | m-umap components |

batch effects in systems biology (Goh et al., 2017), the foreground signal in cosmology (Chen, 2020), and the global signal — the mean brain-activity signal — in imaging neuroscience (Liu et al., 2017). The removal of the global signal, or global-signal regression, is an especially prominent example that plays an important role in the analysis of brain-activity data.

In **Box 1** and the Methods section, we show that first-component removal is approximately equivalent to degree correction more generally, and to global-signal regression in correlation networks more specifically. We do this by noting that, in many networks of interest, the vector of node degrees is approximately equivalent to the first component, while in correlation networks, it is also exactly equivalent to the correlation of nodes with the global signal. Accordingly, **Figure 2A** shows that in our example data, connection weights of structural networks after first-component removal and degree correction are essentially identical, with correlations of 1.000 (1.000–1.000). Similarly, **Figure 2B** shows that the connection weights of correlation networks after first-component removal and global-signal regression have similarly strong correlations of 0.996 (0.995–0.997). Note that He and Liu (2012) previously described a distinct variant of this latter result for rows of correlation matrices.

More generally, it is well-known that the first component usually aligns with the vector of degrees, but sometimes deviates from this vector in networks that are sparse or have heterogeneous degrees (Chung et al., 2003; Goltsev et al., 2012; Martin et al., 2014; Nadakuditi and Newman, 2013). Indeed, we found some evidence of this deviation in a supplementary analysis of 24 biological, social, information, and economic networks that were diverse but lacked extreme sparseness or degree heterogeneity (**Figure S1**). Specifically, we found that the degree and first component in these networks generally aligned but also sometimes deviated, with correlations of 0.894 (0.282–0.995). Interestingly, however, the relationship between first-component removal and degree correction in these networks was robust to such deviations (**Figure S2**), with corresponding correlations of 0.974 (0.883–0.999). This shows that even a rough correspondence between the first component and the vector of degrees can bind variants of global residualization across diverse networks.

**Box 1. First-component removal ≅ degree correction ≅ global-signal regression.**

*Data and correlation matrix.* We use $\mathbf{X}$ to denote a data matrix of $n$ columns. Each column of this matrix, $\mathbf{x}_i$, denotes an individual data point. Correspondingly, we use $\mathbf{C}$ to denote the correlation network,

$$\mathbf{C} = \mathbf{X}^\top \mathbf{X},$$

and denote the elements of $\mathbf{C}$ by $c_{ij}$. We normalize $\mathbf{X}$ in a way that makes $c_{ij} = \mathbf{x}_i^\top \mathbf{x}_j$ equivalent to the Pearson correlation coefficient between $\mathbf{x}_i$ and $\mathbf{x}_j$, but rescaled to lie in the $[0, 1]$ interval (see Methods for details and justification of this rescaling).

*First-component removal.* We define the eigendecomposition of $\mathbf{C}$ as

$$\mathbf{C} = \mathbf{U}\boldsymbol{\Psi}\mathbf{U}^\top = \sum_{i=1}^{n} \psi_i \mathbf{u}_i \mathbf{u}_i^\top.$$

We thus denote the first eigenvector of $\mathbf{C}$ by $\mathbf{u}_1$, the first eigenvalue by $\psi_1$, and the first-component (best rank-one) approximation by $\psi_1 \mathbf{u}_1 \mathbf{u}_1^\top$. Correspondingly, we use $\mathbf{C}^*$ to denote the residual network after first-component removal,

$$\mathbf{C}^* = \mathbf{C} - \psi_1 \mathbf{u}_1 \mathbf{u}_1^\top.$$

*Degree correction.* We denote the degree of $\mathbf{C}$ by $\mathbf{d} = \mathbf{C}\mathbf{1}$, and the sum of all network connection weights by $\sum(c)$. Correspondingly, we use $\mathbf{C}^\diamond$ to denote the residual network after degree correction,

$$\mathbf{C}^\diamond = \mathbf{C} - \frac{1}{\sum(c)} \mathbf{d}\mathbf{d}^\top.$$

In the Methods section, we show why the relationship $\mathbf{d} \cong \mathbf{u}_1$ holds for many networks of interest. This, together with our knowledge that $\mathbf{u}_1^\top \mathbf{u}_1 = 1$ allows us to establish that $\mathbf{C}^\diamond \approx \mathbf{C}^*$.

*Global-signal regression.* We define the global signal of $\mathbf{X}$ as $\bar{\mathbf{x}} = \frac{1}{n}\mathbf{X}\mathbf{1}$, where $\mathbf{1}$ is a column vector of ones. We know that $\mathbf{d}$ is equivalent to the nodal correlation of $\mathbf{X}$ with the global signal, $\mathbf{d} = n\mathbf{X}^\top\bar{\mathbf{x}}$. We can use this equivalence to define $\mathbf{X}'$ as the residual data matrix after global-signal regression,

$$\mathbf{X}' = \left[\mathbf{X} - \frac{n}{\sum(c)} \bar{\mathbf{x}}\mathbf{d}^\top\right]\boldsymbol{\eta},$$

where the diagonal matrix $\boldsymbol{\eta}$ rescales each column of $\mathbf{X}'$ to have unit norm. Correspondingly, we can use $\mathbf{X}'$ to define $\mathbf{C}'$ as the residual network after global-signal regression,

$$\mathbf{C}' = \mathbf{X}'^\top \mathbf{X}' = \boldsymbol{\eta}\left[\mathbf{C} - \frac{1}{\sum(c)} \mathbf{d}\mathbf{d}^\top\right]\boldsymbol{\eta}.$$

In the Methods section, we show that the rescaling by $\boldsymbol{\eta}$ will, in general, have a modest effect on the structure of $\mathbf{C}'$, which allows us to establish that $\mathbf{C}' \cong \mathbf{C}^\diamond$.

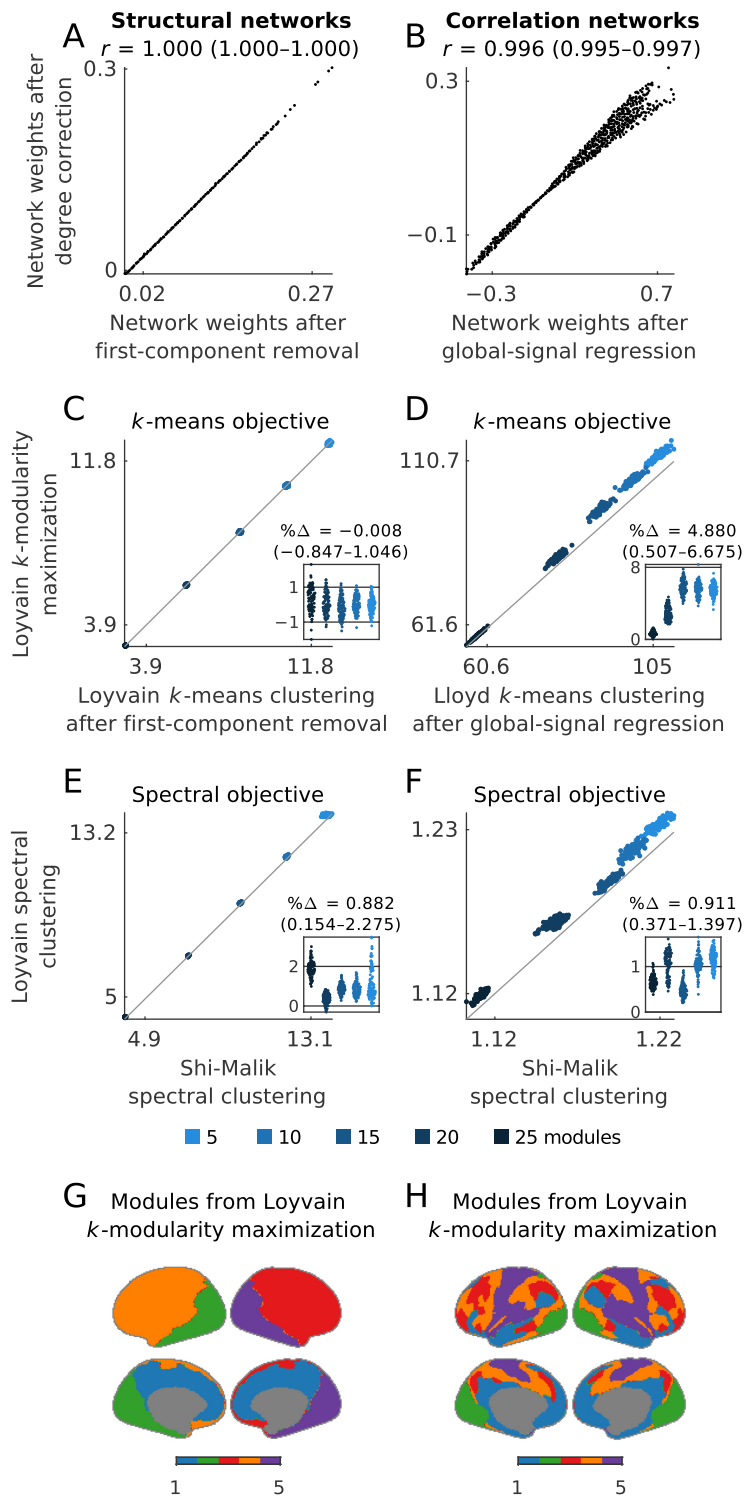

**Figure 2. Global residualization, *k*-modularity maximization and Loyvain performance.**

**A–B.** Comparison of degree correction, first-component removal, and global-signal regression. **A.** Scatter plot of structural weights after first-component removal and degree correction. **B.** Scatter plot of correlation weights after global-signal regression and degree correction.

**C–D.** Comparison of optimized *k*-means objectives. **C.** Scatter plot of optimized values on structural networks after first-component removal. The horizontal axis shows values optimized with Loyvain *k*-means clustering after first-component removal. The vertical axis shows values optimized with Loyvain *k*-modularity maximization. **D.** Scatter plot of optimized values in correlation networks after global-signal regression. The horizontal axis shows values optimized with Lloyd *k*-means clustering of data after global-signal regression. The vertical axis shows values optimized with Loyvain *k*-modularity maximization. Insets show the % improvement of Loyvain *k*-modularity maximization over the other approaches.

**E–F.** Comparison of optimized spectral clustering objectives on structural and correlation networks. The horizontal axis shows the spectral objective (normalized cut) optimized with Shi-Malik spectral clustering. The vertical axis shows this objective optimized with the Loyvain method. Insets show the % improvement of Loyvain over the Shi-Malik algorithm.

**G–H.** Maps of illustrative five-module partitions of structural and correlation networks detected with Loyvain *k*-modularity maximization.

## *k*-modularity: unified *k*-means objective and modularity

We now build on the equivalences in the last section to unify *k*-means clustering and modularity maximization. These approaches are probably the two most popular methods in unsupervised learning and network science, respectively (Fortunato, 2010; Jain, 2010). It is difficult to precisely quantify the number of all studies that have used these methods. Nonetheless, citation numbers of canonical references (Blondel et al., 2008; Lloyd, 1957/1982; MacQueen, 1967; Newman and Girvan, 2004) suggest a conservative estimate in the tens of thousands of studies for each method.

The two methods have strong conceptual similarities insofar as both seek to find groups of similar elements in data or networks. Thus, *k*-means clustering seeks to find data clusters that have relatively high within-cluster similarity (Lloyd 1957/1982). Similarly, modularity maximization seeks to find network modules that have relatively high within-module density (Newman, 2006a; Newman and Girvan, 2004).

Despite these similarities, studies usually view *k*-means clustering and modularity maximization to be distinct methods, or relate them only in some narrow regimes (Li et al., 2008). By contrast, here we show that these methods form closely related variants of a single general approach. We do so by showing an equivalence between a normalized variant of the modularity and a corrected variant of the *k*-means objective.

We will sketch out this equivalence by focusing on the modularity. For some network and module partition, we can define this objective, up to a rescaling constant, as

$$\left(\begin{array}{c}\text{(sum of within-module weights)} - \text{(sum of expected within-module weights)}\\ \text{summed over all modules}\end{array}\right).$$

We can now describe the basis of the equivalence in the following way.

First, we consider the role of the "expected within-module weights" term in the above definition. This term is often viewed as a baseline, or null model, for assessing the quality of the module partition. Such a view is problematic, however, because a null model is usually defined by a null distribution, rather than merely by the expectation of this distribution. This difference is important because the null distribution allows us to estimate statistical uncertainty and significance for each partition, while the expectation alone does not. Correspondingly, treatment of the expectation as a type of null model in network science can lead to paradoxical reports of modular structure in seemingly non-modular networks (Guimerà et al., 2004; Peixoto, 2023).

In this study, by contrast, we view the expectation term from an alternative and more interpretable perspective. We note, specifically, that subtraction of this term is usually equivalent to a data processing step that removes some global, and presumably uninteresting, pattern of variation from the network. Indeed, in the Methods section, we show that the most common choice for this term reduces this subtraction to a variant of global residualization. Here, we adopt this choice to re-express the modularity more directly as

$$\text{(modularity)} \equiv \left(\begin{array}{c}\text{(sum of residual within-module weights)}\\ \text{summed over all modules}\end{array}\right).$$

Next, we adopt a variant of the modularity that normalizes the contribution of each module by its size, or the number of its constituent nodes. This normalization gives more importance to dense modules, rather than merely to large modules, and thus intuitively aligns with the general aim of the modularity.

We term this normalized variant of the modularity the $k$-modularity, and define it as

$$(k\text{-modularity}) \equiv \begin{pmatrix}\text{(normalized sum of residual within-module weights)}\\ \text{summed over all modules}\end{pmatrix}.$$

In the Methods section, we reduce this normalized modularity to the $k$-means objective by showing that

$$(k\text{-modularity}) \equiv (k\text{-means objective in residual networks}).$$

Similarly, we build on the results in the last section to show that in imaging neuroscience

$$(k\text{-modularity}) \cong (k\text{-means objective in time series after global-signal regression}).$$

Separately, we also consider a simpler, although less common, choice for the expectation as a network density (mean connection weight). This choice leads to a so-called "density-corrected" variant of the modularity (Newman, 2006b). In the Methods section, we show that this variant simply reduces to the standard $k$-means objective:

$$(\text{density-corrected } k\text{-modularity}) \equiv (k\text{-means objective}).$$

Together, these results show that the $k$-means objective and the modularity form special cases of a single general objective. The main difference between these objectives lies in data transformation or feature constraints, rather than in more fundamental distinctions. The modularity removes a global pattern of variation, while $k$-means normalizes the sum of within-module weights. The $k$-modularity unifies $k$-means and modularity by combining both these properties in one objective. **Table 3** and **Box 2** summarize these relationships and mathematical details.

**Table 3. Unified $k$-means objective and modularity.**

Columns show equivalent representation of the $k$-means objective (left), $k$-modularity (center), and modularity (right). Adjacent columns differ by a single processing step and thus show closely related variants of a single general objective.

| $k$-means objective | $k$-means objective with global residualization | |
|---|---|---|
| $\equiv$ | $\equiv$ | |
| $k$-modularity without global residualization ($\equiv$ density-corrected $k$-modularity) | **$k$-modularity** | $k$-modularity without module normalization |
| | $\equiv$ | $\equiv$ |
| | modularity with module normalization | **modularity** |

**Box 2. Unified *k*-means objective and modularity.**

*k-means objective and k-modularity.* We can define the two *k*-means objectives as

$$(k\text{-means objective}) \equiv \sum_{h=1}^{k} \frac{1}{N_h} \mathbf{m}_h^\top \mathbf{C} \mathbf{m}_h \text{ and}$$

$$(k\text{-means objective with global residualization}) \equiv \sum_{h=1}^{k} \frac{1}{N_h} \mathbf{m}_h^\top \mathbf{C}^\diamond \mathbf{m}_h.$$

In both cases, the binary vector $\mathbf{m}_h$ indicates the nodes present in cluster $h$, while $N_h$ denotes the total number of nodes in this cluster.

In the Methods section, we show that density-corrected *k*-modularity reduces to

$$(\text{density-corrected } k\text{-modularity}) \equiv \sum_{h=1}^{k} \frac{1}{N_h} \mathbf{m}_h^\top \mathbf{C} \mathbf{m}_h.$$

Correspondingly, we show that the (degree-corrected) *k*-modularity reduces to

$$(k\text{-modularity}) \equiv \sum_{h=1}^{k} \frac{1}{N_h} \mathbf{m}_h^\top \mathbf{C}^\diamond \mathbf{m}_h,$$

which, together with our knowledge of $\mathbf{C}^* \cong \mathbf{C}^\diamond \cong \mathbf{C}'$ from **Box 1** establishes the equivalences in the main text.

**Loyvain: unified *k*-means clustering and modularity maximization**

We can now build on the equivalences of the last section to unify classic algorithms for data clustering and module detection into a single general approach. We specifically consider Lloyd and Louvain methods, probably the two most popular algorithms for doing *k*-means clustering and modularity maximization, respectively (Lloyd 1957/1982; Blondel et al. 2008). These iterative methods have strong conceptual similarities. Lloyd does *k*-means clustering by placing, at each iteration, all data points into their closest clusters. Similarly, the first phase of Louvain maximizes the modularity by placing, at each iteration, a randomly chosen node inside its optimal module.

In this section, we fuse Lloyd and Louvain methods into a unified algorithm that we dub the Loyvain method. Much as the *k*-modularity interpolates between *k*-means and modularity, so the Loyvain interpolates between Lloyd and Louvain. **Box 3** summarizes the details of these three methods, while **Table 4** situates these methods in a unified framework.

**Box 3. Unified *k*-means clustering and modularity maximization.**

*Lloyd method.* This method aims to find an optimal cluster partition in five steps:

1. Place all data points into $k$ clusters using a custom initialization.
2. Define cluster centroids by the means of within-cluster data points.
3. Find the closest cluster centroids for all data points.
4. Move all data points to clusters of their closest centroids.
5. Repeat steps 2 to 4 until no more moves can be made.

*Louvain method.* This method consists of two phases. Here, we focus only on the first phase. Much like Lloyd, this phase aims to find an optimal module partition in five steps:

1. Initially place all nodes into *n* singleton modules.
2. Examine all nodes in random order, one node at a time.
3. Find an optimal module for each examined node (such that the placement of a node into its optimal module will lead to the greatest increase in the modularity).
4. Move each examined node to its optimal module.
5. Repeat steps 2 to 4 until no more moves can be made.

*Loyvain method.* Much like Lloyd and Louvain, the Loyvain method also aims to find an optimal partition in five steps:

1. Place all nodes (or data points) into *k* modules (or clusters) using a custom initialization.
2. Examine all nodes in random batches (groupings of nodes), one batch at a time.
3. Find the optimal module placement for all nodes in each examined batch.
4. Move all nodes in each batch to their optimal modules.
5. Repeat steps 2 to 4 until no moves can be made.

The size of the batches effectively makes Loyvain more Lloyd-like or more Louvain-like (**Table 4**). The placement of all nodes in one batch makes Loyvain identical to Lloyd, except for the update rule (Step 3). Correspondingly, the placement of nodes into *n* singleton batches makes Loyvain identical to the first phase of Louvain, except for the initialization rule (Step 1). Separately from these parallels, Loyvain shares similarities with other variants of *k*-means clustering (Hartigan and Wong, 1979; MacQueen, 1967; Sculley, 2010).

In practice, the Loyvain method combines several benefits of Lloyd and Louvain.

First, and like Lloyd, Loyvain can find clusters directly in data without needing to compute similarity or network matrices. This can make clustering memory efficient, especially in large datasets. It can also simultaneously move many nodes at each iteration, which can increase convergence speed (Sculley, 2010). Finally, it can make use of accurate initialization rules, which can help improve clustering results (Methods, Celebi et al. 2013).

**Table 4. Unified *k*-means clustering and modularity maximization.**

Columns show equivalent algorithms. Adjacent columns differ by a single algorithmic step and thus show closely related variants of a single general algorithm.

| **Lloyd method** | Lloyd method with objective-based update | | |
|---|---|---|---|
| ≡ | ≡ | | |
| Loyvain method (1 batch) with distance-based update | **Loyvain method (1 batch)** | | |
| | **Loyvain method (*n* batches)** | Loyvain method (*n* batches) with singleton initialization | |
| | ≡ | ≡ | |
| | Louvain method (Phase 1) with custom initialization | **Louvain method (Phase 1)** | |

Second, and like Louvain, Loyvain can directly cluster networks, without needing to access data points. This allows it to do $k$-means directly on arbitrary network or similarity matrices, a property we will make use of below. For example, **Figure 2C** shows two equivalent applications of Loyvain on network matrices, neither of which can be done with Lloyd. Loyvain also uses a local update rule that directly optimizes the objective function at each iteration (**Box 3**). **Figure 2D** shows that this property helped Loyvain outperform Lloyd by about ~5% in our example data.

Finally, and like Louvain, Loyvain can use other update rules to optimize diverse objectives (Blondel et al., 2024). For example, the so-called normalized cut is another widely used clustering objective, usually solved with spectral clustering methods (Shi and Malik, 2000). Previous work has shown that the normalized cut is exactly equivalent to a modularity normalized by module degree, or the sum of degrees of all nodes within a module (Yu and Ding, 2010). Here, we extend Loyvain to optimize this objective and thus do spectral clustering. **Figure 2E–F** shows that Loyvain outperformed the standard spectral clustering algorithm by ~1% in our example data.

**Figure 2G–H** shows example modules detected with Loyvain maximization of the $k$-modularity. Below, we will consider close variants of these modules from several other perspectives.

### $k$-modularity maximization ≡ detection of binary components

A well-known equivalence in linear algebra links $k$-means clustering with the detection of leading binary components, or eigenvectors (**Box 4** Fan 1949; Bhatia 1997). This equivalence implies that

$$(\text{detection of } k\text{-means clusters}) \equiv (\text{detection of first to } k\text{-th binary components}).$$

Correspondingly, we can use the relationship between $k$-means clustering and $k$-modularity to adjust the above expression to

$$(\text{detection of } k\text{-modularity modules}) \cong (\text{detection of second to } (k+1)\text{-th binary components}).$$

These equivalences are useful because many methods use leading components to do dimensionality reduction. It follows that we can use $k$-means clustering, or $k$-modularity maximization, to define binary variants of many such methods. In practice, these binary variants will often produce simpler, more interpretable, and more robust results (Ye and Liu, 2012). On the flip side, these results will also often be less rich and may depend on the number of specified clusters. The choice to adopt these binary variants should thus be guided by the specific question of interest.

In the next two sections, we show the utility of this approach in two distinct ways. First, we extend the Loyvain method to do binary canonical covariance analysis (partial least squares correlation), a popular multivariate dimensionality reduction method. Second, we show that the components of co-neighbor networks are equivalent to co-activity gradients (diffusion-map embeddings), popular low-dimensional representations in imaging neuroscience. We combine these insights to show that $k$-modularity modules of co-neighbor networks are equivalent to binary co-activity gradients. This latter equivalence unifies co-activity gradients and modules in imaging neuroscience (**Table 5**).

**Table 5. Unified modules and co-activity gradients in imaging neuroscience.**

Columns show equivalent objectives (the Rayleigh quotient is an objective for component detection, **Box 4**). Adjacent columns differ by a single processing step and thus show closely related special cases of a single general objective.

| $k$-modularity of correlation networks | | |
|---|---|---|
| ≡ | ≡ | |
| Sum of $k$ Rayleigh quotients of residual correlation networks with binary constraints | **Sum of $k$ Rayleigh quotients of residual correlation networks** | Sum of $k$ Rayleigh quotients of residual co-neighbor networks |
| | | ≡ |
| | | **Co-activity gradient objective** |

**Box 4. $k$-modularity maximization ≡ detection of binary components.**

The equivalences in this section directly follow from the maximum principle of Ky Fan (1949), a well-known result in linear algebra (Bhatia, 1997). This result states that the $k$ leading eigenvectors $\mathbf{u}_1, \ldots \mathbf{u}_k$ of a symmetric matrix $\mathbf{C}$ maximize the sum

$$\sum_{h=1}^{k} \mathbf{u}_h^\top \mathbf{C} \mathbf{u}_h$$

relative to any other set of $k$ orthonormal vectors. Equivalently, we can say that the $k$ leading eigenvectors of $\mathbf{C}$ maximize the sum

$$\sum_{h=1}^{k} \frac{\mathbf{u}_h^\top \mathbf{C} \mathbf{u}_h}{\mathbf{u}_h^\top \mathbf{u}_h}$$

relative to any other set of $k$ orthogonal vectors. Each individual sum in this latter expression is known as a so-called Rayleigh quotient (Yu et al., 2011).

We can define a variant of this latter sum by adding binary constraints on the values of the vectors. Denoting the resulting binary vectors by $\mathbf{m}_h$, and noting that the number of nonzero elements in $\mathbf{m}_h$ is given by $N_h = \mathbf{m}_h^\top \mathbf{m}_h$, we find that

$$(k\text{-means objective}) = \sum_{h=1}^{k} \frac{1}{N_h} \mathbf{m}_h^\top \mathbf{C} \mathbf{m}_h = \sum_{h=1}^{k} \frac{\mathbf{m}_h^\top \mathbf{C} \mathbf{m}_h}{\mathbf{m}_h^\top \mathbf{m}_h}$$

and therefore

$$(k\text{-modularity}) \cong \sum_{h=1}^{k} \frac{1}{N_h} \mathbf{m}_h^\top \mathbf{C}^* \mathbf{m}_h = \sum_{h=1}^{k} \frac{\mathbf{m}_h^\top \mathbf{C}^* \mathbf{m}_h}{\mathbf{m}_h^\top \mathbf{m}_h},$$

which establishes our equivalence of interest.

### $k$-modularity co-maximization ≅ binary canonical covariance analysis

As mentioned in the last section, we now describe a binary variant of canonical covariance analysis (partial least squares correlation, Abdi 2010). This method finds linear relationships between pairs of distinct datasets that have the same number of data points but usually have different numbers of features. It specifically finds $k$ pairs of coefficient vectors that transform the original data points into $k$ pairs of components, in a way that maximizes the total covariance over all pairs of components. This method is closely related to canonical correlation analysis, which works similarly but maximizes the total correlation, rather than the total covariance. Variants of both methods are popular across diverse scientific fields, including social science (Mateos-Aparicio, 2011), systems biology (Boulesteix and Strimmer, 2006), and imaging neuroscience (Mihalik et al., 2022). At the same time, the application of these methods to data with many features often leads to unstable estimates of coefficients (Helmer et al., 2024; Lambert and Durand, 1975). This problem can be ameliorated, to some extent, through the adoption of sparse variants of these methods (Lin et al., 2013).

A common formulation of both methods centers on the detection of (principal) components of cross-covariance matrices (**Box 5**, Abdi 2010; Press 2011). Here, we use this formulation to describe binary variants of both methods. We do this by extending the Loyvain algorithm to independently cluster the rows and columns of cross-covariance matrices, or any other bipartite (two-part) networks for that matter. This process thus simultaneously finds pairs of modules from both datasets. In the Methods section, we show that this process is equivalent to canonical covariance analysis with binary coefficients. We also show a similar equivalence for canonical correlation analysis, although additional transforms in that method ultimately result in non-binary coefficients.

**Figure 3** compares the results of canonical covariance analysis applied to the cross-correlation of structural and correlation matrices in our example data. These results confirm that binary coefficients can be sparse and more interpretable than their weighted counterparts. Moreover, binary coefficients lead to a particularly simple definition of canonical components, as sums of data points over the non-zero features. Our results show that the correlation between the components defined by standard and binary coefficients was moderately high, with 0.818 (0.478–0.971) for the first component. We note, however, that this high correlation is not necessarily guaranteed by the equivalence because binary constraints can, in principle, result in considerably different representations.

Together, our results suggest that binary canonical covariance analysis can provide a simpler and more interpretable alternative to the standard weighted variant of this method.

**Box 5. $k$-modularity co-maximization $\cong$ binary canonical covariance analysis.**

The general formulation of canonical correlation analysis considers two data matrices $\mathcal{X}$ and $\mathcal{Y}$, each with $n$ centered columns. We can define the cross-covariance matrix as $\mathcal{Z} \equiv \mathcal{X}\mathcal{Y}^\top$, and the canonical covariance objective as

$$\text{(canonical covariance objective)} \equiv \sum_{h=1}^{k} \frac{\boldsymbol{a}_h^\top \mathcal{Z} \boldsymbol{b}_h}{\left(\boldsymbol{a}_h^\top \boldsymbol{a}_h\right)^{1/2}\left(\boldsymbol{b}_h^\top \boldsymbol{b}_h\right)^{1/2}}.$$

Canonical covariance analysis seeks to maximize this objective over all sets of $k$ orthogonal vectors $\boldsymbol{a}_1, \dots \boldsymbol{a}_k$ and simultaneously over all sets of $k$ orthogonal vectors $\boldsymbol{b}_1, \dots \boldsymbol{b}_k$.

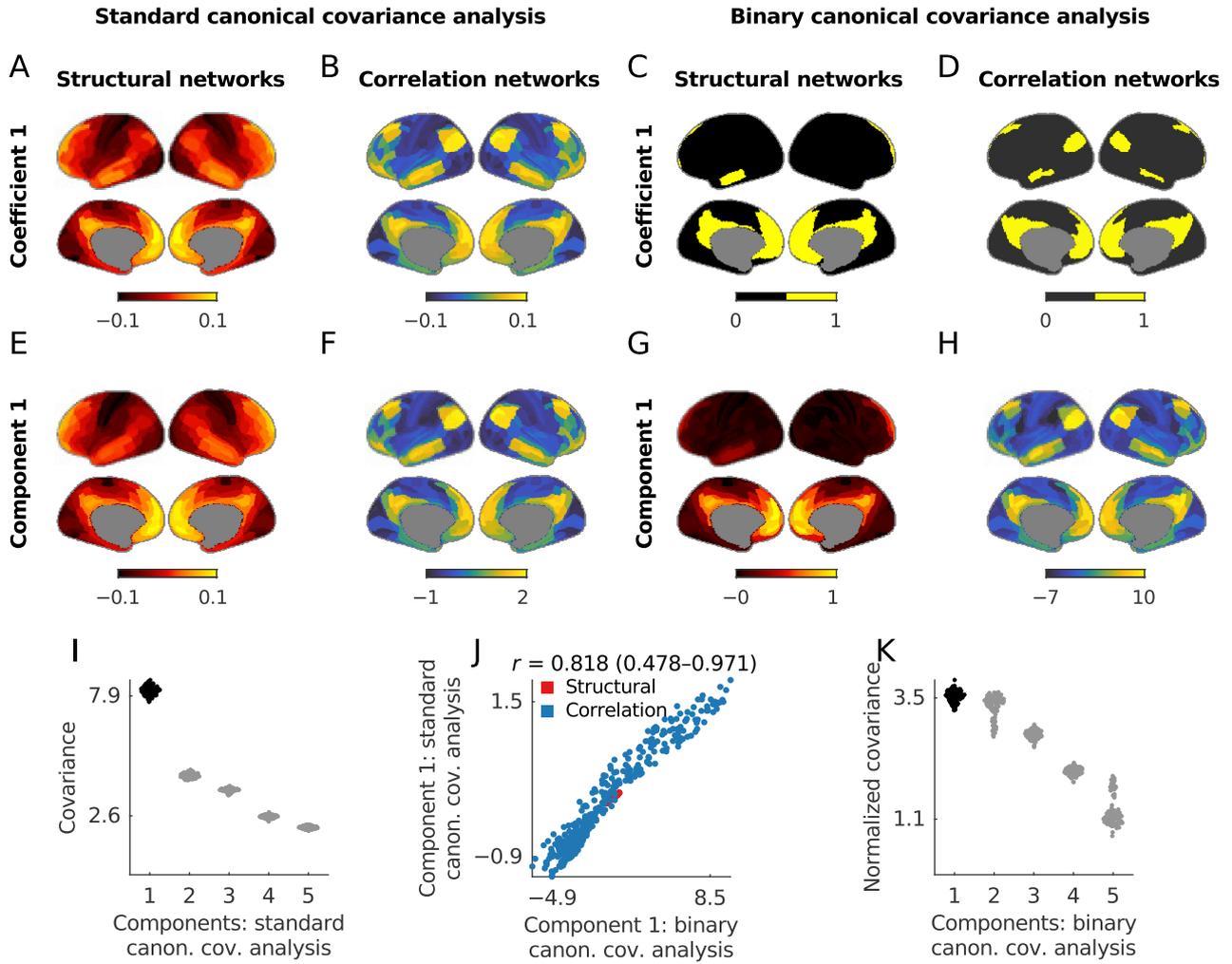

**Figure 3. *k*-modularity co-maximization of cross-covariance matrices ≡ canonical covariance analysis with binary constraints.**

**A–D.** Maps of coefficient vectors for the weighted (**A–B**) and binary (**C–D**) first canonical covariance analysis pair.

**E–H.** Maps of component vectors for the weighted (**E–F**) and binary (**G–H**) first canonical covariance analysis pair.

**J.** Scatter plot of component vectors for the weighted and binary canonical first covariance analysis pair.

**I, K.** Swarm plots of covariances between the first five canonical covariance analysis pairs. **I.** Covariances from the weighted canonical covariance analysis. **K.** Normalized covariances from the binary canonical covariance analysis. Covariance between the first pairs (**E–F** and **E–H**) is in black. The values of the covariances in panels **I** and **K** are not directly comparable due to different normalizations of the weighted and binary analyses.

In practice, we know that pairs $\boldsymbol{a}_h$ and $\boldsymbol{b}_h$ form the $h$-th leading eigenvectors of $\mathcal{Z}\mathcal{Z}^\top$ and $\mathcal{Z}^\top\mathcal{Z}$ respectively. In this way, canonical covariance analysis reduces to the eigendecomposition of the cross-covariance matrices.

Here, we adopt a binary variant of this method by imposing binary constraints on all coefficient vectors $\boldsymbol{a}_h$ and $\boldsymbol{b}_h$. This effectively converts these vectors into module indicators $\mathbf{m}_{\boldsymbol{a}h}$ and $\mathbf{m}_{\boldsymbol{b}h}$. We denote the sizes of the corresponding modules by $N_{\boldsymbol{a}h}$ and $N_{\boldsymbol{b}h}$, and thus define a binary canonical covariance analysis as a $k$-means co-clustering objective,

$$(k\text{-means co-clustering objective}) = \sum_{h=1}^{k} \frac{\mathbf{m}_{\boldsymbol{a}h}^\top \mathcal{Z} \mathbf{m}_{\boldsymbol{b}h}}{\left(\mathbf{m}_{\boldsymbol{a}h}^\top \mathbf{m}_{\boldsymbol{a}h}\right)^{1/2} \left(\mathbf{m}_{\boldsymbol{b}h}^\top \mathbf{m}_{\boldsymbol{b}h}\right)^{1/2}} = \sum_{h=1}^{k} \frac{\mathbf{m}_{\boldsymbol{a}h}^\top \mathcal{Z} \mathbf{m}_{\boldsymbol{b}h}}{(N_{\boldsymbol{a}h} N_{\boldsymbol{b}h})^{1/2}}.$$

Similarly, we use $\mathcal{Z}^*$ to denote $\mathcal{Z}$ after global residualization, and thus express a corresponding $k$-co-modularity as

$$(k\text{-co-modularity}) \cong \sum_{h=1}^{k} \frac{\mathbf{m}_{\boldsymbol{a}h}^\top \mathcal{Z}^* \mathbf{m}_{\boldsymbol{b}h}}{\left(\mathbf{m}_{\boldsymbol{a}h}^\top \mathbf{m}_{\boldsymbol{a}h}\right)^{1/2} \left(\mathbf{m}_{\boldsymbol{b}h}^\top \mathbf{m}_{\boldsymbol{b}h}\right)^{1/2}} = \sum_{h=1}^{k} \frac{\mathbf{m}_{\boldsymbol{a}h}^\top \mathcal{Z}^* \mathbf{m}_{\boldsymbol{b}h}}{(N_{\boldsymbol{a}h} N_{\boldsymbol{b}h})^{1/2}}.$$

In both cases, we devise fast update rules to efficiently maximize these objectives with co-Loyvain, our extension of the Loyvain algorithm to bipartite networks (see Methods for details).

**Components or modules of co-neighbor networks $\cong$ co-activity gradients**

We now consider a simple class of integer networks that, together with our preceding discussion, allows us to simplify a variant of diffusion-map embedding — a versatile method for nonlinear dimensionality reduction (Coifman and Lafon, 2006). The specific variant we study has transformed analyses in much of imaging neuroscience over the last decade (Margulies et al., 2016). It has done so by allowing investigators to robustly detect co-activity (or functional) gradients, low-dimensional representations that capture important properties of functional brain organization, including graded transitions in co-activity patterns between brain areas (Huntenburg et al., 2018).

In practice, a standard application of this variant comprises the following steps: construction of a correlation network, conversion of this network to a $\kappa$-nearest-neighbor network weighted by the original correlations, conversion of the latter network to a sparse similarity association network, two consecutive degree-based normalizations, detection of the second to $(k+1)$-th components of the resulting network, and a final normalization of these components (Margulies et al., 2016).

Here, we describe equivalences that considerably simplify many of these steps and indeed obviate the need to use diffusion maps for this analysis (**Box 6**). Our description centers on a simple class of co-neighbor networks. These networks, as their name suggests, encode the number of shared $\kappa$-nearest, or strongest correlated, neighbors between pairs of nodes (**Table 1**). The networks thus provide a simple and typically sparse representation of node similarity (**Figure 1**).

In the Methods section, we establish the following approximate equivalence:

(detection of $k$ co-activity gradients with diffusion-map embedding) $\cong$
(detection of second to $(k+1)$-th components of co-neighbor networks).

The maps in the first two columns of **Figure 4** and the scatter plot in **Figure 4J** show that this equivalence holds in our example data, with correlation coefficients of 0.988 (0.958–0.993) between the first five co-activity gradients and corresponding components of co-neighbor networks.

The effective exclusion of the first component in this equivalence, together with our $k$-modularity result above, allows us to describe binary co-activity gradients in even simpler terms:

(detection of $k$ binary co-activity gradients with diffusion-map embedding) $\cong$
(detection of $k$-modularity modules of co-neighbor networks).

The maps in the third column of **Figure 4** show that this equivalence also holds in our example data. More quantitatively, the swarm plots in **Figure 4K–L** show that within-module weights of co-activity gradients (and co-neighbor components) were strongly positive, while between-module weights were weakly negative. The sole exception was the last module. This exception confirms the rule, however, insofar as the last module "absorbs" all remaining nodes in the network, simply because all nodes must be assigned to one module.

**Box 6. Components or modules of co-neighbor networks $\cong$ co-activity gradients.**

The variant of diffusion-map embedding we consider comprises the following steps:

1. Construct a sparse similarity network from a $\kappa$-nearest neighbor network.
2. Normalize this similarity network by the product of the node degrees.
3. Convert the normalized matrix to a transition probability matrix.
4. Define the first $k$ co-activity gradients as the normalized second to $(k+1)$-th eigenvectors of the transition probability matrix.

In the Methods section, we show three results. First, we show that the sparse similarity network in step 1 is approximately equivalent to a co-neighbor network. Second, we show that steps 2 and 3 primarily affect the first eigenvector of this matrix, but not the other $k$ leading eigenvectors. Third, we show that step 4 amounts to the removal of the first eigenvector (up to rescaling by constants). Together, these results imply that this variant of diffusion-map embedding reduces to the eigendecomposition, or $k$-modularity maximization, of co-neighbor networks.

**m-umap (modularity with Cauchy components) $\equiv$ first-order approximation of UMAP**

We can make things even simpler by doing away with co-neighbor similarities and getting co-activity gradients directly from symmetric $\kappa$-nearest-neighbor networks. The networks connect pairs of nodes when at least one of these nodes is a $\kappa$-nearest neighbor of another (**Table 1**). In our example data, we find that such networks can have similar components to co-neighbor networks, with correlation coefficients of 0.957 (0.905, 0.9680). Since there is no principled *a priori* reason to use co-neighbor similarity, we will adopt $\kappa$-nearest-neighbor similarity in this section.

This seemingly modest change in effect converts our approach to a vanilla method for nonlinear dimensionality reduction or "manifold learning". This method first defines local representations of data through $\kappa$-nearest-neighbor networks and then compresses these representations with components. This general approach subsumes, as special cases, several popular and more specific methods (Belkin and Niyogi, 2003; Coifman and Lafon, 2006; Donoho and Grimes, 2003; Izenman, 2011; Roweis and Saul, 2000; Tenenbaum et al., 2000).

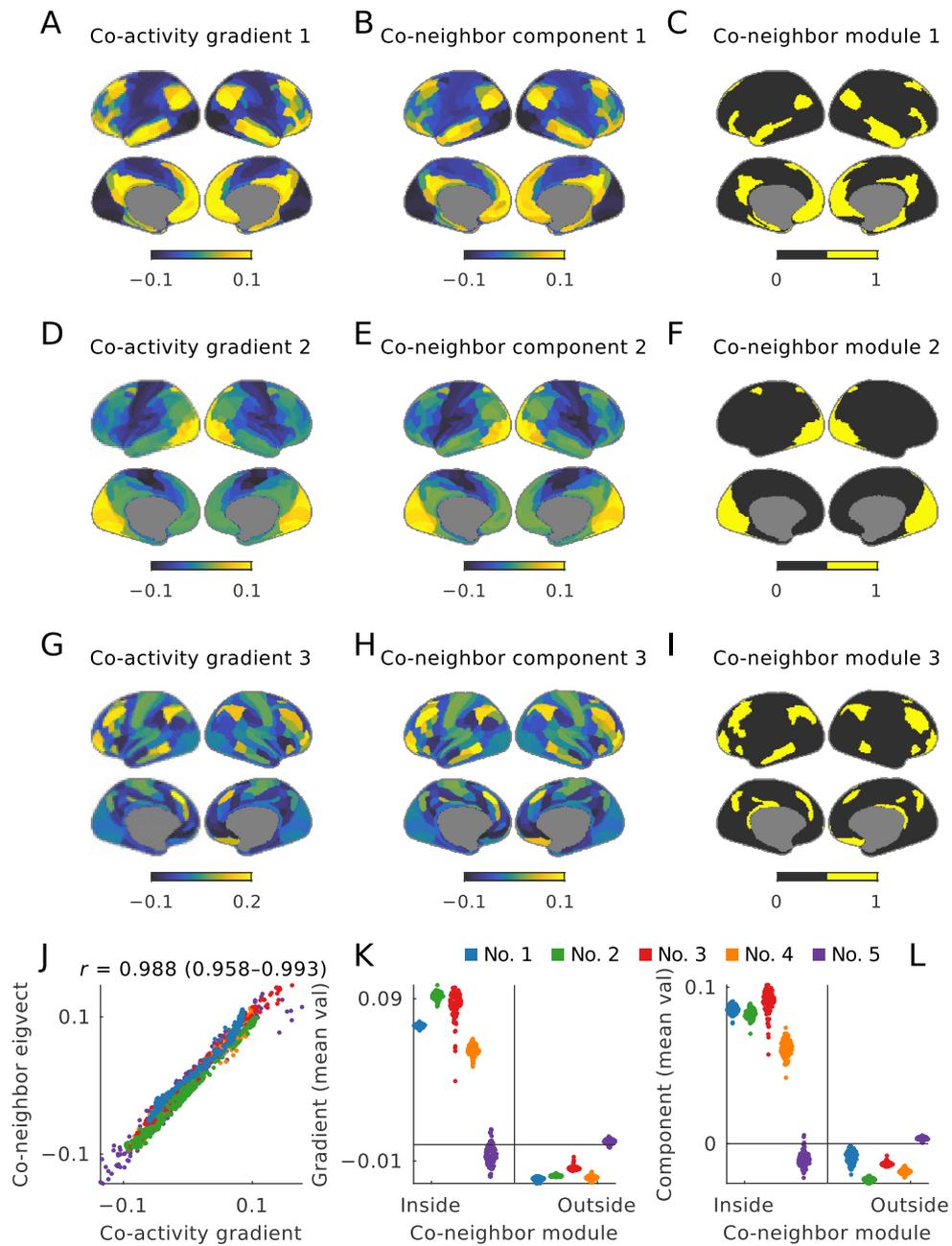

**Figure 4. Components or modules of co-neighbor networks ≅ co-activity gradients.**

**A–I.** Maps of co-activity gradients and their equivalent weighted or binary representations.

**Column 1 (A, D, G).** Maps of the first three co-activity gradients (diffusion-map embeddings).

**Column 2 (B, E, H).** Maps of corresponding components (leading eigenvectors) of co-neighbor networks.

**Column 3 (C, F, I).** Maps of corresponding modules of co-neighbor networks, estimated using Loyvain $k$-modularity maximization with a five-module partition.

**J.** Scatter plots of the first five co-activity gradients and corresponding co-neighbor components.

**K–L.** Swarm plots of mean values of (**K**) co-activity gradients and (**L**) co-neighbor components, within and between co-neighbor modules.

In this and the next section, we describe an extension of this basic method that we term m-umap, and that reflects a simplification of UMAP (McInnes et al., 2018a), a prominent manifold-learning method especially popular in the analysis and visualization of single-cell, population-genetic, and other biological data (Becht et al., 2019; Diaz-Papkovich et al., 2021). We first derive m-umap as a first-order approximation of UMAP, then show that the binary variant of m-umap reduces to standard modularity maximization, and finally note that m-umap unifies modularity with more traditional network (spring) layout methods. We also describe the conceptual advantages of m-umap over UMAP, show that m-umap embeds our example data better than UMAP, and finally note some of its limitations.

The high performance of UMAP was originally attributed to its deep theoretical underpinnings (McInnes et al., 2018a). More recently, however, a sequence of intriguing empirical and analytical studies (Böhm et al., 2022; Damrich et al., 2023; Damrich and Hamprecht, 2021) has shown that the standard implementation of UMAP deviates from its motivating theory. This work has shown that the true UMAP objective — the objective actually optimized by the standard implementation — essentially seeks to find low-dimensional representations that accurately approximate symmetric $\kappa$-nearest-neighbor networks. It does so by nonlinearly aligning these representations to the nearest-neighbor networks, using a so-called Cauchy similarity, a measure of similarity that pushes nodes away even at long ranges and thus prevents the formation of amorphous clouds (or network hairballs).

Here, we focus on a parametric implementation of UMAP (Sainburg et al., 2021). Damrich and Hamprecht (2021) have shown that this implementation optimizes an objective that distantly resembles the modularity. We build on this observation to define m-umap as a modularity with Cauchy similarity (**Box 7**) and note that

$$\text{(m-umap objective)} \equiv \text{(first-order approximation of UMAP objective)}.$$

Both m-umap and UMAP thus seek to approximate symmetric $\kappa$-nearest-neighbor representations and coincide in their theoretical optimum. We can only get to this optimum, however, if we place disconnected nodes infinitely far apart. In more realistic settings, m-umap is simpler than UMAP and places more emphasis on fidelity over aesthetics. For example, m-umap always pulls connected nodes together and pushes disconnected nodes apart. By contrast, UMAP pulls connected nodes together if they are far apart but also pushes them apart if they are too close together (and thus prevents their collapse into single points).

The simplicity of m-umap can lead to runaway solutions, such as the infinite repulsion of disconnected nodes. In what follows, we check this outcome by embedding m-umap solutions on ($k$-dimensional) spheres. These embeddings have an additional nice property of reducing m-umap with binary constraints to the standard modularity (Methods). This equivalence between m-umap embeddings and modularity modules thus parallels the equivalence between components and $k$-modularity modules we described above. Correspondingly, we also note that m-umap without global residualization reduces to the purely attractive objective of Cauchy graph embedding (Luo et al., 2011). This equivalence somewhat parallels our earlier equivalence between the modularity and the $k$-means objective.

**Table 6. Unified spring layouts and UMAP objective.**

Columns show equivalent representations of the spring-layout objective (left), m-umap (center), and UMAP objective (right). Adjacent columns show related variants of a more general objective.

| **Spring-layout objective** | Spring-layout objective with Cauchy similarity | |
|---|---|---|
| ≡ | ≡ | |
| m-umap with Euclidean similarity | **m-umap** | |
| | ≡ | ≡ |
| | first-order approximation of UMAP | **UMAP** |

Finally, we note that seminal work (Noack, 2009, 2007) has previously unified the modularity with a family of spring-layout methods. These methods have a longer history than UMAP (Fruchterman and Reingold, 1991) and remain popular today, including for network visualizations (Jacomy et al., 2014). The methods work by treating networks as physical systems in Euclidean space — with nodes that have a baseline level of repulsion and connections that act as mechanical springs to drive attraction. The methods seek layouts that minimize the mechanical energy by balancing these attractive and repulsive forces.

We can use this result to alternatively define m-umap as a modified spring-layout method,

$$(\text{m-umap objective}) \equiv (\text{spring-layout objective with Cauchy similarity}).$$

In this way, we can use m-umap to unify spring layouts with UMAP (**Table 6**).

### Box 7. m-umap ≅ first-order approximation of UMAP.

We denote the symmetric $\kappa$-nearest-neighbor matrix by $\mathfrak{C}$ and its elements by $\mathfrak{c}_{ij}$. Correspondingly, we denote the degree of node $i$ in this network by $\mathfrak{d}_i$.

The main equivalence in this section directly builds on the work of Damrich and Hamprecht (2021), who showed that, up to a rescaling constant, the true parametric UMAP objective can essentially be written as

$$(\text{UMAP objective}) = -\sum_{i,j} \left( \mathfrak{c}_{ij} \log(\phi_{ij}) + \gamma \frac{\mathfrak{d}_i \mathfrak{d}_j}{\sum(\mathfrak{c})} \log(1 - \phi_{ij}) \right),$$

where the Cauchy similarity

$$\phi_{ij} = \left( 1 + \alpha \left\| \mathbf{u}_{i:} - \mathbf{u}_{j:} \right\|^{2\beta} \right)^{-1}$$

is a function of the Euclidean distance $\|\cdot\|$ between low-dimensional row vectors $\mathbf{u}_{i:}$ and $\mathbf{u}_{j:}$, while $\alpha$, $\beta$ and $\gamma$ are parameters (note that we use $\mathbf{u}_{i:}$ to denote row vectors).

We define m-umap as the first-order Taylor expansion of the UMAP objective around 1/2. After doing the algebra and dropping all constant terms, we can write

$$(\text{m-umap}) = -\sum_{i,j}\left(\mathfrak{c}_{ij} - \gamma \frac{\mathfrak{d}_i \mathfrak{d}_j}{\sum(\mathfrak{c})}\right)\phi_{ij} = -\sum_{i,j}\mathfrak{c}^{\diamond}_{ij}\phi_{ij},$$

where $\mathfrak{c}^{\diamond}_{ij}$ is the degree-corrected version of $\mathfrak{c}_{ij}$. Clearly, therefore, m-umap is just the modularity (**Box 1–2**) with Cauchy similarity and a resolution parameter $\gamma$ (Reichardt and Bornholdt, 2006) that corresponds exactly to the so-called negative-sampling rate of UMAP (Damrich and Hamprecht, 2021). Here, we set $\gamma = 1$ to preserve the alignment of degree correction with global residualization (**Table 2**). In the Methods section, we discuss cases when we may need to increase $\gamma$.

The spherically embedded m-umap forces each row vector $\mathbf{u}_{i:}$ to have unit-norm and implies that $\left\|\mathbf{u}_{i:} - \mathbf{u}_{j:}\right\|^2 = 2(1 - \mathbf{u}_{i:}^\top \mathbf{u}_{j:})$. Here, we set $\alpha = \beta = 1$ and thus simplify

$$\phi_{ij} = \left(1 + 2(1 - \mathbf{u}_{i:}^\top \mathbf{u}_{j:})\right)^{-1} = \left(1 + \left\|\mathbf{u}_{i:} - \mathbf{u}_{j:}\right\|^2\right)^{-1}.$$

Finally, under the additional assumptions of binary $\mathbf{u}_{i:}$ we have $\phi_{ij} = 1$ when $\mathbf{u}_{i:} = \mathbf{u}_{j:}$ and some constant $0 < \phi_{ij} < 1$ otherwise. In the Methods section, we show that this additional constraint reduces the m-umap objective to the standard modularity.

**m-umap optimization and performance**

We developed a simple algorithm to optimize m-umap. First, we optimized the binary m-umap via modularity maximization of the symmetric $\kappa$-nearest-neighbor network. Next, we used the resulting module-indicator matrix to initialize the continuous embedding. Finally, we optimized the continuous embedding directly on the sphere using manifold optimization methods (Boumal et al., 2014). **Box 8** describes the details of our method, as well as its main advantages.

**Box 8. m-umap optimization via modularity maximization.**

We used the equivalence between binary m-umap and the modularity to aid the optimization and interpretability of m-umap.

First, the common use of UMAP to evaluate existing clusters, or detect new clusters (Healy and McInnes, 2024) can be problematic due to the intrinsic distortions of this algorithm (Chari and Pachter, 2023). By contrast, the equivalence of binary m-umap and modularity makes the interpretation of clusters more direct by disconnecting clustering from the continuous embedding.

Second, the $k \times k$ module-connectivity matrix can be used to effectively initialize the continuous embedding by placing modules near-uniformly on a sphere (Methods). In our experiments, this starting point sped up convergence by a factor of ~2 and produced similar embeddings to more standard initializations (Kobak and Linderman, 2021).

Third, a major bottleneck in the optimization of UMAP, and related objectives such as $t$-SNE (van der Maaten and Hinton, 2008), is the need to compute all pairwise forces at every step of the optimization. This computation becomes intractable in large datasets. In sparse networks, the vast majority of pairwise similarities are repulsive, and algorithms commonly approximate these repulsive forces with a variety of methods (McInnes et al., 2018a; van der Maaten, 2014). Here, we use the binary m-umap as an intuitive framework for such approximations. Specifically, we computed the exact attractive and repulsive forces of each node to all nodes in its module and approximated the remaining forces using node-to-module degrees and module centers of mass (Methods). This simple approximation sped up our computations roughly $k$-fold and allowed us to optimize m-umap in about a minute for networks of tens of thousands of nodes, on a modern graphics processing unit. For much larger networks, more speed-ups will likely be required.

In practice, Louvain and related methods are fast, especially in sparse networks (Traag et al., 2019), and thus contribute relatively little to m-umap computation. The methods are stochastic, but we obtained relatively stable partitions by choosing the best result from 100 replicates. We also note that apart from this minor stochasticity, our algorithm is fully deterministic, unlike most UMAP optimizations.

We next used m-umap and UMAP to derive cartographic representations of the 59,412 vertex-level correlation network, averaged over all the data in our sample. Despite the general popularity of UMAP, we could not find a published example of such a representation in the imaging-neuroscience literature. This hints perhaps at a basic challenge of using UMAP on the smooth and sluggish functional MRI data. We have sought to partly overcome this challenge by limiting each brain region to provide at most one nearest neighbor, and by experimenting with UMAP parameters (Methods). Despite these efforts, however, UMAP generally produced cloud representations with little discernible intrinsic morphology (**Figure 5A–E**).

By contrast, m-umap produced much better embeddings of these data (**Figure 5F–K**). It clearly separated neurobiological modules that were derived with modularity maximization of symmetric $\kappa$-nearest-neighbor networks (binary m-umap, cf. **Figure 4**). It also placed these modules on a neurobiological gradient between primary and association cortical areas (**Figure 5K**, inset). Quantitatively, m-umap produced embeddings with ~8% higher fidelity than UMAP (**Figure 5A–J**), although it is unlikely that our metric of fidelity captures the profound visual differences between the two embeddings. Together, these results suggest that m-umap's mix of discrete module detection and continuous embedding allows it to successfully capture the similar mix of discrete and continuous aspects of cortical organization.

Finally, and for completeness, we also compared the performance on m-umap and UMAP on a well-known benchmark MNIST dataset of images of handwritten digits (**Figure S3**). UMAP performs well on this dataset by clearly separating distinct examples of the 10 digits into low-dimensional clusters. By contrast, while m-umap still accurately embeds its binary modules, these modules subtly differed from the ground-truth labels and led to less accurate embeddings, with ~11% decrease in fidelity relative to UMAP. This example illustrates that m-umap embeddings are tied to m-umap modules and may be less accurate when the modules do not recover the ground truth.

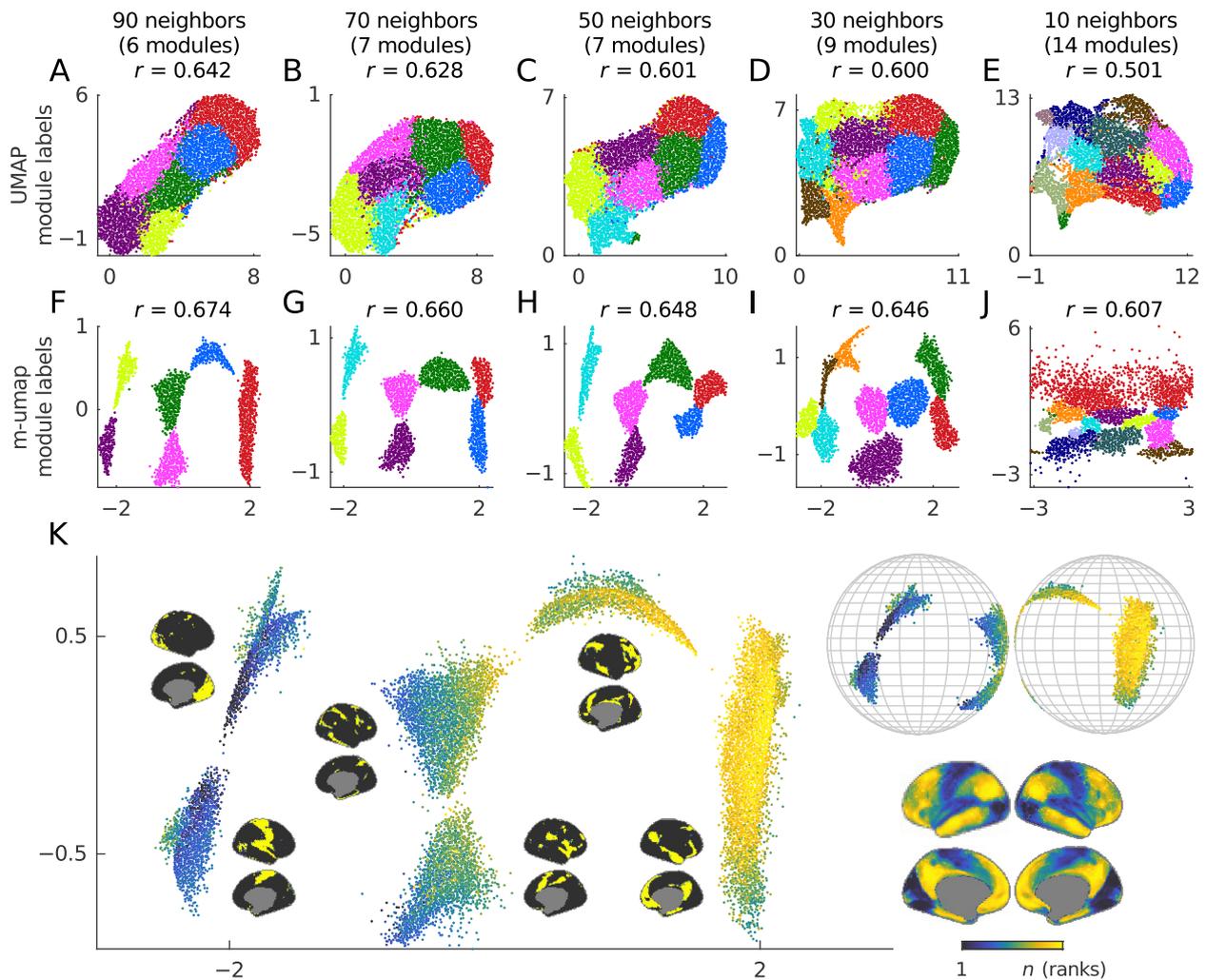

**Figure 5. UMAP and m-umap performance on highly resolved correlation networks.**

**A–J.** UMAP and m-umap embeddings constructed from the symmetric $\kappa$-nearest-neighbor network of the 59,412 cortical-vertex correlation matrix, estimated as an average over all the data. To counter the somewhat artifactual dominance of spatially close correlations in these networks, we imposed a limit of at most one nearest neighbor for each of the 360 brain regions. Columns show results across a range of $\kappa$-nearest-neighbor values. Colors in all panels denote module identities estimated with binary m-umap (modularity maximization). **Row 1 (A–E).** Two-dimensional UMAP embeddings. **Row 2 (F–J).** Mercator ("classic map") projection of spherical m-umap embeddings onto a plane. $r$ values represent Pearson correlation coefficients between distances of nodes to module centroids, in native and embedding space.

**K.** A detailed view of the m-umap embedding from panel **F**. Brain maps next to each module in the main panel show the binary representations of that module. The top right corner shows two-hemisphere views of the spherical embedding. All data points are colored by rank-transformed values of the second component — the leading component after first-component removal — of the symmetric $\kappa$-nearest-neighbor network (bottom right corner).

# Part 2. Network centrality and dynamics

This part of the study, and the next one, mostly equate basic statistics of network organization with popular, but often speculative, theoretical analyses of network dynamics. An overarching theme of both parts is the need to avoid redundant theoretical explanations, irrespective of any conceptual merit these explanations entail. We recently made this point in depth (Rubinov, 2023) and now describe analytical examples of it in network science and network neuroscience.

In this part, specifically, we equate basic measures of connectional magnitude and dispersion with six measures of communication, control, and diversity. Since many of these measures assume that activity propagates on physical networks, we primarily study them on structural networks (rather than correlation networks).

**Degree ≅ eigenvector centrality ≅ diffusion efficiency**

Our discussion in **Part 1** equated the degree with the first component. Here, we consider this equivalence from a complementary perspective. Specifically, we approximately equate the degree with two theoretical measures of network communication, the eigenvector centrality and the diffusion efficiency. Both of these measures assume that activity diffuses on network connectivity (such that the spread of activity from one node to another depends, at any point in time, only on the relative weight of the connection between these nodes). We now consider each of these measures in turn.

*Eigenvector centrality.* Eigenvector centrality is a popular measure of nodal centrality (Borgatti, 2005). The measure has a common self-referential interpretation: a node with a high eigenvector centrality is likely to be strongly connected to other nodes with high eigenvector centrality. This interpretation is intriguing but also somewhat opaque. Here, by contrast, we focus on a more transparent definition that assumes diffusion dynamics. Under this assumption, nodes with high eigenvector centrality are more likely to be reached by other nodes through diffusion dynamics.

In practice, the eigenvector centrality is simply the first component (leading eigenvector) of the network, and thus a first-order approximation of the degree (**Box 9**). In our above discussion of global residualization, we noted that this approximation is usually accurate but can break down in sparse or heterogeneous networks (**Figure S1**). In our example structural networks, we find that the degree and the eigenvector centrality have correlations of 0.922 (0.921–0.924) (**Figure 6E**).

*Diffusion efficiency.* Diffusion efficiency is a related measure of centrality that assumes communication through random walks, a specific type of diffusion that conserves the transition probabilities from a node to all other nodes. Formally, diffusion efficiency is defined as the inverse of the so-called mean first passage time (Grinstead and Snell, 1998), or the mean number of steps it takes to reach one node from another through random walks. Nodes with high diffusion efficiency are therefore easy to reach from other nodes through random walks.

In practice, we can show that the inverse of the degree is approximately equivalent to the mean first passage time (**Box 9**). In the Methods section, we describe that this approximation is accurate in networks with few single-neighbor nodes. From this equivalence, we can deduce that the diffusion efficiency is approximately equivalent to the degree. Accordingly, in our example data (**Figure 6G**), we find that the degree and the diffusion efficiency have a correlation of 0.961 (0.960–0.962).

Together, these approximate equivalences show that diffusion-based interpretations of network dynamics are often redundant with basic network statistics.

**Box 9. Degree $\cong$ eigenvector centrality $\cong$ diffusion efficiency.**

*Structural network.* We denote a structural connectivity matrix by $\mathbf{W}$ and its elements by $w_{ij}$. In the following discussion, we assume that this matrix is non-negative, weighted, symmetric, and connected. We denote the eigendecomposition of this matrix by

$$\mathbf{W} = \sum_{i=1}^{n} \lambda_i \mathbf{v}_i \mathbf{v}_i^\top.$$

*Degree.* We define the (structural) degree as

$$\mathbf{s} = \mathbf{W}\mathbf{1} = \sum_{i=1}^{n} \lambda_i \mathbf{v}_i \mathbf{v}_i^\top \mathbf{1}.$$

*Eigenvector centrality.* The eigenvector centrality is simply the leading eigenvector or first component $\mathbf{v}_1$. This quantity is therefore a first-order approximation of the degree. The Methods section describes the accuracy of this approximation when it equates the degree to the first component in **Part 1**.

*Diffusion efficiency.* The diffusion efficiency of node $j$ is given by $\sum_{j=1}^{n}(\mathrm{MFPT})_{ij}^{-1}$ where $(\mathrm{MFPT})_{ij}$ is the mean first passage time between distinct nodes $i$ and $j$. We can formally define $(\mathrm{MFPT})_{ij}$ as

$$(\mathrm{MFPT})_{ij} \equiv \frac{1}{s_j}\left(1 + \sum_{\tau=1}^{\infty}\left[(\mathbf{Y}^\tau)_{jj} - (\mathbf{Y}^\tau)_{ij}\right]\right),$$

where $(\mathbf{Y}^\tau)_{ij}$ is the $i,j^{\mathrm{th}}$ element of the transition-probability matrix $\mathbf{Y} = \mathrm{diag}(\mathbf{s})^{-1}\mathbf{W}$ raised to the $\tau$-th power (and thus modeling a random walk of length $\tau$).

In the Methods section, we show that, under the assumption of $w_{ij} \ll s_j$, the contribution of the sum term in the above equation is negligible, and the expression thus approximately reduces to the inverse of the degree. From this, we can conclude that the diffusion efficiency, or the inverse of the mean first passage time over columns, is approximately equivalent to the degree.

**Second degree $\cong$ communicability $\cong$ average controllability $\cong$ modal controllability**

We now describe the second degree, a complementary measure to the degree that we define as the sum of squared nodal connection weights. The second degree reflects both the magnitude and the dispersion of nodal connection weights. In practice, the degree and the second degree are equivalent to (respectively) the first and second raw moments of nodal connectivity and thus capture complementary connectional statistics. **Figure 6A–B** shows the maps of the first and second degree, while **Figure 6C** shows the relationship between these two metrics in our example data.

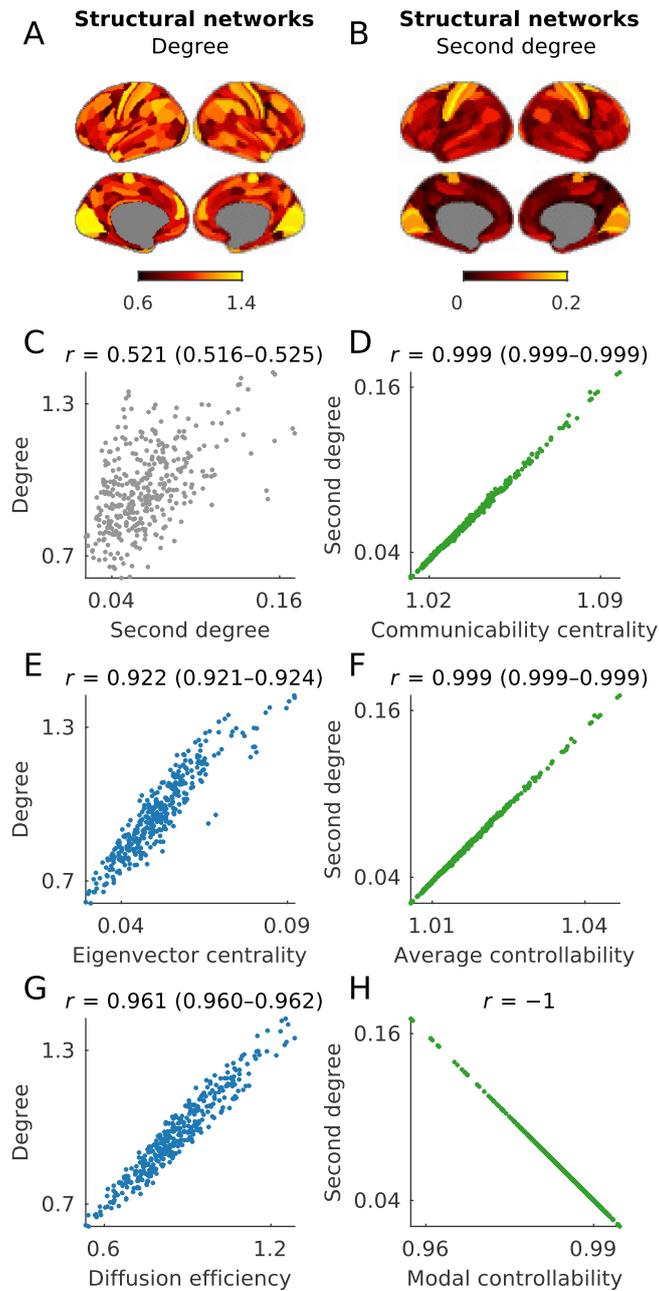

**Figure 6. Degree and second degree ≅ measures of communication and control.**

**A–B.** Maps of the degree and the second degree of structural networks.

**C.** Scatter plot of the degree and the second degree.

**Column 1 (E, G).** Scatter plots of the degree, the eigenvector centrality, and the diffusion efficiency.

**Column 2 (D, F, H).** Scatter plots of the second degree, the communicability centrality, the average controllability, and the modal controllability.

In this section, we show that the second degree is exactly or approximately equivalent to three popular measures of communication and control. Much like the measures of the last section, these measures assume diffusion dynamics. Unlike the measures of the last section, these measures also consider the number and length of all possible walks between two pairs of nodes and thus, in principle, provide more sophisticated estimates of network communication. We now describe each of these measures in turn.

*Communicability and average controllability.* Communicability centrality (also known as subgraph centrality) is a popular measure of network communication (Estrada et al., 2012). The measure is defined as a weighted average over all possible walks from a node to itself, with progressively less weighting of longer walks. Nodes with high communicability centrality are thus likely to be connected to the rest of the network by many and/or short walks, and can, in theory, efficiently communicate with the rest of the network by this virtue.

Separately from communicability, measures of network controllability quantify the theoretical propensity of specific, externally stimulated, nodes to facilitate transitions between activity states, under the assumption of linear and time-invariant dynamics (Pasqualetti et al., 2014). In this subsection, we specifically consider the measure of average controllability. Under standard assumptions, nodes with high average controllability can, in theory, facilitate transitions to activity states with relatively little external stimulation (Gu et al., 2015).

Despite these outwardly distinct motivations and interpretations, in practice, both measures are approximately equivalent to the second degree (**Box 10**). In the Methods section, we show that this approximation is generally highly accurate and, in the case of average controllability, can be made arbitrarily accurate by tuning a free parameter. Accordingly, in our example data (**Figure 6D, F**), we find that the second degree has correlations of 0.999 (0.999–0.999) with both communicability centrality and average controllability.

*Modal controllability.* Modal controllability is a complementary measure that quantifies the theoretical propensity by which specific, externally stimulated nodes can help move systems to difficult-to-reach activity states (Gu et al., 2015; Pasqualetti et al., 2014). The measure is defined in terms of the eigenvectors and eigenvalues of the structural network and has an outwardly different formulation from the average controllability. Despite these differences, in the Methods section we show that modal controllability is exactly equivalent to the second degree (**Figure 6H**) and is therefore approximately equivalent to both communicability centrality and average controllability. These relationships imply that the average controllability and modal controllability can be made arbitrarily strongly correlated (up to sign) by tuning a free parameter.

**Box 10. Second degree $\cong$ communicability $\cong$ average controllability $\cong$ modal controllability.**

In this study, we assume for convenience (and without loss of generality) that all structural connectivity matrices are normalized to have a leading eigenvalue of 1.

*Second degree.* We define the second degree of node *i* as

$$\mathbb{s}_i = \mathbf{w}_i^\top \mathbf{w}_i = \sum_{j=1}^{n} w_{ij}^2,$$

where $\mathbf{w}_i$ denotes the $i^{\text{th}}$ column of $\mathbf{W}$.

*Communicability centrality.* The communicability centrality is defined as the main diagonal of the so-called communicability matrix,

$$\sum_{\tau=0}^{\infty} \frac{1}{\tau!} \mathbf{W}^{\tau}.$$

The second-order approximation of this matrix is given by $\mathbf{I} + \mathbf{W} + \mathbf{W}^2/2$, where $\mathbf{I}$ is the identity matrix.

*Average controllability.* Correspondingly, the average controllability is defined as the main diagonal of the controllability Gramian matrix,

$$\sum_{\tau=0}^{\infty} \left(\frac{1}{\gamma}\mathbf{W}\right)^{\tau} \mathbf{B}\mathbf{B}^{\top} \left(\frac{1}{\gamma}\mathbf{W}^{\top}\right)^{\tau},$$

where, $\gamma > \lambda_1$ is a free normalization parameter that enables convergence, while $\mathbf{B}$ is a matrix that encodes the structure of external stimulation. Under the standard choices of $\gamma = 1$ and and $\mathbf{B} = \mathbf{I}$, the second-order approximation of this matrix is given by $\mathbf{I} + \mathbf{W}^2/2$.

It follows that, for some node $i$, the second-order approximations of communicability centrality and average controllability both reduce to $\left(1 + \mathbf{w}_i^{\top}\mathbf{w}_i/2\right)$ and are, in other words, equivalent to the second degree. In the Methods section, we show that both approximations are highly accurate and, in the case of average controllability, can be made arbitrarily accurate by choosing a sufficiently large $\gamma$.

*Modal controllability.* The modal controllability of node $i$ is defined as

$$\sum_{j=1}^{n} v_{ij}^2 (1 - \lambda_j^2)$$

where $v_{ij}$ is the $j^{\text{th}}$ element of the eigenvector $\mathbf{v}_i$ and $\lambda_j$ is the $j^{\text{th}}$ eigenvalue of $\mathbf{W}$.

In the Methods section, we first simplify this expression to $1 - \sum_{j=1}^{n} v_{ij}^2 \lambda_j^2$ and then show that the second degree of node $i$ can be expressed as $\sum_{j=1}^{n} v_{ij}^2 \lambda_j^2$. Together, these derivations establish the exact equivalence, up to sign, of the modal controllability with the second degree.

### Squared coefficient of variation $\cong$ $k$-participation coefficient

The squared coefficient of variation is a basic measure of normalized dispersion. It is defined as the ratio of the variance over the squared mean, or equivalently, the ratio of the first and second degrees:

$$(\text{coefficient of variation})^2 = \frac{(\text{variance})}{(\text{mean})^2} \equiv \frac{(\text{second degree})}{(\text{degree})^2}.$$

In this section, we show that in some types of networks, the squared coefficient of variation is equivalent to a variant of the participation coefficient, a popular module-based measure of connectional diversity (Guimerà and Amaral, 2005). The participation coefficient is also known as the Gini-Simpson diversity index (Caso and Angeles Gil, 1988; Simpson, 1949), and is itself a first-order approximation of the Shannon entropy of node-module connectivity (Cajic et al., 2024). Nodes with low participation coefficients have most of their connections in one module, while nodes with high participation coefficients have similarly strong connections to many modules.

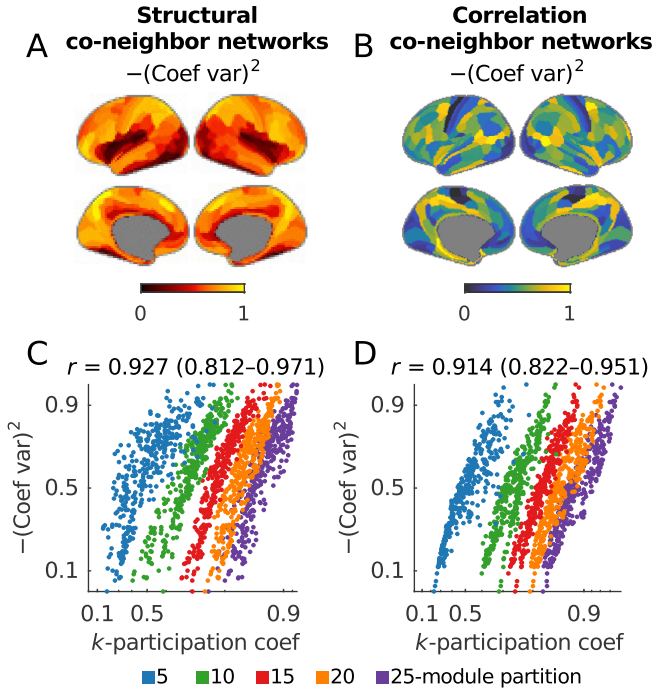

**Figure 7. Squared coefficient of variation $\cong$ k-participation coefficient.**
**A–B.** Maps of the negative squared coefficient of variation of structural and correlation co-neighbor networks (rescaled to the range [0,1]). **C–D.** Scatter plot of the *k*-participation coefficient, and the negative coefficient of variation (rescaled as in panels **A–B**), plotted for five module partitions.

In parallel with our discussion of the *k*-modularity, we focus on the *k*-participation coefficient, or, in other words, on participation coefficient normalized by module size. We show that the *k*-participation coefficient is approximately equivalent to the squared coefficient of variation (**Box 11** and Methods), in cases when network nodes have relatively homogeneous connections within modules. In practice, we often find such homogeneous connectivity in networks with transitive properties, such as correlation networks. Separately, these equivalences emerge with an increase in the number of modules, as node-to-module connectivity automatically becomes more homogeneous.

Here, we illustrate these equivalences in structural and correlation co-neighbor networks, which have high connectional homogeneity by construction. **Figure 7** shows that the squared coefficient and the *k*-participation coefficient in these networks have respective correlations of 0.927 (0.812–0.971) and 0.914 (0.822–0.951) across five module partitions. In line with our considerations, these relationships become stronger with more highly resolved module partitions.

**Box 11. Squared coefficient of variation $\cong$ *k*-participation coefficient.**

*Participation coefficient.* We define the *k*-participation coefficient as

$$(k\text{-participation coefficient})_i = 1 - \sum_{h=1}^{k} \frac{1}{N_h}\left(\frac{s_{i(h)}}{s_i}\right)^2,$$

where $s_i$ is the node degree and $s_{i(h)}$ is the degree of node $i$ to module $h$. Under the assumption that the connection weights of node $i$ to module $h$ are relatively homogeneous, we can accurately approximate the above expression as

$$(k\text{-participation coefficient})_i \approx 1 - \frac{\sum_{i=1}^{n} w_{ij}^2}{s_i^2}.$$

Separately, we can express the squared coefficient of variation as

$$(\text{coefficient of variation})_i^2 \equiv \sum_{j=1}^{n} \frac{w_{ij}^2}{s_i^2},$$

which establishes the equivalence up to sign.

## Part 3. Semi-analytical vignettes in imaging and network neuroscience

In this final part of the study, we describe three vignettes that combine analytical derivations with numerical results to probe more intricate relationships between analyses (degree and diffusion-map embeddings), datasets (structural and proximity networks), and representations (average correlations and dynamic affinities). We begin this section by describing a simple interpretation of the primary co-activity gradient. We continue by reducing dynamical models of network growth to statistical models of network proximity. We conclude by equating average node-module correlations with dynamic module affinities. This discussion primarily focuses on analyses in imaging and network neuroscience.

**Degree after global-signal regression ≅ primary co-activity gradient**

The primary co-activity gradient plays an especially important role in imaging neuroscience because it represents a transition between primary and association cortical areas and thus forms part of the foundational knowledge of cortical organization (Mesulam, 1998). We already showed variants of this gradient across several analyses in **Part 1** (**Figure 4A–B**, **Figures 3B–F**; **Figures 5K**).

This section describes a particularly simple, approximately equivalent definition of this gradient. This definition relies on a specific property of functional MRI data and may not generalize to other datasets.

We show this equivalence in three steps. First, we note that the primary co-activity gradient is approximately equivalent to the second component of the correlation network. Second, we show that the first and second components in functional-MRI correlation networks have a uniquely high Pearson correlation coefficient. This well-known property reflects the propensity of cortical association areas (in the second component) to strongly correlate with the global signal (in the first component). Third, we use this result to show that the degree after global-signal regression is approximately equivalent to the second component and, by extension, to the primary co-activity gradient (**Box 12**).

In line with our observations, **Figure 8** shows a map of the negative degree after global regression and correlations of 0.922 (0.911–0.929) between this degree and the primary co-activity gradient.

We conclude this section by noting a distinction between variants of residual networks that is subtle but important to this vignette. In **Part 1**, we established an approximate equivalence between global-signal regression or first-component removal on the one hand, and degree-correction on the other hand. We built on the near-exact nature of this relationship (**Figure 2B**) to show approximate equivalences between several clustering and dimensionality reduction methods (**Table 2**).

Despite the accuracy of these approximations, the subtle difference between degree correction and global-signal regression (or first-component removal) makes a dramatic difference to this specific vignette. Degree correction subtracts the contribution to the degree of the first and all other components, so that the residual degree of all nodes after this process is 0 by definition. By contrast, global-signal regression (or first-component removal) preserves the contributions of the second and higher-order components, so the residual degrees, while small in absolute terms, are nonzero and drive the relationship with the primary co-activity gradient. A similar difference drives the negative sign of the correlation in **Figure 8** (Methods). This distinction underscores the need for caution in generalizing approximate equivalences across analyses, even in the same dataset.

**Box 12. Degree after global-signal regression $\cong$ primary co-activity gradient.**

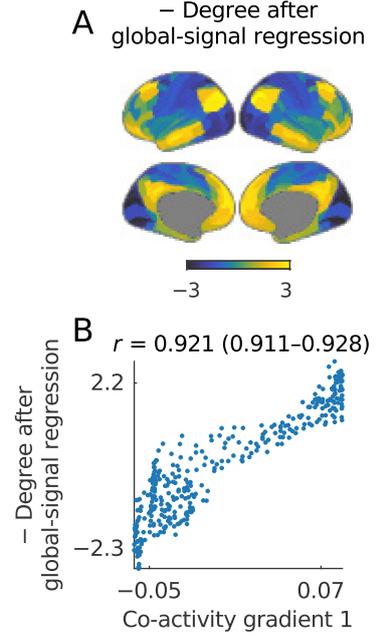

**Figure 8. Degree after global-signal regression $\cong$ primary co-activity gradient.**

**A.** Map of the (negative) degree after global-signal regression in correlation networks.

**B.** Scatter plot of the (negative) degree after global-signal regression and the primary co-activity gradient.

In **Part 1**, we noted that the second co-neighbor eigenvector is approximately equivalent to the first co-activity gradient. Since this eigenvector captures a dominant and largely density-invariant network structure, it is likely to be approximately equivalent to $\mathbf{u}_2$, the second eigenvector of the correlation matrix $\mathbf{C}$. (This relationship is present in our data, as we show in the Methods section.)

It follows that the equivalence in this section rests on establishing a relationship between $\mathbf{u}_2$ and the degree after global-signal regression $\mathbf{d}' = \mathbf{C}'\mathbf{1}$.

In the Methods section, we show that we can express $\mathbf{d}'$ as

$$\mathbf{d}' \cong \sum_{i=2}^{n} \psi_i^2 \text{corr}(\mathbf{u}_1, \mathbf{u}_i)\mathbf{u}_i$$

where $\psi_i$ and $\mathbf{u}_i$ are the eigenvectors and eigenvalues of $\mathbf{C}$. In functional MRI data, we know that $\mathbf{u}_1$ is highly correlated with $\mathbf{u}_2$ even as these vectors are orthogonal by definition, $\mathbf{u}_1^\top \mathbf{u}_2 = 0$.

This correlation is unique, insofar as the correlations of $\mathbf{u}_1$ with other eigenvectors are generally low. These relationships imply that $\text{corr}(\mathbf{u}_1, \mathbf{u}_2) \gg \text{corr}(\mathbf{u}_1, \mathbf{u}_{i>1})$ which, together with $\psi_2 > \psi_{i>2}$, establishes that $\mathbf{d}' \cong \mathbf{u}_2$, and therefore the equivalence in the main text.

**Shrunken proximity matrix ⇒ structural connectivity network**

Our second vignette seeks to simplify popular models of network growth in network neuroscience. We specifically consider a class of studies that model network growth as a function of spatial proximity and connectional similarity (Akarca et al., 2021; Betzel et al., 2016; Oldham et al., 2022; Vértes et al., 2012). We first describe an analytically tractable variant of these models and then show that the assumption of network growth in this variant can be subsumed by a type of shrinkage, or the weakening of dominant structural patterns in the network (**Figure 9A–B**).

The specific models we consider typically start with a sparse binary "seed" network of strong, and usually empirical, connections. The models then proceed to simulate growth by adding connections to this network in discrete time steps. Each step first defines a pairwise probability of connection formation as a function of physical proximity and connectional similarity (essentially, the number of co-neighbors). It then forms a connection between a pair of unconnected nodes with this probability. This process repeats until the number of connections matches the number of empirical network connections.

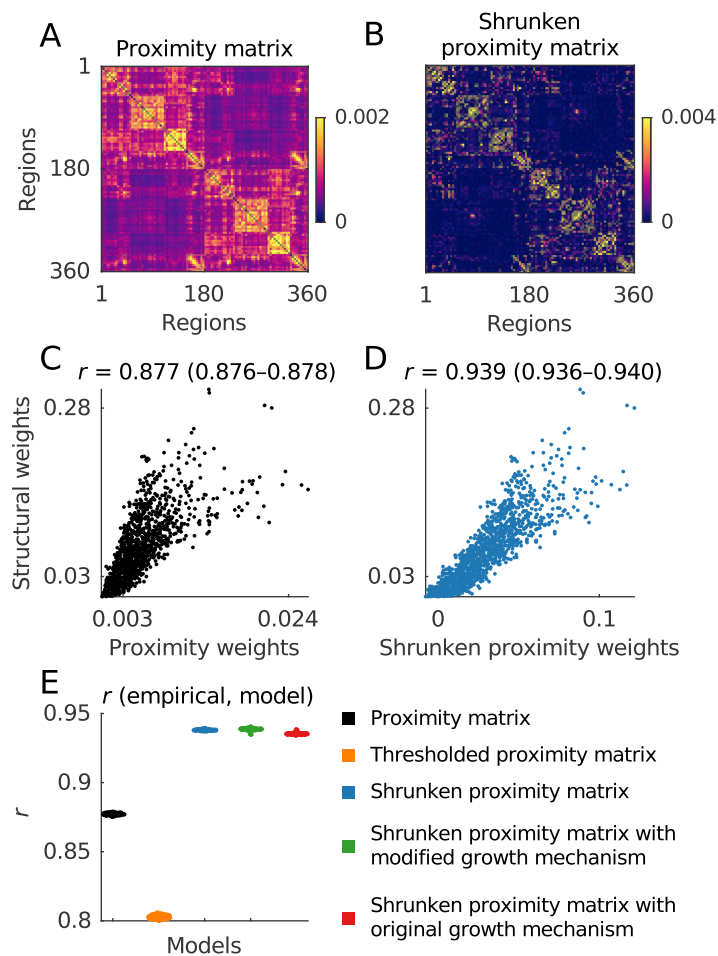

**Figure 9. Shrunken proximity matrix ⇒ structural connectivity network.**

**A–B.** Proximity matrices before and after shrinkage, ordered as in Figure 1.

**C–D.** Scatter plots of proximity weights (before and after shrinkage) and structural connectivity weights.

**E.** Swarm plots of correlations between statistical and growth models computed on individual networks. The black and blue swarms show the performance of the statistical models in **A–D**. The green and red swarms show the performance of the modified and original growth models. For completeness, the orange swarm shows the performance of a proximity matrix, thresholded to preserve the strongest 1% of weights (other thresholds resulted in similar performance). See the main text and the Methods section for additional details on definitions of models.

The standard formulation of these models is hard to treat analytically, so here we study a modified but broadly faithful variant (**Box 13**). Our analysis of this variant shows the following. First, we find that, in line with extensive previous work (Bullmore and Sporns, 2012; Markov et al., 2013), proximity matrices accurately approximate structural networks. In our example data, we find correlation coefficients of 0.877 (0.876–0.878) for this relationship (**Figure 9C**). Next, we sparsify these matrices by weakening the contribution of their first several components. This weakening is formally known as shrinkage, and is commonly used to clean covariance or correlation matrices across diverse fields, from economics to biology and neuroscience (Honnorat and Habes, 2022; Ledoit and Wolf, 2003; Schäfer and Strimmer, 2005). In our analyses, we find that the shrinkage of the proximity network raises its correlation with the structural network to 0.939 (0.936–0.940) (**Figure 9D**). Finally, we show that the inclusion of growth mechanisms does not further increase the correlation above this baseline (**Figure 9E**).

**Box 13. Shrunken proximity matrix ⇒ structural connectivity network.**

The original formulation of network growth models defines a binary structural network $\widetilde{\mathbf{W}}$ at time step $\tau + 1$ by

$$\widetilde{\mathbf{W}}_{\tau+1} = f(\mathbf{\Pi}_\tau) + \widetilde{\mathbf{W}}_\tau.$$

Here, the function $f$ generates a new binary connection for a disconnected pair of distinct nodes from the probability matrix

$$\mathbf{\Pi}_\tau \propto \mathbf{\Phi}^{\circ\alpha} \odot \mathcal{W}_\tau^{\circ\beta},$$

where ∘ denotes elementwise power, ⊙ denotes an elementwise product, $\mathbf{\Phi}^{\circ\alpha}$ is a matrix of physical proximities, $\mathcal{W}_\tau^{\circ\beta}$ is a co-neighbor matrix at time $\tau$, and $\alpha$ and $\beta$ are parameters. Note that, unlike our preceding co-neighbor matrices that were computed from the $\kappa$-nearest neighbor networks, this matrix is computed for all node neighbors, but at a specific time point $\tau$.

In this study, we work with an analytically tractable variant of this growth model. This variant simulates growth through the strengthening of weighted connections, rather than the addition of binary connections. We specifically consider a weighted structural network $\widetilde{\mathbf{W}}$ at time step $\tau$ by

$$\widehat{\mathbf{W}}_{\tau+1} = \gamma\left[\rho\widehat{\mathbf{W}}_\tau^2 + (1-\rho)\mathbf{\Phi}^{\circ\alpha}\right] + \widehat{\mathbf{W}}_\tau,$$

where $\alpha$, $\gamma$ and $\rho$ are parameters.

In the Methods section, we semi-analytically motivate a shrinkage of the proximity matrix $\mathbf{\Phi}^{\circ\alpha}$. We show that this shrinkage raises the correlation of $\widehat{\mathbf{W}}$ with $\mathbf{W}$, and show that the inclusion of additional network growth mechanisms does not provide additional improvements over this basic transformation.

**Node-module correlations ⇒ node-module dynamical affinities**

The somewhat idiosyncratic nature of analytical derivations can make it hard to systematically equate or unify arbitrary analyses. In this section, we show how the combination of analytical motivations with numerical results can extend our approach to many more analyses than would be possible with analytical derivations alone.

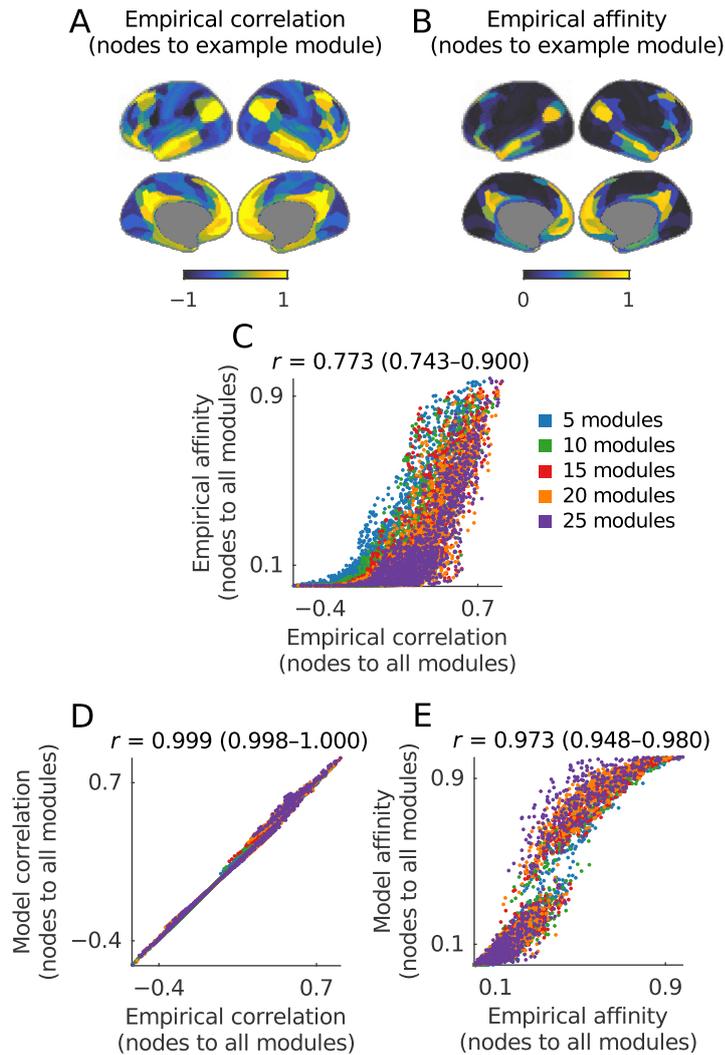

**Figure 10. Node-module correlation ⇒ node-module dynamical affinity.**

**A.** Map of average node correlations to an example module from a five-module partition.

**B.** Corresponding map of average dynamic node affinities to the module in **A**. Values denote the fraction of times a node was placed within the module during ~2-minute (160-frame) recording time window (the results were robust to the choice of 1–5-minute windows).

**C.** Scatter plot of average node correlation and dynamic affinities across all five module partitions.

**D.** Scatter plot of empirical and model average correlations of nodes to modules (validation of the modeling constraints).

**E.** Scatter plot of dynamic module affinities in empirical and model data across all five module partitions.

---

This vignette specifically establishes a relationship between the correlation of nodes to modules averaged over the full time series and the dynamic placement of nodes in modules at specific time windows. In network neuroscience, a popular variant of this latter measure is known as flexibility or network switching (Bassett et al., 2011; Pedersen et al., 2018).

We start with a simple analytical observation: nodes with similarly strong correlations to different module centroids are more likely to switch between these modules (**Box 14**). This observation is intuitive: for example, **Figure 10A** shows average correlations between all nodes and a single module, while **Figure 10B** shows the corresponding fraction of times nodes were placed in this module across time windows. **Figure 10C** shows a moderately strong correlation of 0.773 (0.743–0.900) between these quantities. Nonetheless, the inherent nonlinearities of dynamic module detection make it difficult to establish a fully analytical basis for this relationship.

Despite this difficulty, we can build on our analytical foundation to test this relationship numerically. Here, we do so by adopting an algorithm for generating time series with constrained correlations of nodes to module-like centroids (Nanda et al., 2023; Nanda and Rubinov, 2023). The approach allows us to directly test if average correlations between nodes and modules are sufficient to determine the time-dependent dynamic affinity of nodes for modules (**Figure 10D**). Our numerical results show this to be generally the case, with correlations of 0.973 (0.948–0.980) between dynamical module affinities in empirical and model data (**Figure 10E**). This result illustrates the effectiveness of combining analytical and numerical methods to equate more diverse and less tractable analyses.

**Box 14. Node-module correlation ⇒ node-module dynamical affinity.**

In this vignette, we adopt the $k$-modularity maximization of correlation matrices $\mathbf{C}$. As we have shown above, this objective is approximately equivalent to $k$-means clustering of correlation matrices after global-signal regression $\mathbf{C}'$.

It is easy to show that the difference in $k$-modularity after moving node $i$ from one module $g$ to another module $h$, is approximately equivalent to

$$\Delta(k\text{-modularity})_{i:g \to h} \approx \text{corr}(\mathbf{x}_i, \bar{\mathbf{x}}_h) - \text{corr}(\mathbf{x}_i, \bar{\mathbf{x}}_g).$$

where $\mathbf{x}_i$ is the time series of node $i$ and $\bar{\mathbf{x}}_h$ is the mean activity of module $h$. It follows that, without additional assumptions, nodes that have similar within-module correlations are more likely to switch between modules across time windows.

Here, we used numerical methods to generate synthetic data with approximately preserved correlations of individual nodes to module centroids and, as a consequence, with effectively constrained $\text{corr}(\mathbf{x}_i, \bar{\mathbf{x}}_h)$ for all nodes and module combinations. Analysis of dynamic switching in these data allowed us to assess the dependence of this switching on our analytically motivated constraints.

## abct

To accompany our main results, we provide *abct*, a MATLAB and Python toolbox that implements our main analyses (github.com/mikarubi/abct/). The toolbox includes the Loyvain and co-Loyvain methods (for $k$-means, $k$-modularity, or spectral clustering of data or network inputs), three variants of global residualization, binary and weighted canonical covariance analysis, co-activity gradient detection, and m-umap optimization. It also includes computation of degrees (first, second, and residual), squared coefficient of variation, as well as our network shrinkage method. Finally, the toolbox shows our analyses on example data that can be run online without a local computational environment. Overall, we hope that these tools will make it easier for people to use and build on our results.

## Discussion

We made three general contributions to scientific integration. First, we showed equivalences between variants of several foundational objectives. We also developed fused methods that outperformed standard methods on example brain-imaging data. Second, and relatedly, we described equivalent simplifications of popular analyses across unsupervised learning and imaging neuroscience. Third, we combined analytical and semi-analytical approaches to equate speculative interpretations of network dynamics with basic statistical properties of network structure. We conclude by discussing some overarching themes of these results.

We showed that many methods for clustering and dimensionality reduction blend into each other through changes in data transformations (residual, co-neighbor, and $\kappa$-nearest-neighbor networks), and feature definitions (normalization, binarization, and choice of similarity, **Table 2**). Many studies that unified methods in unsupervised learning made this point before, but usually from the perspective of prediction or utility (Alshammari et al., 2025; Damrich et al., 2023; Dhillon et al., 2004; Ding and He, 2004; Ham et al., 2004; Ravuri and Lawrence, 2024; Williams, 2024). These studies have shown, for example, that a judicious combination of individual constraints can lead to more useful methods, as we have also found to some extent.

In contrast to these studies, however, we were primarily motivated by scientific integration and thus approached unification from the perspective of explanation or truth (relative trueness). From this perspective, the choice of a constraint should primarily rest on its veracity rather than utility (Peel et al., 2022; Reid et al., 2019; Rubinov, 2023). In this way, our results imply that a principled motivation for choosing constraints is more important than the choice of outwardly distinct but inwardly equivalent optimization methods. Gould and Lewontin (1979) made a related point from a different perspective of evolutionary biology, "organisms must be analysed as integrated wholes, with [body plan] so constrained […] that the constraints themselves become more interesting and more important […] than the selective force" of optimization.

Our focus on constraints helps to clarify discussions and debates across unsupervised learning, network science, and imaging neuroscience. We give three examples below:

1. Scientists across fields have ongoing debates about the need for global residualization, whether as first-component removal (Chen, 2020; Goh et al., 2017), degree correction (Lambiotte et al., 2014; Traag et al., 2011), or global-signal regression (Murphy and Fox, 2017). These debates usually center on competing views of the relevance or utility of the global component across fields. Views that this component is relevant or useful support its preservation, while opposing views support its removal. Here, we do not adopt a position in these debates, but note that specific transformations of the data — for example, through $\kappa$-nearest neighbor, co-neighbor, or related similarities — lead to trivial global components that are removed uncontroversially (Von Luxburg, 2007). In this way, we suggest that some of these debates can more fruitfully focus on the veracity of more testable data transformations, rather than on the more ambiguous structure of global components.
2. Network scientists have criticized the modularity from several perspectives, including the failure of its null model, its resolution limit (the inability to detect small modules), and its narrow focus on assortative structure (the structure of densely-connected modules) (Fortunato and Barthélemy, 2007; Guimerà et al., 2004; Lancichinetti and Fortunato, 2011; Peixoto, 2023; Traag et al., 2011). Here, we have shown that normalized modularity maximization has no null model or resolution limit and, with specific data transforms, can detect diverse network structures. Perhaps more importantly, we reduced the maximization of this normalized modularity to the detection of binary components with residualization. In this way, we have shown that the normalized modularity is a binary variant of the main workhorse of unsupervised learning (**Table 2**). These results again suggest that a focus on data transforms or feature constraints — rather than specific optimization objectives — can help fields more fruitfully converge on unified low-dimensional representations.

3. Imaging neuroscientists have engaged in a prominent debate about the primacy of binary components ("networks") or weighted components ("gradients") as building blocks of cortical organization (Bolt et al., 2022; Damoiseaux et al., 2006; Huntenburg et al., 2018; Margulies et al., 2016; Nanda and Rubinov, 2023; Petersen et al., 2024). We have shown that networks and gradients are equivalent representations of the data, up to a binarization constraint (**Figure 4**, **Table 5**). In this way, our results can reduce this debate to the debate over the neurobiological veracity of binarization (for example, as a developmental mechanism).

From another perspective, our treatment of many unsupervised learning methods as special cases of component detection roughly parallels the treatment of many association measures as special cases of the dot product (Deza and Deza, 2013). We can extend this rough analogy by relating global residualization to mean subtraction, and component-normalization to norm rescaling. These relationships align the modularity with the covariance, the $k$-means objective with the cosine similarity, and the $k$-modularity with the Pearson correlation coefficient. Such alignments can also help motivate future methodological development. For example, they suggest that extensions of the modularity to networks with negative weights should adopt a more principled residualization centered on the removal of components, rather than more ad-hoc extensions of degree-based null models (Gómez et al., 2009; Rubinov and Sporns, 2011; Traag and Bruggeman, 2009). Similarly, the number of removed components can be formalized as a field-specific parameter, and could be 0, 1, or many (often $\geq 10$ in genomics, for example (Price et al., 2006)). Alternatively, component removal can altogether be replaced with shrinkage. Separately, the use of middle-ground constraints, such as non-negativity, can balance richness and interpretability and naturally interpolate between binary and weighted representations in the unified framework of **Table 2** (Ding et al., 2005; Li and Ding, 2013).

The focus on constraints likewise motivates our discussion of equivalences between basic network statistics and intricate network dynamics. One common view in the literature, for example, treats equivalences of this type as analytical insights into important dynamical processes (see Borgatti (2005), Gu et al. (2015), and Zamora-López and Gilson (2024) for important expositions and examples). For such a view to survive stringent scrutiny, however, it must establish the existence and importance of the proposed dynamical processes in individual systems. To give an analogy from statistics, the mean of a data sample is analytically equivalent to the maximum-likelihood estimate of the mean of a normal distribution. This equivalence, however, does not imply that the sample mean gives us any evidence for the importance of the normal distribution to our data. In the case of networks, and without such evidence, we must fall back on the robust existing knowledge that simultaneously explains basic network structure and explains away speculative network dynamics.

We conclude by reiterating the distinction between exact and other equivalences. As we saw above, the assumptions that underpin some approximate or semi-analytical equivalences do not always generalize and must thus be checked in individual datasets. On the other hand, the idiosyncratic nature of individual analytical derivations makes it unlikely that we can establish a fully analytical framework to unify all future results across fields. In this context, the mix of analytical motivation with numerical validation that we adopted at the end of the study can provide a promising way for future integration of diverse analyses into unified frameworks.

# Methods

We define equivalence ($\equiv$) as equality up to addition and multiplication by a constant. Thus $\mathbf{y} \equiv \mathbf{x}$ implies that $\mathbf{y} = \alpha \mathbf{x} + \beta$. It follows that equivalent results have a maximal Pearson correlation coefficient, $r = \mathrm{corr}(\mathbf{x}, \mathbf{y}) = \pm 1$.

## Data matrices and correlation networks

We use $\mathbf{X}$ to represent a $p \times n$ data matrix. The columns of this matrix represent data points, while the rows represent features. Correspondingly, we use $\mathbf{C} = \mathbf{X}^\top \mathbf{X}$ to represent an $n \times n$ correlation network. In this study, $\mathbf{X}$ usually denotes an activity matrix of $n$ nodes and $p$ time points, and $\mathbf{C}$ correspondingly denotes a node-node correlation network. Of course our discussion encompasses other types of data matrices and correlation networks.

We use $\mathbf{x}_i$ to denote the columns of $\mathbf{X}$. In what follows, we linearly scale each $\mathbf{x}_i$ to have a mean of $(2p)^{-1/2}$ and a Euclidean norm of 1, through a transform of the raw data vector $\chi_i$,

$$\mathbf{x}_i = \frac{1}{2^{1/2}}\left(\mathrm{normalize}(\chi_i) + \frac{1}{p^{1/2}}\mathbf{1}\right),$$

where $\mathbf{1}$ is a vector of ones and $\mathrm{normalize}(\chi_i)$ rescales $\chi_i$ to have mean 0 and norm 1.

This transform is convenient because it makes $c_{ij}$ equivalent to the Pearson correlation coefficient of $\mathbf{x}_i$ and $\mathbf{x}_j$, $\mathrm{corr}(\mathbf{x}_i, \mathbf{x}_j)$, but rescales it to the range $[0, 1]$ and thus excludes negative values:

$$c_{ij} = \mathbf{x}_i^\top \mathbf{x}_j = \frac{1}{2}\left(\mathrm{corr}(\mathbf{x}_i, \mathbf{x}_j) + 1\right) \equiv \mathrm{corr}(\mathbf{x}_i, \mathbf{x}_j)$$

In practice, this transform helps us equate degree-based corrections or normalizations, but has no effect on the other equivalences in this study.

## First component $\cong$ degree $\equiv$ correlation with the global signal

We define the eigendecomposition of $\mathbf{C}$ as

$$\mathbf{C} = \mathbf{U}\mathbf{\Psi}\mathbf{U}^\top = \sum_{i=1}^{n} \psi_i \mathbf{u}_i \mathbf{u}_i^\top, \tag{1}$$

where the $n \times n$ matrix $\mathbf{U} = [\mathbf{u}_1, \mathbf{u}_2, \ldots \mathbf{u}_n]$ has the eigenvectors of $\mathbf{C}$ as its columns, while the diagonal $n \times n$ matrix $\mathbf{\Psi} = \mathrm{diag}(\psi_1, \psi_2, \ldots \psi_n)$ has the eigenvalues of $\mathbf{C}$ on its main diagonal. In the following, we assume that $\mathbf{U}$ is orthonormal, such $\mathbf{u}_i^\top \mathbf{u}_i = 1$ and $\mathbf{u}_i^\top \mathbf{u}_{j \neq i} = 0$. We also assume that the eigenvalues in $\mathbf{\Psi}$ satisfy $\psi_1 > \psi_2 > \cdots \psi_n \geq 0$.

We denote the degree of $\mathbf{C}$ as

$$\mathbf{d} = \mathbf{C}\mathbf{1} = \mathbf{U}\mathbf{\Psi}\mathbf{U}^\top \mathbf{1} = \sum_{i=1}^{n} \psi_i \mathbf{u}_i \mathbf{u}_i^\top \mathbf{1} = \sum_{i=1}^{n} \alpha_i \mathbf{u}_i, \tag{2}$$

where $\alpha_i = \psi_i \sum(u_i)$ is the product of $\psi_i$ and $\sum(u_i) = \mathbf{u}_i^\top \mathbf{1}$, the sum of the elements of $\mathbf{u}_i$.

Equation 2 shows that $\mathbf{u}_1$ is a first-order approximation of the degree. This approximation is accurate for many networks of interest. Specifically, the non-negativity of $\mathbf{C}$ allows us to make use of the Perron-Frobenius theorem, a basic result in linear algebra (Pillai et al., 2005). We can use this result to assume that $\mathbf{u}_1$ contains only positive elements. Moreover, since $\mathbf{U}$ is orthonormal, all $\mathbf{u}_{i>1}$ must contain at least one negative element (in order to satisfy $\mathbf{u}_1^\top \mathbf{u}_{i>1} = 0$). It follows that $\mathbf{u}_{i>1}$ will, in most cases, contain a mixture of positive and negative elements, from which we can presume that $|\sum(u_1)| > |\sum(u_{i>1})|$. This, together with knowledge that $|\psi_1| > |\psi_{i>1}|$, will generally imply that $|\alpha_1| \gg |\alpha_{i>1}|$ and therefore suggests that $\mathbf{u}_1 \cong \mathbf{d}$.

Finally, we define the global (mean-activity) signal of $\mathbf{X}$ as

$$\bar{\mathbf{x}} = \frac{1}{n}\mathbf{X}\mathbf{1}.$$

Clearly, the degree is equivalent to the nodal correlation with the global signal (MacMahon and Garlaschelli, 2015). Specifically,

$$\mathbf{d} = \mathbf{C}\mathbf{1} = \mathbf{X}^\top\mathbf{X}\mathbf{1} = \mathbf{X}^\top(\mathbf{X}\mathbf{1}) = n\mathbf{X}^\top\bar{\mathbf{x}}.$$

**First-component removal $\cong$ degree correction $\cong$ global-signal regression**

*First-component removal.* We use $\mathbf{C}^*$ to denote the network after first-component removal,

$$\mathbf{C}^* = \mathbf{C} - \psi_1 \mathbf{u}_1 \mathbf{u}_1^\top.$$

*Degree correction.* We use $\mathbf{C}^\diamond$ to denote the network after degree correction,

$$\mathbf{C}^\diamond = \mathbf{C} - \frac{1}{\sum(c)}\mathbf{d}\mathbf{d}^\top,$$

where $\sum(c)$ is the sum of all elements of $\mathbf{C}$.

Recalling that $\mathbf{u}_1 \cong \mathbf{d}$ and $\mathbf{u}_1^\top \mathbf{u}_1 = 1$, we can write

$$\mathbf{u}_1 \approx \frac{\mathbf{d}}{(\mathbf{d}^\top\mathbf{d})^{1/2}}. \tag{3}$$

Separately, noting that $\psi_1 = \mathbf{u}_1^\top \mathbf{C} \mathbf{u}_1$ by nature of the eigendecomposition (Equation 1), we can write

$$\psi_1 = \mathbf{u}_1^\top \mathbf{C} \mathbf{u}_1 \approx \frac{\mathbf{d}^\top \mathbf{C} \mathbf{d}}{\mathbf{d}^\top \mathbf{d}} = \frac{\mathbf{1}^\top \mathbf{C}^3 \mathbf{1}}{\mathbf{d}^\top \mathbf{d}} = \frac{\sum_{i=1}^n \psi_i^3 \mathbf{1}^\top \mathbf{u}_i \mathbf{u}_i^\top \mathbf{1}}{(\mathbf{d}^\top\mathbf{d})} \approx \frac{\psi_1^3 \mathbf{1}^\top \mathbf{u}_1 \mathbf{u}_1^\top \mathbf{1}}{(\mathbf{d}^\top\mathbf{d})}.$$

where the latter approximation follows the same reasoning we used to link $\mathbf{d}$ with $\mathbf{u}_1$.

This approximation, together with Equation 3, allows us to further approximate $\psi_1$ as

$$\psi_1 \approx \frac{(\mathbf{d}^\top\mathbf{d})^{1/2}}{\mathbf{u}_1^\top \mathbf{1}} = \frac{\mathbf{d}^\top\mathbf{d}}{\sum(c)}, \tag{4}$$

where we made use of $\sum(c) = \mathbf{d}^\top \mathbf{1}$.

We can now use Equations 3 and 4 to establish that

$$\mathbf{C}^* = \mathbf{C} - \psi_1 \mathbf{u}_1 \mathbf{u}_1^\top \approx \mathbf{C} - \frac{1}{\sum(c)}\mathbf{d}\mathbf{d}^\top = \mathbf{C}^\diamond. \tag{5}$$

*Global-signal regression.* We use $\mathbf{X}'$ to denote a matrix of residuals of $\mathbf{X}$ after global-signal regression,

$$\mathbf{X}' = \left[\mathbf{X} - \frac{n^2}{\sum(c)}\bar{\mathbf{x}}(\bar{\mathbf{x}}^\top \mathbf{X})\right]\eta = \left[\mathbf{X} - \frac{n}{\sum(c)}\bar{\mathbf{x}}\mathbf{d}^\top\right]\eta.$$

where the diagonal matrix $\eta$ rescales each column of $\mathbf{X}'$ to have norm 1.

We can use this matrix to define the residual correlation network after global-signal regression,

$$\mathbf{C}' = \mathbf{X}'^\top \mathbf{X}' = \eta\left[\mathbf{X} - \frac{n}{\sum(c)}\bar{\mathbf{x}}\mathbf{d}^\top\right]^\top \left[\mathbf{X} - \frac{n}{\sum(c)}\bar{\mathbf{x}}\mathbf{d}^\top\right]\eta = \eta\left[\mathbf{C} - \frac{1}{\sum(c)}\mathbf{d}\mathbf{d}^\top\right]\eta = \eta\mathbf{C}^\diamond\eta. \quad (6)$$

It follows, therefore, that $\mathbf{C}'$ is equivalent to the degree-corrected network rescaled by $\eta$.

To understand the nature of this rescaling, we first formally define $\eta$ as

$$\eta = \mathrm{diag}\left(\mathbf{1} - \frac{1}{\sum(c)}\mathbf{d}^{\circ 2}\right)^{-1/2}.$$

We can now investigate the contribution of this term to $\mathbf{C}'$, by decomposing the degree as

$$\mathbf{d} = n(\bar{c}\mathbf{1} + \delta),$$

where $\bar{c} = \sum(c)/n^2 \in [0, 1]$ is the mean correlation value and $\delta$ is a vector of deviations of the node-wise mean correlation values around $\bar{c}$. With some algebra, we can simplify $\eta$ as

$$\eta = \mathrm{diag}\left(\mathbf{1} - \frac{n^2}{\sum(c)}[\bar{c}\mathbf{1} + \delta]^{\circ 2}\right)^{-1/2}$$
$$= \mathrm{diag}\left(\mathbf{1} - \frac{1}{\bar{c}}[\bar{c}^2\mathbf{1} + 2\bar{c}\delta + \delta^{\circ 2}]\right)^{-1/2}$$
$$= \mathrm{diag}\left([1-\bar{c}]\mathbf{1} - 2\delta - \frac{1}{\bar{c}}\delta^{\circ 2}\right)^{-1/2}$$
$$= \frac{1}{(1-\bar{c})^{1/2}}\mathrm{diag}\left(\mathbf{1} - \frac{1}{1-\bar{c}}\left[2\delta + \frac{1}{\bar{c}}\delta^{\circ 2}\right]\right)^{-1/2}$$
$$\approx \frac{1}{(1-\bar{c})^{1/2}}\mathrm{diag}\left(\mathbf{1} - \frac{2\delta}{1-\bar{c}}\right)^{-1/2},$$

where the last approximation assumes that $2\bar{c} \gg |\delta_i|$. This is a reasonable assumption in dense correlation networks with bounded connection values. In our data, for example, we find $\bar{c} \approx 0.57$, while $|\delta_i| = 0.019$ (0.002–0.043) (**Figure 11A**). Reasoning similarly, we can assume that $(1-\bar{c}) \gg |2\delta_i|$, which allows us to use an accurate first-order Taylor expansion of the diagonal term around 0, and thus write

$$\eta \approx \frac{1}{(1-\bar{c})^{1/2}}\mathrm{diag}\left(\mathbf{1} + \frac{\delta}{1-\bar{c}}\right) = \frac{1}{(1-\bar{c})^{1/2}}\left(\mathbf{I} + \frac{1}{1-\bar{c}}\mathrm{diag}(\delta)\right),$$

where $\mathbf{I}$ is the identity matrix. Substituting this approximation into Equation 6, we can finally approximate $\mathbf{C}'$ as

$$\mathbf{C}' \approx \frac{1}{1-\bar{c}}\left(\mathbf{I} + \frac{1}{1-\bar{c}}\mathrm{diag}(\delta)\right)\mathbf{C}^\diamond\left(\mathbf{I} + \frac{1}{1-\bar{c}}\mathrm{diag}(\delta)\right)$$

$$= \frac{1}{1-\bar{c}}\left(\mathbf{C}^\diamond + \frac{1}{1-\bar{c}}\text{diag}(\boldsymbol{\delta})\mathbf{C}^\diamond + \frac{1}{1-\bar{c}}\mathbf{C}^\diamond\text{diag}(\boldsymbol{\delta}) + \mathbf{C}^\diamond\frac{1}{(1-\bar{c})^2}\text{diag}(\boldsymbol{\delta})\mathbf{C}^\diamond\text{diag}(\boldsymbol{\delta})\right). \qquad (7)$$

We can write out each term of this matrix as

$$c'_{ij} \approx \frac{1}{1-\bar{c}}c^\diamond_{ij}\left[1 + \frac{1}{1-\bar{c}}\left(\delta_i + \delta_j + \frac{1}{1-\bar{c}}\delta_i\delta_j\right)\right] \approx \frac{1}{1-\bar{c}}c^\diamond_{ij}\left(1 + \frac{\delta_i + \delta_j}{1-\bar{c}}\right),$$

where in the last approximation we dropped the negligible quadratic term.

It follows, therefore, that under $(1-\bar{c}) \gg |\delta_i + \delta_j|$, an assumption similar to the one we made above, the effect of the rescaling term on the residual correlations will be relatively small, which supports the approximate equivalence between $\mathbf{C}'$ and $\mathbf{C}^\diamond$. In our example data, we find that these terms center around 1.001 and lie in the range (0.868–1.125), where 1 implies no rescaling (**Figure 11B**).

**Degree after global-signal regression $\cong$ primary co-activity gradient**

We sketch out this equivalence in the following way. First, we assume that co-neighbor and correlation networks will have similar second eigenvectors, because both eigenvectors capture the dominant and largely density-invariant structure in the data. It follows, therefore (as we also discuss below), that $\mathbf{u}_2$, the second eigenvector of the correlation network, will be similar to the primary co-activity gradient. Accordingly, in our example data, we find that $\mathbf{u}_2$ and the primary co-activity gradient have a correlation of 0.921 $(0.911-0.929)$ (**Figure 11C**).

Our task, therefore, reduces to showing that the degree after global-signal regression $\mathbf{d}'$ is approximately equivalent to $\mathbf{u}_2$ (**Figure 11D**). Here, we describe why this equivalence holds in functional MRI correlation networks.

We begin to do so by using Equation 7 to define $\mathbf{d}'$ as

$$\mathbf{d}' = \mathbf{C}'\mathbf{1} \cong \frac{1}{1-\bar{c}}\left(\mathbf{I} + \frac{1}{1-\bar{c}}\text{diag}(\boldsymbol{\delta})\right)\mathbf{C}^\diamond\left(\mathbf{I} + \frac{1}{1-\bar{c}}\text{diag}(\boldsymbol{\delta})\right)\mathbf{1}$$

$$= \frac{1}{1-\bar{c}}\left(\mathbf{I} + \frac{1}{1-\bar{c}}\text{diag}(\boldsymbol{\delta})\right)\left(\mathbf{C}^\diamond\mathbf{1} + \frac{1}{1-\bar{c}}\mathbf{C}^\diamond\boldsymbol{\delta}\right)$$

$$= \frac{1}{(1-\bar{c})^2}\left(\mathbf{I} + \frac{1}{1-\bar{c}}\text{diag}(\boldsymbol{\delta})\right)\mathbf{C}^\diamond\boldsymbol{\delta}$$

where we made use of the fact that $\mathbf{C}^\diamond\mathbf{1} = \mathbf{0}$. Assuming that $(1-\bar{c}) \gg |\delta_i|$, as we did above, and recalling that $\mathbf{d} = n(\bar{c}\mathbf{1} + \boldsymbol{\delta}) \equiv \boldsymbol{\delta}$ we can drop the rescaling constant and simply approximate $\mathbf{d}'$ as

$$\mathbf{d}' \cong \mathbf{C}^\diamond\boldsymbol{\delta} \equiv \left(\mathbf{C} - \frac{1}{\sum(c)}\mathbf{d}\mathbf{d}^\top\right)\mathbf{d}. \qquad (8)$$

Now, with the help of Equations 1 and 5, we can further simplify

$$\mathbf{d}' \cong \left(\mathbf{U}\boldsymbol{\Psi}\mathbf{U}^\top - \psi_1\mathbf{u}_1\mathbf{u}_1^\top\right)\mathbf{U}\boldsymbol{\Psi}\mathbf{U}^\top\mathbf{1} = \left(\mathbf{U}\boldsymbol{\Psi}^2\mathbf{U}^\top - \psi_1^2\mathbf{u}_1\mathbf{u}_1^\top\right)\mathbf{1} = \sum_{i=2}^n \psi_i^2\mathbf{u}_i\mathbf{u}_i^\top\mathbf{1} = \sum_{i=2}^n n\bar{u}_i\psi_i^2\mathbf{u}_i, \qquad (9)$$

where $\bar{u}_i$ denotes the mean of the elements in $\mathbf{u}_i$.

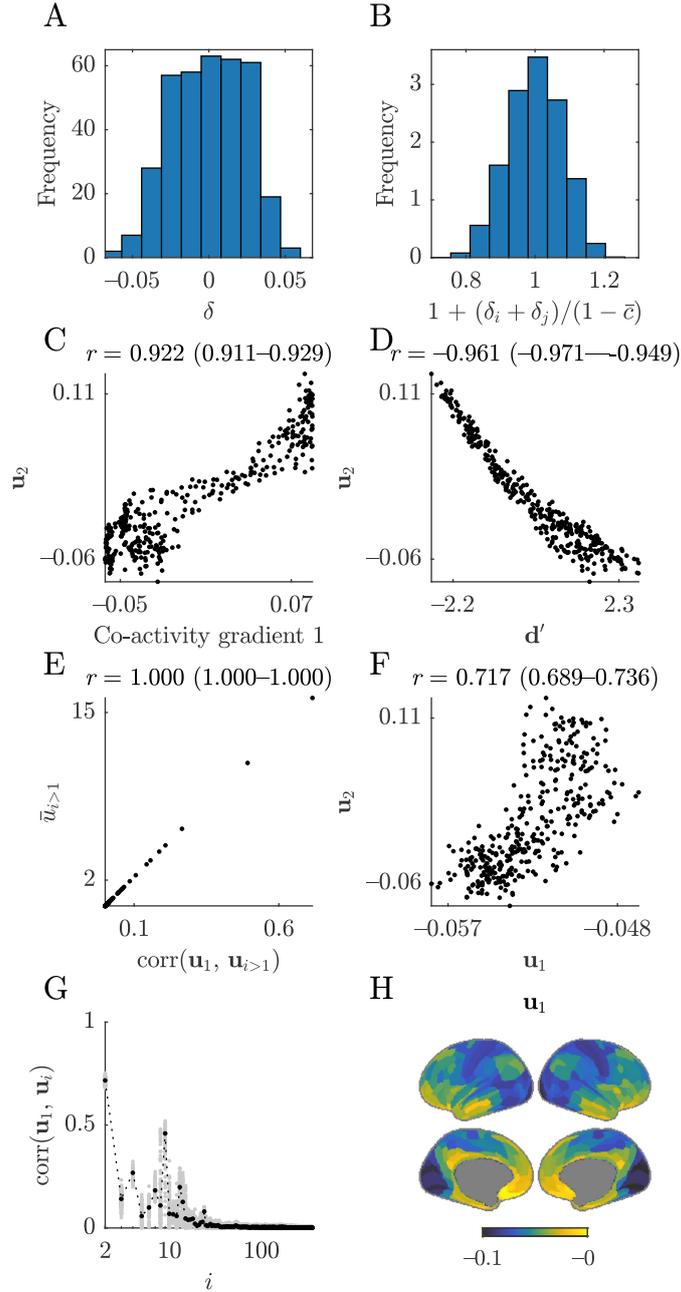

**Figure 11.**

**A.** Distribution of the relative degree deviation around the degree mean.

**B.** Distribution of the approximate rescaling term on individual correlations after global-signal regression (1 implies no rescaling).

**C.** Scatter plot of the primary co-activity gradient and the second eigenvector of the correlation network.

**D.** Scatter plot of the degree after global-signal regression and the second eigenvector of the correlation network.

**E.** Scatter plot of the correlation between the first and second to $n$-th eigenvectors, and the mean value of elements in the second to $n$-th eigenvectors.

**F.** Scatter plots of the correlation between the first and second eigenvectors of the correlation network.

**G.** Correlation coefficients between the first and second to $n$-th eigenvectors. Shaded area shows the 5–95% interval.

**H.** A map of the first eigenvector ($\cong$ degree) of the correlation network.

Equation 9 shows that $\mathbf{u}_2$ can be thought of as a first-order approximation of $\mathbf{d}'$. In this vignette, we describe that this approximation is generally accurate in functional MRI data.

We do so by first defining the Pearson correlation between $\mathbf{u}_1$ and $\mathbf{u}_{i>1}$. We know that $\mathbf{u}_1^\top \mathbf{u}_{i>0} = 0$ and can therefore express the correlation between $\mathbf{u}_1$ and $\mathbf{u}_i$ as

$$\mathrm{corr}(\mathbf{u}_1, \mathbf{u}_i) = \frac{(\mathbf{u}_1 - \bar{u}_1 \mathbf{1})^\top (\mathbf{u}_i - \bar{u}_i \mathbf{1})}{\left[(\mathbf{u}_1 - \bar{u}_1 \mathbf{1})^\top (\mathbf{u}_1 - \bar{u}_1 \mathbf{1})\right]^{1/2} \left[(\mathbf{u}_i - \bar{u}_i \mathbf{1})^\top (\mathbf{u}_i - \bar{u}_i \mathbf{1})\right]^{1/2}} = \frac{-n \bar{u}_1 \bar{u}_i}{(1 - n\bar{u}_1^2)^{1/2} (1 - n\bar{u}_i^2)^{1/2}}.$$

We now note that $-n\bar{u}_1 (1 - n\bar{u}_1^2)^{-1/2}$ is constant, while $(1 - n\bar{u}_i^2)^{-1/2} \approx 1$ when $n\bar{u}_i^2 \ll 1$, which is generally the case for $i > 1$, in large part as a consequence of the Perron-Frobenius theorem (as we noted above). We now drop all constants and write

$$\mathrm{corr}(\mathbf{u}_1, \mathbf{u}_i) \cong \bar{u}_i.$$

In our example data, we find this equivalence to be highly accurate, with correlations between $\mathrm{corr}(\mathbf{u}_1, \mathbf{u}_i)$ and $\bar{u}_i$ of 1.000 (1.000, 1.000) (**Figure 11E**).

This equivalence allows us to further simplify $\mathbf{d}'$ as

$$\mathbf{d}' \cong \sum_{i=2}^{n} \psi_i^2 \mathrm{corr}(\mathbf{u}_1, \mathbf{u}_i) \mathbf{u}_i.$$

Now, we know that in functional MRI data, the first or "global-signal correlation" eigenvector $\mathbf{u}_1$ is strongly correlated with the second or "primary-gradient" eigenvector $\mathbf{u}_2$, in part because areas that have high correlations with the global signal also tend to be association areas. By contrast, $\mathbf{u}_1$ has weaker correlations with the other leading eigenvectors (**Figure 11F–H**). Noting also that $\psi_2 > \psi_{i>2}$, it follows that $\psi_2^2 \mathrm{corr}(\mathbf{u}_1, \mathbf{u}_2) \gg \psi_{i>2}^2 \mathrm{corr}(\mathbf{u}_1, \mathbf{u}_{i>2})$, which establishes the accuracy of this approximation.

Finally, we note that the negative sign of the correlation in **Figure 8** and **Figure 11d** arises from subtle differences between the vector $\psi_1^2 \mathbf{u}_1 \mathbf{u}_1^\top \mathbf{1}$ on the one hand, and the vectors $\mathbf{d}\mathbf{d}^\top \mathbf{d}/\sum(c)$ and $\eta \mathbf{d}\mathbf{d}^\top \eta \mathbf{1}/\sum(c)$ on the other hand. Here, we can illustrate this effect by comparing vectors $\psi_1^2 \mathbf{u}_1 \mathbf{u}_1^\top \mathbf{1}$ and $\mathbf{d}\mathbf{d}^\top \mathbf{d}/\sum(c)$. These two vectors are strongly correlated with each other ($r \approx 0.995$ in our data) and with $\mathbf{Cd}$ ($r > 0.995$ for both vectors). Despite this, minute differences in the slope of these vectors lead to negatively correlated, but otherwise approximately equivalent, residuals $(\mathbf{C} - \mathbf{d}\mathbf{d}^\top \mathbf{d}/\sum(c))\mathbf{d}$ from Equation 8 and $(\mathbf{C} - \psi_1 \mathbf{u}_1 \mathbf{u}_1^\top)\mathbf{d}$ from Equation 9 ($r \approx -0.987$ in our data). This negative correlation, in turn, propagates to the final result.

### $k$-modularity: unified $k$-means objective and modularity

We use an $n \times k$ binary indicator matrix $\mathbf{M}$ to represent a partition of the data or network into $k$ clusters or modules. The binary columns of this matrix $\mathbf{m}_h$ thus indicate the presence of nodes in module $h$. We denote the number of these nodes by $N_h = \mathbf{m}_h \mathbf{1}$, which implies that $n = \sum_{h=1}^{k} N_h$.

*$k$-means objective.* $k$-means clustering seeks to find a partition that minimizes the squared Euclidean distance between all nodes and their within-cluster centroids. Denoting by $\mu_h$ the set of nodes in module $h$, and thus defining the centroid of module $h$ by $1/N_h \sum_{j \in \mu_h} \mathbf{x}_j$, we can write:

$$(k\text{-means objective}) = \sum_{h=1}^{k} \sum_{i \in \mu_h} \left( \mathbf{x}_i - \frac{1}{N_h} \sum_{j \in \mu_h} \mathbf{x}_j \right)^{\top} \left( \mathbf{x}_i - \frac{1}{N_h} \sum_{j \in \mu_h} \mathbf{x}_j \right)$$

$$= \sum_{h=1}^{k} \sum_{i \in \mu_h} \left( \mathbf{x}_i^{\top} \mathbf{x}_i - \frac{1}{N_h} \mathbf{x}_i^{\top} \sum_{j \in \mu_h} \mathbf{x}_j \right).$$

Noting that $\mathbf{x}_i^{\top} \mathbf{x}_j = c_{ij}$ and $\mathbf{x}_i^{\top} \mathbf{x}_i = 1$, we can further simplify the $k$-means objective as

$$(k\text{-means objective}) = \sum_{i=1}^{n} \mathbf{x}_i^{\top} \mathbf{x}_i - \sum_{h=1}^{k} \frac{1}{N_h} \sum_{i,j \in \mu_h} \mathbf{x}_i^{\top} \mathbf{x}_j = n - \sum_{h=1}^{k} \frac{1}{N_h} \mathbf{m}_h^{\top} \mathbf{C} \mathbf{m}_h,$$

where $\mathbf{m}_h^{\top} \mathbf{C} \mathbf{m}_h$ represents the sum of all connection weights in module $h$. Note that this representation allows us to replace $\mathbf{C}$ with an arbitrary similarity or network matrix.

*Density-corrected k-modularity.* We can define the density-corrected modularity as

$$(\text{density-corrected modularity}) = \frac{1}{\sum(c)} \sum_{h=1}^{k} \mathbf{m}_h^{\top} \left( \mathbf{C} - \frac{\sum(c)}{n^2} \mathbf{1}\mathbf{1}^{\top} \right) \mathbf{m}_h,$$

where $\sum(c)/n^2\, \mathbf{1}\mathbf{1}^{\top}$ denotes the (constant) density expectation matrix.

Correspondingly, we can define the density-corrected $k$-modularity as

$$(\text{density-corrected } k\text{-modularity}) = \frac{1}{\sum(c)} \sum_{h=1}^{k} \frac{1}{N_h} \mathbf{m}_h^{\top} \left( \mathbf{C} - \frac{\sum(c)}{n^2} \mathbf{1}\mathbf{1}^{\top} \right) \mathbf{m}_h,$$

and simplify this equation to

$$(\text{density-corrected } k\text{-modularity}) = \frac{1}{\sum(c)} \sum_{h=1}^{k} \frac{1}{N_h} \mathbf{m}_h^{\top} \mathbf{C} \mathbf{m}_h - \frac{1}{n^2} \sum_{h=1}^{k} \frac{1}{N_h} \mathbf{m}_h^{\top} \mathbf{1}\mathbf{1}^{\top} \mathbf{m}_h$$

$$= \frac{1}{\sum(c)} \sum_{h=1}^{k} \frac{1}{N_h} \mathbf{m}_h^{\top} \mathbf{C} \mathbf{m}_h - \frac{1}{n} \equiv (k\text{-means objective}).$$

*(Degree-corrected) k-modularity.* We can define the (degree-corrected) modularity as

$$(\text{modularity}) = \frac{1}{\sum(c)} \sum_{h=1}^{k} \mathbf{m}_h^{\top} \mathbf{C}^{\diamond} \mathbf{m}_h.$$

Correspondingly, we can define the (degree-corrected) $k$-modularity as

$$(k\text{-modularity}) = \frac{1}{\sum(c)} \sum_{h=1}^{k} \frac{1}{N_h} \mathbf{m}_h^{\top} \mathbf{C}^{\diamond} \mathbf{m}_h = (k\text{-means objective with degree correction})$$

$$\cong \frac{1}{\sum(c)} \sum_{h=1}^{k} \frac{1}{N_h} \mathbf{m}_h^{\top} \mathbf{C}' \mathbf{m}_h = (k\text{-means objective with global-signal regression})$$

$$\cong \frac{1}{\sum(c)} \sum_{h=1}^{k} \frac{1}{N_h} \mathbf{m}_h^{\top} \mathbf{C}^{*} \mathbf{m}_h = (k\text{-means objective with first-component removal}),$$

where we made use of the equivalences we established above.

**Spectral clustering ≡ *s*-modularity maximization**

For completeness, we also summarize a known result (Yu and Ding, 2010) that reduces the normalized cut to another variant of the normalized modularity.

Normalized cut is a standard objective of spectral clustering (Shi and Malik, 2000), defined as

$$(\text{normalized cut}) = \sum_{h=1}^{k} \frac{1}{D_h} \mathbf{m}_h^\top \mathbf{C} \mathbf{m}_h,$$

where $D_h = \mathbf{m}_h^\top \mathbf{d}$ is the module degree, which implies that $\sum(c) = \sum_{h=1}^{k} D_h$.

In this section, we consider another normalized variant of the modularity that we term the *s*-modularity. This modularity is normalized by module degree and defined as

$$(s\text{-modularity}) = \frac{1}{\sum(c)} \sum_{h=1}^{k} \frac{1}{D_h} \mathbf{m}_h^\top \left( \mathbf{C} - \frac{1}{\sum(c)} \mathbf{d}\mathbf{d}^\top \right) \mathbf{m}_h.$$

As above, we can simplify this equation as

$$(s\text{-modularity}) = \frac{1}{\sum(c)} \sum_{h=1}^{k} \frac{1}{D_h} \mathbf{m}_h^\top \mathbf{C} \mathbf{m}_h - \frac{1}{\sum(c)^2} \sum_{h=1}^{k} \frac{1}{D_h} \mathbf{m}_h^\top \mathbf{d}\mathbf{d}^\top \mathbf{m}_h$$

$$= \frac{1}{\sum(c)} \sum_{h=1}^{k} \frac{1}{D_h} \mathbf{m}_h^\top \mathbf{C} \mathbf{m}_h - \frac{1}{\sum(c)} \equiv (\text{normalized cut}),$$

which establishes the equivalence.

**Co-clustering objectives and binary canonical analyses**

*Co-clustering objectives.* It is straightforward to extend the *k*-means and spectral clustering objectives (and therefore the *k*-modularity and *s*-modularity) to bipartite networks. Here, we define these networks by a rectangular $p_a \times p_b$ matrix $\mathcal{Z}$. The rows of this matrix denote $p_a$ features from group $a$, while the columns denote $p_b$ features from group $b$. In the main text, we set $\mathcal{Z}$ to be the cross-covariance of two centered matrices $\mathcal{X}$ and $\mathcal{Y}$, or $\mathcal{Z} = \mathcal{X}\mathcal{Y}^\top$, but in principle $\mathcal{Z}$ could also represent any arbitrary bipartite network.

We can now define an $p_a \times k$ binary indicator matrix $\mathbf{M}_a$ to represent a partition of nodes in group $a$, and a corresponding $p_b \times k$ binary indicator matrix $\mathbf{M}_b$ to represent a partition of nodes in group $b$. This allows us to generalize the *k*-means objective to bipartite networks,

$$(k\text{-means co-clustering objective}) = \sum_{h=1}^{k} \frac{\mathbf{m}_{ah}^\top \mathcal{Z} \mathbf{m}_{bh}}{(\mathbf{m}_{ah}^\top \mathbf{m}_{ah})^{1/2} (\mathbf{m}_{bh}^\top \mathbf{m}_{bh})^{1/2}} = \sum_{h=1}^{k} \frac{\mathbf{m}_{ah}^\top \mathcal{Z} \mathbf{m}_{bh}}{(N_{ah} N_{bh})^{1/2}},$$

where $N_{ah}$ and $N_{bh}$ denote the corresponding sizes of modules $\mathbf{m}_{ah}$ and $\mathbf{m}_{bh}$. In the main text, we noted that this objective is a binary variant of the canonical covariance objective,

$$(\text{canonical covariance objective}) = \sum_{h=1}^{k} \frac{a_h^\top \mathcal{X} \mathcal{Y}^\top b_h}{(a_h^\top a_h)^{1/2} (b_h^\top b_h)^{1/2}}.$$

Correspondingly, we can define the spectral co-clustering objective as

$$\text{(spectral co-clustering objective)} = \sum_{h=1}^{k} \frac{\mathbf{m}_{ah}^\top \mathcal{Z} \mathbf{m}_{bh}}{\left(\mathbf{m}_{ah}^\top \mathcal{Z} \mathbf{1}\right)^{1/2} \left(\mathbf{1}^\top \mathcal{Z} \mathbf{m}_{bh}\right)^{1/2}} = \sum_{h=1}^{k} \frac{\mathbf{m}_{ah}^\top \mathcal{Z} \mathbf{m}_{bh}}{\left(D_{ah} D_{bh}\right)^{1/2}}.$$

where $D_{ah}$ and $D_{bh}$ denote the corresponding degrees of modules $\mathbf{m}_{ah}$ and $\mathbf{m}_{bh}$.

*Canonical correlation analysis.* We can write the canonical correlation objective as

$$\text{(canonical correlation objective)} = \sum_{h=1}^{k} \frac{\boldsymbol{a}_h^\top \mathcal{X} \mathcal{Y}^\top \boldsymbol{b}_h}{\left(\boldsymbol{a}_h^\top \mathcal{X} \mathcal{X}^\top \boldsymbol{a}_h\right)^{1/2} \left(\boldsymbol{b}_h^\top \mathcal{Y} \mathcal{Y}^\top \boldsymbol{b}_h\right)^{1/2}}.$$

This objective resembles the spectral co-clustering objective, although it is not exactly equivalent to it. Instead, we can convert this objective to a canonical covariance objective through the following transforms:

$$\tilde{\boldsymbol{a}}_h = \left(\mathcal{X} \mathcal{X}^\top\right)^{1/2} \boldsymbol{a}_h.$$

$$\tilde{\boldsymbol{b}}_h = \left(\mathcal{Y} \mathcal{Y}^\top\right)^{1/2} \boldsymbol{b}_h.$$

$$\tilde{\mathcal{Z}} = \left(\mathcal{X} \mathcal{X}^\top\right)^{-1/2} \left(\mathcal{X} \mathcal{Y}^\top\right) \left(\mathcal{Y} \mathcal{Y}^\top\right)^{-1/2}.$$

and thus express it as

$$\text{(canonical correlation objective)} = \sum_{h=1}^{k} \frac{\tilde{\boldsymbol{a}}_h^\top \tilde{\mathcal{Z}} \tilde{\boldsymbol{b}}_h}{\left(\tilde{\boldsymbol{a}}_h^\top \tilde{\boldsymbol{a}}_h\right)^{1/2} \left(\tilde{\boldsymbol{b}}_h^\top \tilde{\boldsymbol{b}}_h\right)^{1/2}}.$$

As above, we can use *k*-means co-clustering to find binary-equivalent vectors of $\tilde{\boldsymbol{a}}_h$ and $\tilde{\boldsymbol{b}}_h$. Note, however, that the back transform of these vectors will generally result in non-binary solutions $\boldsymbol{a}_h$ and $\boldsymbol{b}_h$.

**Components or modules of co-neighbor networks $\cong$ co-activity gradients.**

We start this section by describing the variant of diffusion-map embedding popular in imaging neuroscience. This variant comprises the following steps:

*1. Kernel matrix.* We first define $\widetilde{\mathbf{C}}$ to be a thresholded version of $\mathbf{C}$, a matrix of the $\kappa \ll n$ strongest correlation neighbors of each node, excluding self-correlations. We denote the individual columns of $\widetilde{\mathbf{C}}$ by $\tilde{\mathbf{c}}_i$ and thus, for $i \neq j$, formally define its elements as

$$\tilde{c}_{ij} = c_{ij} [\![ c_{ij} > \text{quantile}(\mathbf{c}_i, 1 - \kappa/n) ]\!] \tag{10}$$

where $[\![ \cdot ]\!]$ evaluates to 0 or 1 based on the truth of the enclosed condition.

We now define a so-called cosine-similarity kernel matrix $\mathbf{A}$,

$$\mathbf{A} = \tilde{\eta} \, \widetilde{\mathbf{C}}^\top \widetilde{\mathbf{C}} \tilde{\eta}, \tag{11}$$

where the diagonal matrix $\tilde{\eta}$ rescales the norms to 1,

$$\tilde{\eta}_{ii} = \left(\tilde{\mathbf{c}}_i^\top \tilde{\mathbf{c}}_i\right)^{-1/2}. \tag{12}$$

*2. Laplacian normalization.* We define a normalization vector $\boldsymbol{\nu} = (\mathbf{A1})^{\circ(-\alpha)}$ to compute a degree-normalized variant of $\mathbf{A}$,

$$\mathbf{R} = \mathrm{diag}(\boldsymbol{\nu})\mathbf{A}\mathrm{diag}(\boldsymbol{\nu}) = \mathbf{A} \odot \left(\boldsymbol{\nu}\boldsymbol{\nu}^\top\right) \tag{13}$$

where $\circ$ denotes elementwise power and $\odot$ denotes elementwise product. The scaling parameter $0 \leq \alpha \leq 1$ controls the effects of the degrees of $\mathbf{A}$ on $\mathbf{R}$. Here, we use the default $\alpha = 1/2$. Note, however, that $\mathbf{A}$ has a relatively homogeneous degree (by construction of $\widetilde{\mathbf{C}}$), which implies that the choice of $\alpha$ will have relatively little effect on our final results.

*3. Transition probability normalization.* We define the transition probability matrix $\mathcal{R}$ by normalizing the rows of $\mathbf{R}$,

$$\mathcal{R} = \mathrm{diag}(\mathbf{R1})^{-1}\mathbf{R}. \tag{14}$$

Diffusion-map embedding uses this transition probability matrix to model a random walk. The construction of this matrix is somewhat unnecessary in our case, however, because typical analyses in imaging neuroscience focus only on co-activity gradients at time 0. (As an aside, a similar transition-probability matrix will reappear in our discussion of diffusion efficiency, below).

*4. Co-activity gradients.* We define $\boldsymbol{q}_1, \boldsymbol{q}_2, \ldots \boldsymbol{q}_{k+1}$ to be the leading (right) $k+1$ eigenvectors of $\mathcal{R}$, and finally define the co-activity gradients as

$$(\text{co-activity gradient})_h \equiv \boldsymbol{q}_{h+1} \oslash \boldsymbol{q}_1, \tag{15}$$

where $\oslash$ denotes elementwise division. Note that the diffusion-map embedding algorithm also rescales the values of each gradient by the eigenvalues of $\mathcal{R}$. We omit this step here, however, since it amounts to normalization by constants.

We now simplify this algorithm in the following way.

First, we know, from Equation 10, that each column of $\widetilde{\mathbf{C}}$ will have $\kappa \ll n$ nonzero elements. Since these nonzero elements represent the top fraction of the strongest correlated neighbors of that node, we can reasonably assume that these strongest correlations will be relatively homogeneous.

We now denote the average correlation value between node $i$ and its $\kappa$ most correlated neighbors by $\omega_i$, and define a diagonal matrix of these average correlations by $\boldsymbol{\omega} = \mathrm{diag}(\omega_1, \ldots \omega_n)$. This allows us to approximate $\widetilde{\mathbf{C}}$ as

$$\widetilde{\mathbf{C}} \approx [\![\widetilde{\mathbf{C}} > 0]\!]\boldsymbol{\omega}, \tag{16}$$

where $[\![\cdot]\!]$ again evaluates to 0 or 1 based on the truth of the enclosed condition.

Similarly, we can approximate the elements of the rescaling matrix $\tilde{\boldsymbol{\eta}}$ from Equation 12 by

$$\tilde{\eta}_{ii} \approx \left(\omega_{ii}^2 [\![\tilde{\mathbf{c}}_i > 0]\!]^\top [\![\tilde{\mathbf{c}}_i > 0]\!]\right)^{-1/2} = \left(\kappa\omega_{ii}^2\right)^{-1/2} = (\kappa^{1/2}\omega_{ii})^{-1}. \tag{17}$$

Equations 16 and 17 allow us to simplify the rescaled matrix $\widetilde{\mathbf{C}}\tilde{\boldsymbol{\eta}}$ as

$$\widetilde{\mathbf{C}}\tilde{\boldsymbol{\eta}} \approx [\![\widetilde{\mathbf{C}} > 0]\!]\boldsymbol{\omega}(\kappa^{1/2}\boldsymbol{\omega})^{-1} = \kappa^{-1/2}[\![\widetilde{\mathbf{C}} > 0]\!] \equiv [\![\widetilde{\mathbf{C}} > 0]\!]. \tag{18}$$

Separately, we can define $\mathcal{C}$ as the co-neighbor, or shared $\kappa$-nearest-neighbor, matrix,

$$\mathcal{C} = [\![\widetilde{\mathbf{C}} > 0]\!]^\top [\![\widetilde{\mathbf{C}} > 0]\!].$$

Equations 11 and 18 show that $\mathcal{C}$ is equivalent to the cosine-similarity kernel matrix $\mathbf{A}$,

$$\mathcal{C} \cong \tilde{\eta}\widetilde{\mathbf{C}}^\top \widetilde{\mathbf{C}}\tilde{\eta} \equiv \mathbf{A}.$$

In our example data, we find that $\mathbf{A}$ and $\mathcal{C}$ have correlations of 1.000 (1.000–1.000) (**Figure 12A**).

We can now define $\boldsymbol{u}_1, \boldsymbol{u}_2, \ldots \boldsymbol{u}_{k+1}$ to be the $k+1$ leading eigenvectors of the co-neighbor network $\mathcal{C}$. We note that the computation of $\mathbf{R}$ in Equation 13, and $\mathcal{R}$ in Equation 14 amounts to a rescaling of $\mathcal{C}$ by its degrees, and thus roughly parallels degree correction. In practice, because the degrees of $\mathcal{C}$ are relatively homogeneous, by construction, this normalization will primarily affect $\boldsymbol{u}_1$, and will have little effect on $\boldsymbol{u}_2$ to $\boldsymbol{u}_{k+1}$ (we can show this more formally with a perturbation analysis). **Figure 12B–E** shows this effect in our example data.

The relatively minor effect of the normalization on the structure of $\boldsymbol{u}_{i>1}$ implies that

$$[\boldsymbol{q}_2, \ldots \boldsymbol{q}_{k+1}] \cong [\boldsymbol{u}_2, \ldots \boldsymbol{u}_{k+1}]. \tag{19}$$

Note that $\mathcal{R}\mathbf{1} = \mathbf{1}$ by definition, from which it follows that $\boldsymbol{q}_1 \equiv \mathbf{1}$, the normalization in Equation 15 is unnecessary, and we can simplify:

$$(\text{co-activity gradient})_h \equiv \boldsymbol{q}_{h+1}.$$

Finally, Equations 15 and 19 allow us to write:

$$(\text{co-activity gradient})_h \equiv \boldsymbol{q}_{h+1} \cong \boldsymbol{u}_{h+1} = (\text{eigenvector of co-neighbor matrix})_{h+1}.$$

**m-umap ≡ first-order approximation of UMAP.**

We denote the symmetric $\kappa$-nearest-neighbor network by $\mathfrak{C}$ and its elements by $\mathfrak{c}_{ij}$. Correspondingly, we denote the degree of node $i$ in this network by $\mathfrak{d}_i$. As noted in the main text, we follow Damrich and Hamprecht (2021) to assume a true parametric UMAP objective of

$$(\text{UMAP objective}) \equiv -\sum_{i,j} \left( \mathfrak{c}_{ij} \log(\phi_{ij}) + \gamma \frac{\mathfrak{d}_i \mathfrak{d}_j}{\sum(\mathfrak{c})} \log(1 - \phi_{ij}) \right) \tag{20}$$

where the Cauchy similarity

$$\phi_{ij} = \left(1 + \alpha \left\| \mathbf{u}_{i:} - \mathbf{u}_{j:} \right\|^{2\beta}\right)^{-1}$$

is a function of the Euclidean distance $\|\cdot\|$ between low-dimensional embeddings (row vectors $\mathbf{u}_{i:}$ and $\mathbf{u}_{j:}$) while $\alpha > 0$, $\beta > 0$, and $\gamma > 0$ are parameters.

We define m-umap as the first-order Taylor expansion of Equation 20 around 1/2,

$$(\text{m-umap}) = -\sum_{i,j} \left( \mathfrak{c}_{ij} - \gamma \frac{\mathfrak{d}_i \mathfrak{d}_j}{\sum(\mathfrak{c})} \right) \phi_{ij} = -\sum_{i,j} \mathfrak{c}_{ij}^\diamond \phi_{ij}.$$

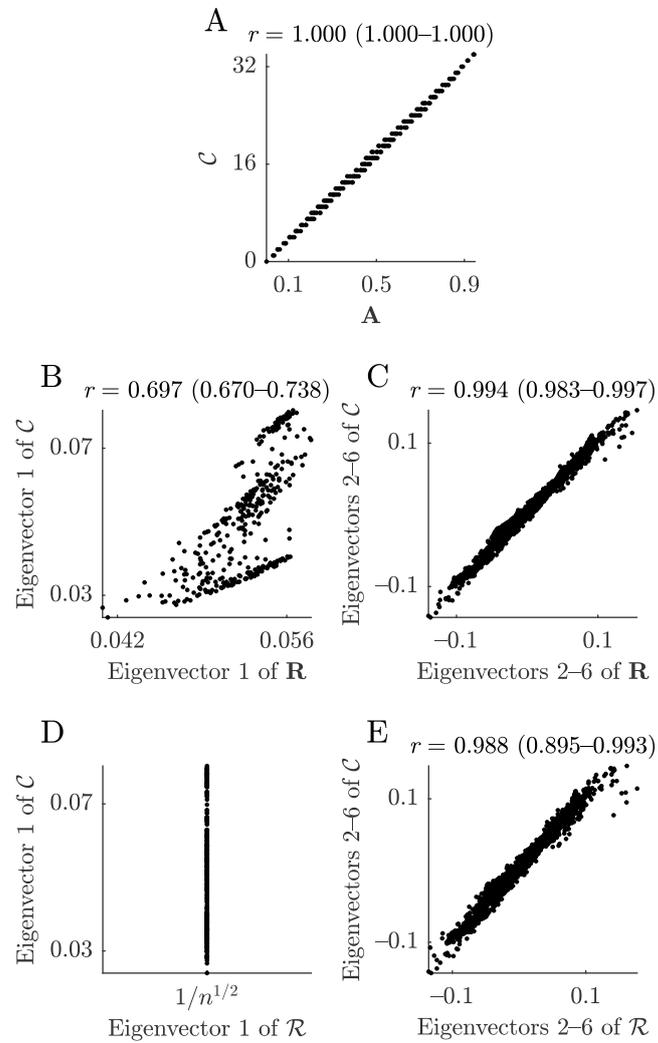

**Figure 12.**

**A.** Scatter plots of the cosine-similarity kernel matrix, and the integer co-neighbor network.

**B, D.** Scatter plots of the first eigenvector from intermediate degree-normalization steps in diffusion-map embedding, and the corresponding first eigenvector of a co-neighbor network.

**C, E.** Scatter plots of the second to sixth eigenvectors from intermediate degree-normalization steps of diffusion-map embedding, and the corresponding eigenvectors of co-neighbor network.

Clearly, m-umap is just the modularity with Cauchy similarity and a modularity resolution parameter $\gamma$ (Reichardt and Bornholdt, 2006) that corresponds exactly to the so-called negative-sampling rate of UMAP's reference implementation (Damrich and Hamprecht, 2021). Here, we always assume $\gamma = 1$. We also note that $\kappa$ indirectly controls the number of modules (**Figure 5**) and somewhat obviates the need to adjust $\gamma$. Nonetheless, the modularity has a resolution limit, which suggests that $\gamma = 1$ can lead to the neglect of small modules when $\kappa \lesssim n^{1/3}$, and can thus become a problem in large networks. (See the original exposition of the resolution limit by Fortunato and Barthélemy (2007) for a more detailed discussion of this phenomenon.)

In this study, we embed m-umap on a sphere and thus constrain the row vectors $\mathbf{u}_{i:}$ to have unit norm. This simplifies the Cauchy similarity to

$$\phi_{ij} = \left(1 + 2\alpha(1 - \mathbf{u}_{i:}^\top \mathbf{u}_{j:})^{2\beta}\right)^{-1}.$$

Under the additional assumptions of binary $\mathbf{u}_{i:}$ we have

$$\phi_{ij} = \begin{cases} 1 & \text{if } \mathbf{u}_{i:} = \mathbf{u}_{j:} \\ \delta & \text{otherwise} \end{cases}$$

where $0 < \delta = \left(1 + 2\alpha^{2\beta}\right)^{-1} < 1$ is a constant.

Our binary module indicators $\mathbf{m}_h$ allow us to define m-umap with binary constraints as

$$\text{(binary m-umap)} = -\sum_h \mathbf{m}_h^\top \mathfrak{C}^\circ \mathbf{m}_h - \delta \sum_{g \neq h} \mathbf{m}_g^\top \mathfrak{C}^\circ \mathbf{m}_h.$$

We can now express the second term in the above equation as

$$\delta \sum_{g \neq h} \mathbf{m}_g^\top \mathfrak{C}^\circ \mathbf{m}_h = \delta \sum_{g,h} \mathbf{m}_g^\top \mathfrak{C}^\circ \mathbf{m}_h - \delta \sum_h \mathbf{m}_h^\top \mathfrak{C}^\circ \mathbf{m}_h \tag{21}$$

The first sum in Equation 21 is over all pairs of nodes and is therefore a constant that we can drop.

It follows that

$$\delta \sum_{g \neq h} \mathbf{m}_g^\top \mathfrak{C}^\circ \mathbf{m}_h \equiv -\delta \sum_h \mathbf{m}_h^\top \mathfrak{C}^\circ \mathbf{m}_h$$

and therefore

$$\text{(m-umap)} = -(1-\delta) \sum_h \mathbf{m}_h^\top \mathfrak{C}^\circ \mathbf{m}_h \equiv -\sum_h \mathbf{m}_h^\top \mathfrak{C}^\circ \mathbf{m}_h,$$

which implies that binary m-umap is exactly equivalent to the modularity.

**Structural network**

We define a structural network by its non-negative and undirected $n \times n$ matrix $\mathbf{W}$. The rows and columns $i$ of this matrix represent nodes, while the elements of this matrix $w_{ij}$ represent weighted, symmetric, and non-negative weights. We assume that $\mathbf{W}$ has no self-connections or $w_{ii} = 0$.

We denote the eigendecomposition of $\mathbf{W}$ as

$$\mathbf{W} = \mathbf{V}\mathbf{\Lambda}\mathbf{V}^\top = \sum_i \lambda_i \mathbf{v}_i \mathbf{v}_i^\top,$$

where the $n \times n$ matrix $\mathbf{V} = [\mathbf{v}_1, \mathbf{v}_2, \ldots \mathbf{v}_n]$ has the eigenvectors of $\mathbf{W}$ as its columns, while the diagonal $n \times n$ matrix $\mathbf{\Lambda} = \text{diag}(\lambda_1, \lambda_2, \ldots \lambda_n)$ has the eigenvalues of $\mathbf{W}$ on its main diagonal. As above, we assume that $\mathbf{V}$ is orthonormal such that $\mathbf{v}_i^\top \mathbf{v}_i = 1$ and $\mathbf{v}_i^\top \mathbf{v}_{j \neq i} = 0$, and similarly $\mathbf{v}_{i:} \mathbf{v}_{i:}^\top = 1$ and $\mathbf{v}_{(j \neq i):} \mathbf{v}_{i:}^\top = 0$. We also assume that the eigenvalues in $\mathbf{\Lambda}$ are unique and ordered as $\lambda_1 > \lambda_2 > \cdots \lambda_n$. Note that, in contrast to $\mathbf{C}$, $\mathbf{W}$ will generally have negative eigenvalues. Here we assume that $\mathbf{W}$ is connected and can again use Perron-Frobenius theorem to state that $\lambda_1 > 0$ and that $\mathbf{v}_1$ will contain only positive elements. For simplicity, and without loss of generality, we will also assume that $\mathbf{W}$ is rescaled to have $\lambda_1 = 1$.

**Degree $\cong$ eigenvector centrality $\cong$ diffusion efficiency**

*Degree.* We denote the structural degree by $\mathbf{s}$ and define it as

$$\mathbf{s} = \mathbf{W}\mathbf{1} = \mathbf{V}\mathbf{\Lambda}\mathbf{V}^\top \mathbf{1} = \sum_{i=1}^n \lambda_i \mathbf{v}_i \mathbf{v}_i^\top \mathbf{1} = \sum_{i=1}^n \alpha_i \mathbf{v}_i, \tag{22}$$

where $\alpha_i$ is the product of an eigenvalue $\lambda_i$ and $\mathbf{v}_i^\top \mathbf{1} = \sum(v_i)$, the sum of elements in $\mathbf{v}_i$.

*Eigenvector centrality.* Eigenvector centrality can be interpreted as the steady state of a diffusion process on $\mathbf{W}$. Recalling that $\lambda_1 = 1$ and denoting by $\mathbf{W}^\tau \mathbf{1}$ the state at time point $\tau$, we can formally define this measure as

$$(\textbf{eigenvector centrality}) \equiv \left(\lim_{\tau \to \infty} \mathbf{W}^\tau\right)\mathbf{1} = \left(\lim_{\tau \to \infty} \mathbf{V}\mathbf{\Lambda}^\tau \mathbf{V}^\top\right)\mathbf{1} = \mathbf{v}_1 \mathbf{v}_1^\top \mathbf{1} = \alpha_1 \mathbf{v}_1, \tag{23}$$

where we made use of $\mathbf{V}^\top \mathbf{V} = \mathbf{I}$, $\lambda_1 = 1$ and $\lambda_{i>1} < 1$.

Equations 22 and 23 show that eigenvector centrality is equivalent to $\mathbf{v}_1$. We can thus reason as above to note that it will often be an accurate approximation of the degree because $|\alpha_1| \gg |\alpha_{i>1}|$. In our example data, we find $|\alpha_1| > |25\alpha_{i>1}|$.

*Diffusion efficiency.* The diffusion efficiency is formally defined as

$$(\text{diffusion efficiency})_i = \sum_{j=1}^n (\text{mean first passage time})_{ij}^{-1}, \tag{24}$$

where the mean first passage time denotes the mean number of steps it takes to reach one node from another through a random walk.

Formally, we can define the mean first passage time via the $n \times n$ transition probability matrix $\mathbf{Y}$,

$$\mathbf{Y} = \text{diag}(\mathbf{s})^{-1}\mathbf{W}.$$

Each element of this matrix $y_{ij}$ represents the probability of moving from node $i$ to node $j$, under the assumption of random-walk dynamics (as in Equation 14).

It is easy to verify that the steady state of a random walk on $\mathbf{Y}$ is given by

$$\lim_{\tau \to \infty} \mathbf{Y}^\tau = \frac{1}{\Sigma(w)} \mathbf{1}\mathbf{s}^\top.$$

where $\Sigma(w)$ is the sum of all values of $\mathbf{W}$. We note, in passing, that $\mathbf{s}$ is a leading left eigenvector (and thus the eigenvector centrality) of $\mathbf{Y}$.

We can now formally define the mean first passage time (Grinstead and Snell, 1998). By definition, the mean first passage time from a node to itself is 0, while the mean first passage time between distinct nodes is

$$(\text{mean first passage time})_{ij} = \frac{\Sigma(w)}{s_j}(z_{jj} - z_{ij}), \tag{25}$$

where $z_{ij}$ is an $(i, j)$-th element of the so-called fundamental matrix $\mathbf{Z}$,

$$\mathbf{Z} = \mathbf{I} + \sum_{\tau=1}^{\infty}\left(\mathbf{Y}^\tau - \frac{1}{\Sigma(w)}\mathbf{1}\mathbf{s}^\top\right).$$

Noting that the matrix $\mathbf{1}\mathbf{s}^\top/\Sigma(w)$ has identical columns, we can simplify Equation 25 to define the elements of the mean first passage time matrix as

$$(\text{mean first passage time})_{ij} = \frac{\Sigma(w)}{s_j}\left(1 + \sum_{\tau=1}^{\infty}\left[(\mathbf{Y}^\tau)_{jj} - (\mathbf{Y}^\tau)_{ij}\right]\right) \approx \frac{\Sigma(w)}{s_j} \equiv \frac{1}{s_j} \tag{26}$$

where the latter approximation is accurate in networks where most nodes have at least several connections, as in our example data. In such networks, the sum terms in Equation 26 are negligible, because the first term is $|w_{ij}/s_j| \ll 1$, and higher-order terms progressively tend to 0 as $\mathbf{Y}$ converges to its steady state.

Equation 24 and 26 finally allow us to conclude that (**diffusion efficiency**) $\cong \mathbf{s}$.

**Second degree $\cong$ communicability $\cong$ average controllability $\cong$ modal controllability**

*Second degree.* We denote the second degree of node $i$ by $\mathbb{s}_i$ and define it as

$$\mathbb{s}_i = \mathbf{w}_i^\top \mathbf{w}_i = \sum_{j=1}^n w_{ij}^2,$$

where $\mathbf{w}_i$ is the $i^{\text{th}}$ column of $\mathbf{W}$.

*Communicability centrality.* The communicability centrality is defined as the main diagonal of the $n \times n$ communicability matrix. Recalling that $\lambda_1 = 1$, we can define this measure as

$$(\textbf{communicability matrix}) = \sum_{\tau=0}^{\infty} \frac{1}{\tau!} \mathbf{W}^\tau \approx \mathbf{I} + \mathbf{W} + \frac{1}{2}\mathbf{W}^2.$$

The above approximation is likely to be accurate because the remaining error terms are $\approx 1/6\, \mathbf{W}^3$. It follows that the communicability centrality of node $i$ is given by

$$(\textbf{communicability})_i \approx 1 + \frac{1}{2}\mathbf{w}_i^\top \mathbf{w}_i \equiv \mathbb{s}_i.$$

*Average controllability.* This average controllability is defined as the main diagonal of the $n \times n$ controllability Gramian matrix. Recalling that $\lambda_1 = 1$, we can define this measure as

$$(\text{controllability Gramian}) = \sum_{\tau=0}^{\infty} \left(\frac{1}{\gamma}\mathbf{W}\right)^{\tau} \mathbf{B}\mathbf{B}^{\top} \left(\frac{1}{\gamma}\mathbf{W}^{\top}\right)^{\tau},$$

where, $\gamma > \lambda_1$ is a free normalization parameter, while the $n \times n$ matrix $\mathbf{B}$ encodes a pattern of external stimulation. Here, we adopt the standard choice of $\mathbf{B} = \mathbf{I}$ (to model an independent external stimulation of every node at every time point) and thus simplify

$$(\text{controllability Gramian}) = \sum_{\tau=0}^{\infty} \frac{1}{\gamma^{2\tau}} \mathbf{W}^{2\tau} \approx \mathbf{I} + \frac{1}{\gamma^2}\mathbf{W}^2.$$

The above approximation can be made arbitrarily accurate through the choice of a sufficiently large $\gamma$. Here, we adopt the standard choice of $\gamma = 1 + \lambda_1 = 2$ which implies that the remaining error terms are $\approx 1/16\,\mathbf{W}^4$.

It follows that the average controllability of node $i$ is given by

$$(\text{average controllability})_i \approx 1 + \frac{1}{\gamma^2}\mathbf{w}_i^{\top}\mathbf{w}_i \equiv \mathbb{s}_i.$$

*Modal controllability.* The modal controllability of node $i$ is defined as

$$(\text{modal controllability})_i = \sum_{j=1}^{n} v_{ij}^2 (1 - \lambda_j^2), \tag{27}$$

where $v_{ij}$ is the $j^{\text{th}}$ element of the eigenvector $\mathbf{v}_i$ and $\lambda_j$ is the $j^{\text{th}}$ eigenvalue of $\mathbf{W}$. Recalling that $\mathbf{v}_{i:}\mathbf{v}_{i:}^{\top} = \sum_{j=1}^{n} v_{ij}^2 = 1$, we can simplify Equation 27 to

$$(\text{modal controllability})_i = \sum_{j=1}^{n} v_{ij}^2 - \sum_{j=1}^{n} v_{ij}^2 \lambda_j^2 = 1 - \sum_{j=1}^{n} v_{ij}^2 \lambda_j^2. \tag{28}$$

Separately, we can write each $w_{ij}$ in terms of the elements of $\mathbf{V}$ and $\mathbf{\Lambda}$, as

$$w_{ij} = \sum_{h=1}^{n} \lambda_h v_{ih} v_{jh}.$$

This, in turn, allows us to express $\mathbb{s}_i$ as

$$\mathbb{s}_i = \sum_{j=1}^{n} w_{ij}^2 = \sum_{j=1}^{n} \left( \sum_{h=1}^{n} \lambda_h v_{ih} v_{jh} \right)^2 = \sum_{j=1}^{n}\sum_{g=1}^{n} \lambda_g v_{ig} v_{jg} \sum_{h=1}^{n} \lambda_h v_{ih} v_{jh}.$$

By rearranging the sums, we can simplify

$$\mathbb{s}_i = \sum_{g=1}^{n}\sum_{h=1}^{n} \lambda_g \lambda_h v_{ig} v_{ih} \sum_{j=1}^{n} v_{jg} v_{jh} = \sum_{g=1}^{n}\sum_{h=1}^{n} \lambda_g \lambda_h v_{ig} v_{ih} \mathbf{v}_g^{\top}\mathbf{v}_h = \sum_{h=1}^{n} \lambda_h^2 v_{ih}^2 \tag{29}$$

where we made use of $\mathbf{v}_g^{\top}\mathbf{v}_g = 1$ and $\mathbf{v}_g^{\top}\mathbf{v}_{h \neq g} = 0$.

Together, Equations 28 and 29 finally imply that

$$(\text{modal controllability})_i = 1 - \mathbb{s}_i \equiv \mathbb{s}_i.$$

**Squared coefficient of variation $\cong$ $k$-participation coefficient**

*Squared coefficient of variation.* The squared coefficient of variation is defined as

$$(\text{coefficient of variation})_i^2 = \frac{\text{var}(\boldsymbol{w}_i)}{\text{mean}(\boldsymbol{w}_i)^2},$$

where $\boldsymbol{w}_i$ is a vector of connection weights that are homogeneous within individual modules.

We first note that $\text{mean}(\boldsymbol{w}_i) = \mathbb{s}_i/n$, where $\mathbb{s}_i = \sum_i w_{ij}$ is the degree of node $i$. This allows us to simplify the squared coefficient of variation as

$$(\text{coefficient of variation})_i^2 = \frac{1}{n}\frac{\sum_{j=1}^n (w_{ij} - \mathbb{s}_i/n)^2}{(\mathbb{s}_i/n)^2} = n\sum_{j=1}^n \left(\frac{w_{ij}}{\mathbb{s}_i} - \frac{1}{n}\right)^2 \equiv \sum_{j=1}^n \frac{w_{ij}^2}{\mathbb{s}_i^2} \quad (30)$$

*$k$-participation coefficient.* The participation coefficient is defined as

$$(\text{participation coefficient})_i = 1 - \sum_{h=1}^k \left(\frac{\mathbb{s}_{i(h)}}{\mathbb{s}_i}\right)^2.$$

where we use $\mathbb{s}_{i(h)}$ to denote the degree of node $i$ to module $h$.

Correspondingly, we define the $k$-participation coefficient as

$$(k\text{-participation coefficient})_i = 1 - \sum_{h=1}^k \frac{1}{N_h}\left(\frac{\mathbb{s}_{i(h)}}{\mathbb{s}_i}\right)^2,$$

where, in parallel with the $k$-modularity, we normalize the contribution of each module by its size.

The assumption of homogeneous within-module connectivity allows us to use $\bar{w}_{i(h)}$, the mean connection weight of $i$ in module $h$, as an accurate representation of all connection weights in that module. This, in turn, allows us to approximate the squared within-module degree as

$$\mathbb{s}_{i(h)}^2 = \left(\sum_{j\in\mu_h} w_{ij}\right)^2 \approx \left(N_h \bar{w}_{i(h)}\right)^2 = N_h^2 \bar{w}_{i(h)}^2 \approx N_h \sum_{j\in\mu_h} w_{ij}^2.$$

This finally allows us to simplify the $k$-participation coefficient,

$$(k\text{-participation coefficient})_i \approx 1 - \frac{1}{\mathbb{s}_i^2}\sum_{h=1}^k \frac{1}{N_h}\left(N_h \sum_{j\in\mu_h} w_{ij}^2\right) = 1 - \frac{1}{\mathbb{s}_i^2}\sum_{i=1}^n w_{ij}^2,$$

which, together with Equation 30, establishes that

$$(\text{coefficient of variation})_i^2 \cong (k\text{-participation coefficient})_i.$$

**Shrunken proximity matrix ⇒ structural connectivity network**

*Original model formulation.* We denote a growing binary network at step $\tau$ by $\widetilde{\mathbf{W}}_\tau$ with elements $(\tilde{w}_{ij})_\tau$. Correspondingly, we denote the probability of forming a new binary connection at step $\tau$ by $\mathbf{\Pi}_\tau$ with elements $(\pi_{ij})_\tau$. The original formulation of the growth models we consider (Vértes et al., 2012) defines these probabilities for individual elements as

$$(\pi_{ij})_\tau \propto \phi_{ij}^\alpha (w_{ij})_\tau^\beta [\![(\tilde{w}_{ij})_\tau = 0]\!]. \tag{31}$$

In this equation, $\phi_{ij}$ denotes physical proximity (spatial closeness) of nodes $i$ and $j$, $(w_{ij})_\tau$ denotes the number of shared neighbors between these nodes at step $\tau$, $\alpha > 0$ and $\beta > 0$ are parameters, and $[\![\cdot]\!]$ effectively sets the probabilities for already connected nodes to 0.

For convenience, we set $\phi_{ii} = 0$ and express Equation 31 in matrix form, as

$$\mathbf{\Pi}_\tau \propto \left(\mathbf{\Phi}^{\circ\alpha} \odot \mathcal{W}_\tau^{\circ\beta} \odot [\![\widetilde{\mathbf{W}}_\tau = 0]\!]\right),$$

where $\odot$ denotes elementwise product and $\circ$ denotes elementwise power.

This formulation allows us to define the growth process as

$$\widetilde{\mathbf{W}}_{\tau+1} = f(\mathbf{\Pi}_\tau) + \widetilde{\mathbf{W}}_\tau, \tag{32}$$

where the function $f$ generates a matrix with a single binary connection according to the probabilities $\mathbf{\Pi}_\tau$.

The assumption of binary connectivity and the elementwise operations make it hard to treat the model in Equation 32 analytically. Here, instead, we develop a variant of this model that makes it more analytically tractable, but that still broadly preserves its growth principles.

We specifically make the following three changes to the original model.

First, we consider weighed, rather than binary, networks, and correspondingly replace the addition of new connections with a strengthening of existing connections. This allows us to replace the binary growth function $f$ with a scalar growth-rate constant $0 < \gamma < 1$, obviates the need for the binarization operator, and thus leads us to modify Equation 32 to

$$\widetilde{\mathbf{W}}_{\tau+1} = \gamma\left(\mathbf{\Phi}^{\circ\alpha} \odot \mathcal{W}_\tau^{\circ\beta}\right) + \widetilde{\mathbf{W}}_\tau. \tag{33}$$

Second, we replace the multiplicative relationship of spatial and similarity matrices with a corresponding additive relationship parameterized by a scalar constant $0 < \rho < 1$. This further modifies Equation 32 to

$$\widetilde{\mathbf{W}}_{\tau+1} = \gamma\left(\rho\mathbf{\Phi}^{\circ\alpha} + (1-\rho)\mathcal{W}_\tau^{\circ\beta}\right) + \widetilde{\mathbf{W}}_\tau.$$

Third, we note that the co-neighbor matrix $\mathcal{W}_\tau$ is simply equal to $\widetilde{\mathbf{W}}_\tau^2$. By omitting the exponent $\beta$, and thus absorbing its effects into $\alpha$ and $\rho$, we finally arrive at the following model variant:

$$\widehat{\mathbf{W}}_{\tau+1} = \gamma\left[\rho\mathbf{\Phi}^{\circ\alpha} + (1-\rho)\widehat{\mathbf{W}}_\tau^2\right] + \widehat{\mathbf{W}}_\tau. \tag{34}$$

We now define the eigendecomposition of $\mathbf{\Phi}^{\circ\alpha}$ as

$$\mathbf{\Phi}^{\circ\alpha} = \widehat{\mathbf{V}}\widehat{\mathbf{\Lambda}}_*\widehat{\mathbf{V}}^\top. \tag{35}$$

In our example data, we find that the off-diagonal elements of $\mathbf{\Phi}^{\circ\alpha}$ and $\mathbf{W}$ strongly correlate (**Figure 9**). Correspondingly, we find that the structure of the leading eigenvectors of $\widehat{\mathbf{V}}$ closely matches the structure of the corresponding eigenvectors of $\mathbf{V}$ (**Figure 13A**).

We now consider $\widehat{\mathbf{W}}_0$, the "seed" matrix that initializes the model at $\tau = 0$ in Equation 34. We will assume that we can express this matrix as a function of $\mathbf{\Phi}^{\circ\alpha}$. This assumption is reasonable because the seed matrix typically denotes a sparse matrix of the strongest or most consistent connections, which in practice will typically be spatially close. In our example data, we find that a proximity-based seed largely preserves the structure of the leading eigenvectors but effectively shrinks the eigenspectrum $\widehat{\mathbf{\Lambda}}_*$ (**Figure 13B–C**).

Accordingly, we can define the seed matrix as a shrunken proximity matrix,

$$\widehat{\mathbf{W}}_0 = \widehat{\mathbf{V}}\widehat{\mathbf{\Lambda}}'_*\widehat{\mathbf{V}}^\top = \widehat{\mathbf{V}}\widehat{\mathbf{\Lambda}}_0\widehat{\mathbf{V}}^\top, \tag{36}$$

which effectively makes $\widehat{\mathbf{W}}_0$ a sparse variant of $\mathbf{\Phi}^{\circ\alpha}$.

Let us now assume that we know $\widehat{\mathbf{W}}_\tau = \widehat{\mathbf{V}}\widehat{\mathbf{\Lambda}}_\tau\widehat{\mathbf{V}}^\top$ at some time point $\tau$. This assumption, together with Equations 34–36, allows us to define $\widehat{\mathbf{W}}_{\tau+1}$ as

$$\widehat{\mathbf{W}}_{\tau+1} = \gamma\big[\rho\widehat{\mathbf{V}}\widehat{\mathbf{\Lambda}}_*\widehat{\mathbf{V}}^\top + (1-\rho)\widehat{\mathbf{V}}\widehat{\mathbf{\Lambda}}_\tau^2\widehat{\mathbf{V}}^\top\big] + \widehat{\mathbf{V}}\widehat{\mathbf{\Lambda}}_\tau\widehat{\mathbf{V}}^\top = \widehat{\mathbf{V}}[\gamma(1-\rho)\widehat{\mathbf{\Lambda}}_\tau^2 + \widehat{\mathbf{\Lambda}}_\tau + \gamma\rho\widehat{\mathbf{\Lambda}}_*]\widehat{\mathbf{V}}^\top,$$

It follows that we can model network growth dynamics using the quadratic map evolution of the eigenvalues in $\widehat{\mathbf{\Lambda}}_\tau$.

This quadratic map does not have a closed-form solution, but we can study it numerically by optimizing the parameters in Equations 33 and 34. Separately, we can also test the effect of the quadratic (connectional similarity) term analytically by considering a variant of the model with $\rho = 1$. For this specific variant, we obtain the following closed-form solution,

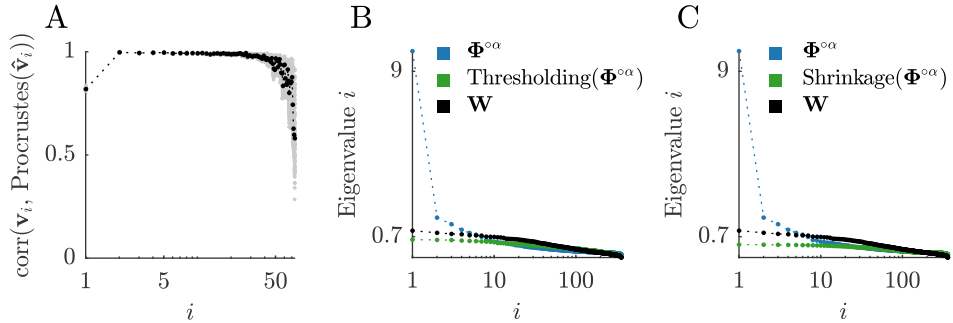

**Figure 13.**
**A.** Correlation between the leading eigenvectors of $\mathbf{W}$, and the corresponding leading eigenvectors of the proximity matrix (aligned via Procrustes rotation).
**B.** Eigenspectra of the proximity matrix before and after thresholding, shown against eigenspectra of the structural network.
**C.** Eigenspectra of the proximity matrix before and after shrinkage, shown against eigenspectra of the structural network.

$$\widehat{\mathbf{W}}'_\tau = \widehat{\mathbf{V}}(\widehat{\mathbf{\Lambda}}_0 + \tau\gamma\widehat{\mathbf{\Lambda}}_*)\widehat{\mathbf{V}}^\top = \widehat{\mathbf{W}}_0 + \tau\gamma\mathbf{\Phi}^{\circ\alpha} \approx \widehat{\mathbf{W}}_0,$$

where the approximation is reasonable because we already know that the seed matrix $\widehat{\mathbf{W}}_0$ aligns with $\mathbf{W}$ much better than a proximity-based approximation $\mathbf{\Phi}^{\circ\alpha}$.

**Node-module correlation ⇒ node-module dynamical affinity**

In this section, we consider the $k$-modularity on a residual correlation matrix. Since this section focuses on time series, it is convenient to define this residual matrix with global-signal regression, as

$$(k\text{-modularity}) \approx \sum_{h=1}^{r} \frac{1}{N_h} \mathbf{m}_h^\top \mathbf{C}' \mathbf{m}_h.$$

We define the average correlation of node $i$ to module $h$ as

$$\bar{c}_{i(h)} = \text{corr}(\mathbf{x}_i, \bar{\mathbf{x}}_h) = \frac{1}{N_h} d'_{i(h)},$$

where $\mathbf{x}_i$ is the time series of node $i$, $\bar{\mathbf{x}}_h$ is the mean activity of module $h$ and $d'_{i(h)}$ is the residual node-module degree.

Correspondingly, we define node-module dynamical affinity as

$$\frac{1}{T}\sum_{\tau=1}^{T} (\widetilde{m}_{ih})_\tau,$$

where $T$ is the number of time windows in the recording, and $(\widetilde{m}_{ih})_\tau$ denotes the presence of node $i$ in module $h$ at time $\tau$.

We can formally motivate the relationship between these quantities by defining the difference in $k$-modularity when node $i$ is moved from module $g$ to module $h$ in the residual network $\mathbf{C}'$,

$$\Delta(k\text{-modularity})_{i:g\to h} = \frac{1}{N_h+1}\left(2d'_{i(h)} + c'_{ii} - \frac{1}{N_h}\mathbf{m}_h^\top\mathbf{C}'\mathbf{m}_h\right) - \frac{1}{N_g-1}\left(2d'_{i(g)} - c'_{ii} - \frac{1}{N_g}\mathbf{m}_g^\top\mathbf{C}'\mathbf{m}_g\right).$$

Note that the terms $\mathbf{m}_h^\top\mathbf{C}'\mathbf{m}_h/[N(N_h+1)]$ and $\mathbf{m}_g^\top\mathbf{C}'\mathbf{m}_g/[N_g(N_g-1)]$ essentially correspond to average within-module correlations, which we assume will be roughly similar across modules of correlation networks. Ignoring the self-correlation values of $c_{ii}$, we can thus approximate

$$\Delta(k\text{-modularity})_{i:g\to h} \approx \frac{2d_{i(h)}}{N_h+1} - \frac{2d_{i(g)}}{N_g-1} \approx \text{corr}(\mathbf{x}_i, \bar{\mathbf{x}}_h) - \text{corr}(\mathbf{x}_i, \bar{\mathbf{x}}_g) = \bar{c}_{i(h)} - \bar{c}_{i(g)}$$

It follows that, without any additional assumptions, nodes with similar correlations to distinct modules are more likely to switch between modules without significantly changing $k$-modularity.

# Details of analysis and modeling

## Additional algorithmic details

*Loyvain and Co-Loyvain initialization.* We considered the effects of three initializations:

1. A uniformly random initialization.
2. A standard (probabilistic) *k*-means++ initialization (Arthur and Vassilvitskii, 2007).
3. Greedy *k*-means++ (or maximin) initialization (Gonzalez, 1985).

We found that, in structural networks, the greedy initialization had the highest values of the normalized modularity, while in correlation networks, the balanced and random initializations had the highest values of the normalized modularity. We adopted the greedy initialization for all analyses in this study.

*Loyvain update rules.* We used the following update rules for Loyvain:

$$\Delta(k\text{-modularity})_{i:g \to h} = \frac{1}{N_h + 1}\left(2d_{i(h)} + c_{ii} - \frac{1}{N_h}\mathbf{m}_h^\top \mathbf{C}\mathbf{m}_h\right) - \frac{1}{N_g - 1}\left(2d_{i(g)} - c_{ii} - \frac{1}{N_g}\mathbf{m}_g^\top \mathbf{C}\mathbf{m}_g\right)$$

$$\Delta(s\text{-modularity})_{i:g \to h} = \frac{1}{D_h + d_i}\left(2d_{i(h)} + c_{ii} - \frac{d_i}{D_h}\mathbf{m}_h^\top \mathbf{C}\mathbf{m}_h\right) - \frac{1}{D_g - d_i}\left(2d_{i(g)} - c_{ii} - \frac{d_i}{D_g}\mathbf{m}_g^\top \mathbf{C}\mathbf{m}_g\right).$$

*Co-Loyvain algorithm.* We optimized the co-Loyvain objectives by using alternating maximization of individual partitions $\mathbf{m}_a$ and $\mathbf{m}_b$ with Loyvain-like update rules at each maximization step.

*m-umap initialization.* We considered three m-umap initializations:

1. *Spectral node initialization.* We used row-normalized eigenvectors 2–($k$ + 1) of $\mathfrak{C}$ to initialize the position of each node. This is essentially equivalent to a Laplacian-eigenmaps-based initialization (Kobak and Linderman, 2021).
2. *Spectral module initialization.* We used row-normalized eigenvectors 1–$k$ of $\mathbf{M}^\top \mathfrak{C}^\diamond \mathbf{M}$ to initialize the position of each module. We then assigned each node to the position of its module.
3. *Greedy spherical k-means++ (maximin) module initialization.* We first used spherical Fibonacci point sets to define $k$ points near-uniformly on the unit sphere (Keinert et al., 2015). We then placed the module with the largest degree on a point that was closest to all other points. We next iteratively chose an unassigned module with the lowest degree to all the assigned modules. We placed this module on a point that was farthest from all the assigned points. Finally, we assigned each node to the position of its module.

We found that all three initializations produced similar final embeddings. We adopted the third method for all analyses in this study because it sped up convergence by a factor of ~2.

*m-umap gradients.* We computed the (Euclidean) m-umap gradients as

$$-4\alpha\beta\left[\mathfrak{C}^\diamond \odot \mathbf{H}^{\circ(\beta-1)} \odot \left(1 + \alpha\mathbf{H}^{\circ\beta}\right)^{\circ(-2)}\right]\mathbf{U},$$

where

$$\mathbf{H} = 2(1 - \mathbf{U}\mathbf{U}^\top).$$

*Nonlinear shrinkage.* We adopted a cubic shrinkage method comprised of the following steps:

1. For each $i = 1, 2 \ldots n - 3$, fit a cubic polynomial to eigenvalues $i, \ldots n$.
2. Compute the root-mean-squared error (RMSE) for each of the fits in step 1.
3. Estimate a knee using a variant of the kneedle objective (Satopaa et al., 2011):
$$\frac{\text{RMSE}_0 - \text{RMSE}_i}{\text{RMSE}_0} - \frac{i}{n-1}.$$
4. Use the cubic fit at optimal $i$ to reconstruct the full eigenspectrum.
5. Compute a shrunken matrix on the basis of this eigenspectrum.

*Synthetic data generation.* We generated synthetic data that approximately preserved correlations of individual nodes to module centroids. We did this in the following two steps:

1. We first generated $k$ module-centroid time series that were constrained by empirical $k \times k$ pairwise module correlation but were maximally random otherwise.
2. We then generated $n$ node time series that were constrained by empirical $n \times k$ node-to-module correlations but were maximally random otherwise.

The centroid constraints in step 1 are important to accurately preserve the node to module correlations. Specifically, the module centroids generated as part of step 1 will generally differ from the centroids generated on the final synthetic data. Despite this, the generation of structured centroids maintains general correspondence between modules and therefore accurately constrains node-module correlations. Both steps used nullspace sampling that we recently developed to solve these and related problems. See our recent studies for a more detailed description of these methods, and for implementational details (Nanda et al., 2023; Nanda and Rubinov, 2023).

**Details of data acquisition and processing**

*Data.* We analyzed diffusion and functional MRI data from the Human Connectome Project (Van Essen et al., 2012). We chose 100 unrelated young adults who each had a low-head movement functional MRI scan (maximum relative root-mean-square movement < 0.25).

*Definitions of nodes.* We defined nodes using the Human-Connectome-Project multimodal parcellation, an atlas of 360 cortical regions (Glasser et al., 2016), except for UMAP and m-umap analyses, which used 59,412 cortical vertices.

*Definition of structural networks.* We used previous estimates of anatomical connectivity (Rosen and Halgren, 2021). These estimates were made using diffusion MRI tractography, a non-invasive method for quantifying anatomical connectivity from imaged patterns of water diffusion.

*Definition of correlation networks.* We computed rescaled correlation coefficients of functional-MRI activity data, processed using a minimal pipeline (Glasser et al., 2013), multimodal surface registration (Robinson et al., 2014), and classifier-based denoising (Salimi-Khorshidi et al., 2014).

*Other datasets.* As part of supplementary analyses, we analyzed 24 networks from the Netzschleuder catalogue (Peixoto, 2020) and the MNIST dataset of 70,000 images of handwritten digits (LeCun, 1998).

**Details of clustering and embedding analyses**

We ran the reference UMAP implementation in Python and did all the other analyses in MATLAB using a mix of built-in, *abct*, and custom scripts.

*Loyvain method.* We used the `loyvain` function in *abct*, with the following parameters: `kmodularity`, `kmeans`, or `spectral` objectives (depending on analysis); `network` similarity, `greedy` initialization, `10` batches, and `100` replicates.

*k-means and spectral clustering.* We compared the Loyvain to MATLAB built-in `kmeans` function (2023). We used this function with the following custom parameters: `correlation` distance, `100` replicates. We also compared Loyvain to MATLAB's built-in `spectralcluster` function (2019), modified to admit a custom number of clustering replicates. We used this function with the following custom parameters: `precomputed` distance, `100` clustering replicates.

*m-umap.* We computed m-umap across a range of $\kappa$ values using the `mumap` function in *abct* with precomputed indices of $\kappa$-nearest neighbors, and a precomputed module partition, and with otherwise default options. We computed the module partition using the `louvains` function in *abct* with `100` clustering replicates, and an enabled `finaltune` option.

*UMAP.* We used *umap-learn* 0.5.9 (McInnes et al., 2018b) across a range of $\kappa$ values with precomputed indices of $\kappa$-nearest neighbors, and with otherwise default options. To improve the quality of the embeddings, we experimented with `min_dist`, `spread`, `repulsion_strength`, `negative_sample_rate`, and `output_metric` parameters, but without much effect on the results (not shown).

*$\kappa$-nearest-neighbor networks.* We computed $\kappa$-nearest-neighbor networks directly from the correlation matrices in our example data, and by using cosine distance on the 70,000 images in the MNIST data.

*Clustering and embedding evaluation.* We defined the % improvement metric as

$$100 \times \frac{\text{(result with new method)} - \text{(result with existing method)}}{\text{(result with existing method)}}$$

In the case of *k*-means and spectral clustering, we computed this metric directly on the optimized objectives. In the case of m-umap and UMAP, the optimized objectives are less meaningful, and so we followed previous work (Becht et al., 2019; Wang et al., 2023) to define our statistic as the Pearson correlation coefficient of two $n \times k$ length vectors of distances between nodes (vertices) and module centroids, in native and embedding space. In native space, these distances were equivalent to the original correlations in our example data and to cosine similarities between images in the MNIST data. In embedding space, they corresponded to the Euclidean distances between points in the m-umap and UMAP embeddings. For our example data, we defined modules using binary m-umap (that is, modularity maximization of the symmetric $\kappa$-nearest-neighbor networks). For the MNIST data, we defined modules as ground-truth labels (the replacement of these labels with binary m-umap modules had a negligible effect on our results, not shown).

## Acknowledgments

I thank Aditya Nanda for helpful discussions; Haatef Pourmotabbed, Lucas Sainburg, Bruno Hidalgo, and Anas Reda for comments on an earlier version of the manuscript; Lindsay Bremner for scientific editing; NIH RF1MH125933, NSF 2207891, and NSF 2422393 for funding.

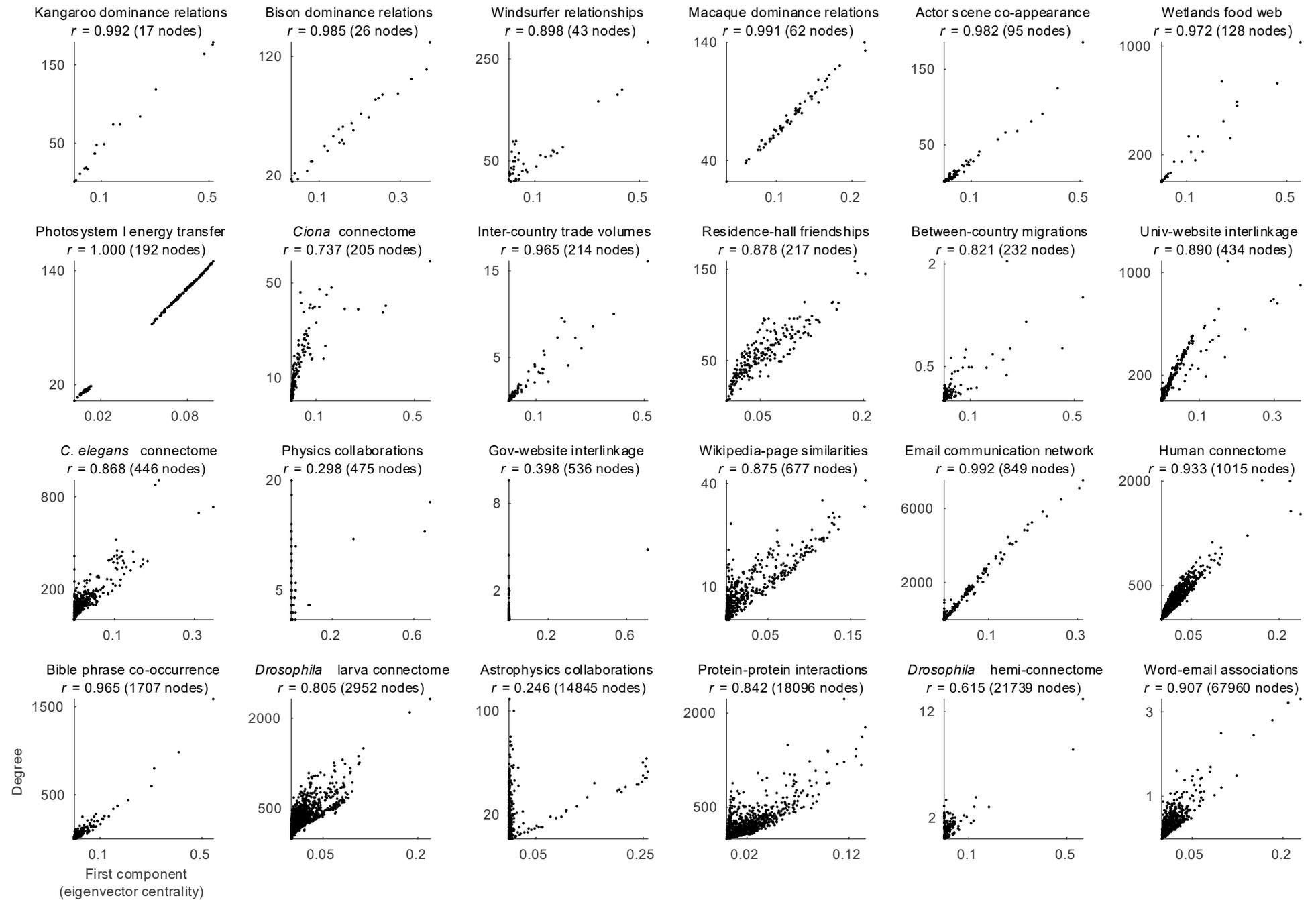

**Figure S1. First component and degree in diverse networks.**

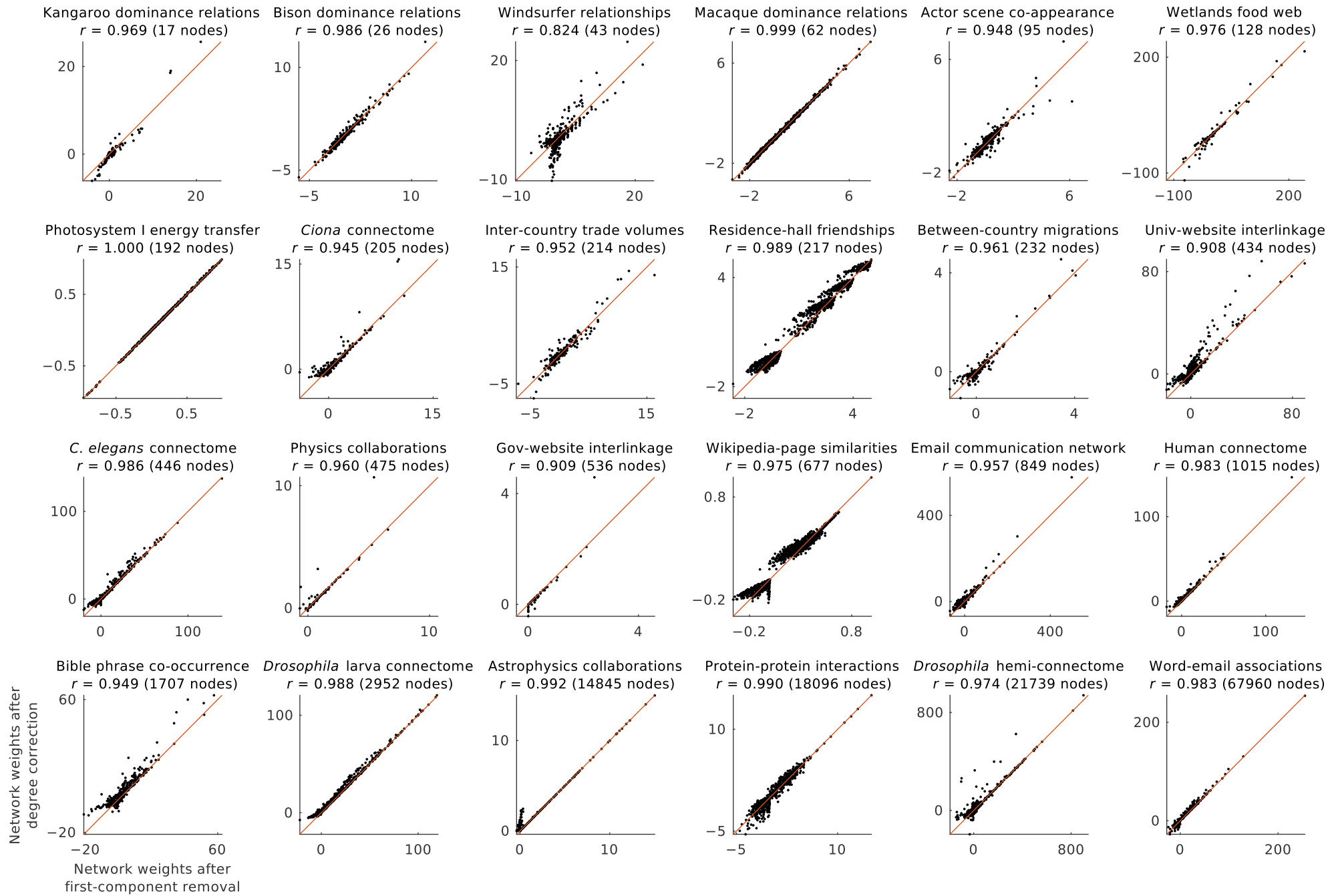

Figure S2. First-component removal and degree correction in diverse networks.

**Figure S1. First component and degree in diverse networks.**

Comparison of the first component and degree in networks from the Netzschleuder catalogue (Peixoto, 2020). From each dataset in this catalogue, we chose one directed or undirected, non-negatively weighted, and non-temporal network with the largest maximal connected component that satisfied (mean binary degree) > 10 to avoid extreme sparseness and (squared coefficient of variation of binary degree) < 10 to avoid extreme degree heterogeneity (Broido and Clauset, 2019). We analyzed the maximal connected components of these networks. We symmetrized all directed networks by adopting the maximum value for pairs of directed connection weights. We excluded datasets that lacked networks with these criteria, as well as the two outlier datasets that each had tens of thousands of networks (human_brains and openstreetmap). For networks with more than 10,000 nodes, we analyzed ~50,000,000 connections, periodically sampled from 1 to the total number of connections. Finally, we ordered the plots by network size.

Here is a full list of dataset/network abbreviations, in alphabetical order: arxiv_collab/astro-ph-1999 (Newman, 2001); bag_of_words/enron (Newman, 2008); bible_nouns (Harrison and Römhild, 2008); bison (Lott, 1979); budapest_connectome/all_20k (Szalkai et al., 2017); celegans_2019/hermaphrodite_chemical (Cook et al., 2019); cintestinalis (Ryan et al., 2016); dnc (Kunegis, 2013); fao_trade (De Domenico et al., 2015b); fly_hemibrain (Scheffer et al., 2020); fly_larva (Winding et al., 2023); foodweb_baywet (Ulanowicz and DeAngelis, 2005); kangaroo (Grant, 1973); macaques (Takahata, 1991); mist/ppi_human (Hu et al., 2018); moviegalaxies/261 (Kaminski et al., 2012); physics_collab/pierreAuger (De Domenico et al., 2015a); psi (Montepietra et al., 2020); residence_hall (Freeman et al., 1998); un_migrations (UN Population Division, 2015); us_agencies/virginia (Kosack et al., 2018); webkb/webkb_washington_cocite (Slattery and Craven, 1998); wiki_science (Calderone, 2020); windsurfers (Freeman et al., 1988).

**Figure S2. First-component removal and degree correction in diverse networks.**

Comparison of network weights after first-component removal and network weights after degree correction in networks from the Netzschleuder catalogue. See Figure S1 for methodological details.

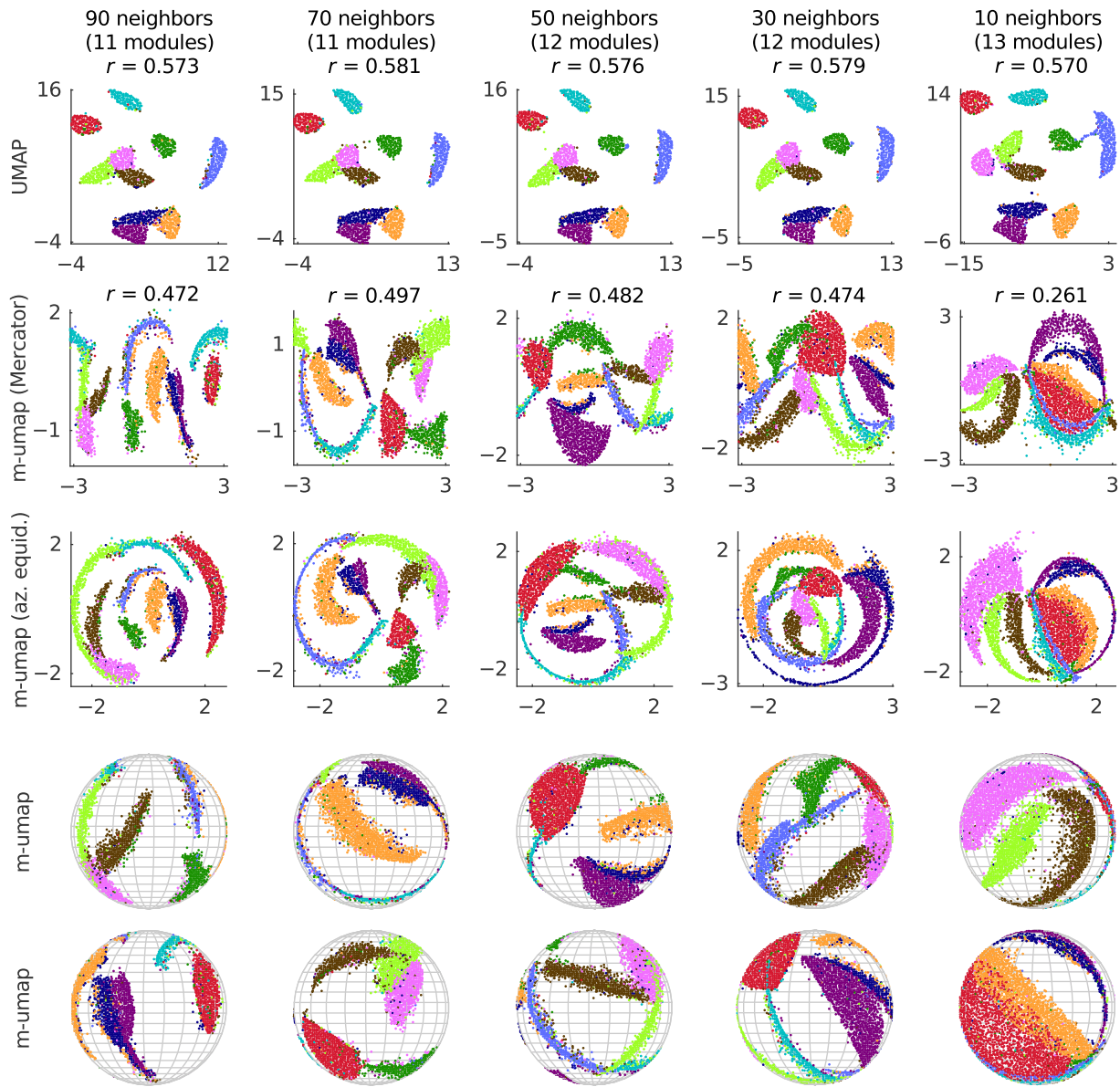

**Figure S3. Performance of UMAP and m-umap on the MNIST handwritten-digit dataset.**

UMAP and m-umap embeddings constructed from the symmetric $\kappa$-nearest-neighbor network of the 70,000 images of handwritten digits, using cosine-similarity distance. Columns show results across a range of $\kappa$-nearest neighbors. Colors in all panels denote the ground-truth MNIST labels for the individual digits. $r$ values represent Pearson correlation coefficients between distances of nodes to each MNIST-label centroid, in native and embedding space.

**Row 1.** Two-dimensional UMAP embeddings.

**Row 2.** Mercator ("classic map") projection of spherical m-umap embeddings onto a plane.

**Row 3.** Azimuthal equidistant ("UN flag") projection of spherical m-umap embeddings onto a plane.

**Rows 4–5.** Hemispheric views of spherical m-umap embeddings.

# Appendix: Summary of main results

See the main text for definitions of variables. Legend: ≡ denotes exact equivalences, ≅ denotes approximate equivalences, ⇒ denotes semi-analytical equivalences or unifications, * denotes the two known results. Unless otherwise specified, the third column shows Pearson correlation coefficients between analyses on our example data. In cases of exact equivalences, these correlations are ±1 by definition, while in other cases, they show the 50 (5–95)% range from 100 bootstrap data samples.

## Part 1. Clustering and dimensionality reduction

| *Relatively more established, simple, or robust analysis* | *Relatively more novel, complex, or speculative analysis* | *Example correlations* |
|---|---|---|
| **First-component removal.** Subtraction of the most dominant structural pattern (best rank-one approximation) from a network or dataset. $$\mathbf{C}^* = \mathbf{C} - \psi_1 \mathbf{u}_1 \mathbf{u}_1^\top$$ | ≅ **Degree correction.** Subtraction of the normalized product of node degrees (total nodal connection weights) from a network. $$\mathbf{C}^\circ = \mathbf{C} - \frac{1}{\sum(c)} \mathbf{d}\mathbf{d}^\top$$ | Structural networks: 1.000 (1.000–1.000) |
| | ≅ **Global-signal regression.** Removal of the contribution of the global (mean-activity) signal from a correlation network. $$\mathbf{C}' = \mathbf{X}'^\top \mathbf{X}' = \eta \left[ \mathbf{C} - \frac{1}{\sum(c)} \mathbf{d}\mathbf{d}^\top \right] \eta$$ | Correlation networks: 0.996 (0.995–0.997) |
| **k-means objective.** The objective function of $k$-means clustering, a popular data-clustering method. $$(k\text{-means objective}) \equiv \sum_{h=1}^{k} \frac{1}{N_h} \mathbf{m}_h^\top \mathbf{C} \mathbf{m}_h$$ | ≡ **$k$-modularity.** The objective function of modularity maximization, a popular module-detection method (normalized by module size, the number of nodes in the module). $$(\text{density-corrected } k\text{-modularity}) \equiv \sum_{h=1}^{k} \frac{1}{N_h} \mathbf{m}_h^\top \mathbf{C} \mathbf{m}_h$$ | 1 |
| $$(k\text{-means objective with global residualization}) \equiv \sum_{h=1}^{k} \frac{1}{N_h} \mathbf{m}_h^\top \mathbf{C}^\circ \mathbf{m}_h$$ | $$(k\text{-modularity}) \equiv \sum_{h=1}^{k} \frac{1}{N_h} \mathbf{m}_h^\top \mathbf{C}^\circ \mathbf{m}_h$$ | 1 |
| **Normalized cut.** The objective function of spectral clustering, a popular network-clustering method. $$(\text{normalized cut}) = \sum_{h=1}^{k} \frac{1}{D_h} \mathbf{m}_h^\top \mathbf{C} \mathbf{m}_h$$ | ≡ **$s$-modularity*.** The objective function of modularity maximization (normalized by module degree, sum of connection weights from a module to the rest of the network). $$(s\text{-modularity}) = \frac{1}{\sum(c)} \sum_{h=1}^{k} \frac{1}{D_h} \mathbf{m}_h^\top \left( \mathbf{C} - \frac{1}{\sum(c)} \mathbf{d}\mathbf{d}^\top \right) \mathbf{m}_h$$ This is a known result by Yu and Ding (2010). | 1 |

## Part 1. Clustering and dimensionality reduction (continued)

**m-umap objective.** (negative) modularity with Cauchy components.

$$(\text{m-umap objective}) = -\sum_{i,j}\left(c_{ij} - \gamma \frac{\mathfrak{d}_i \mathfrak{d}_j}{\sum(\mathfrak{c})}\right)\phi_{ij} = -\sum_{i,j} c_{ij}^\circ \phi_{ij}$$

$$\phi_{ij} = \left(1 + \alpha \|\mathbf{u}_{i:} - \mathbf{u}_{j:}\|^{2\beta}\right)^{-1}$$

$\equiv$ **first-order approximation of (true parametric) UMAP objective.** UMAP is a popular method for nonlinear dimensionality reduction and visualization of large-scale datasets.

$$(\text{UMAP objective}) = -\sum_{i,j}\left(c_{ij}\log(\phi_{ij}) + \gamma \frac{\mathfrak{d}_i \mathfrak{d}_j}{\sum(\mathfrak{c})}\log(1-\phi_{ij})\right)$$

**Co-neighbor components.** Dominant patterns of variation (eigenvectors) of co-neighbor networks.

$$(\textbf{co-neighbor component})_h = \mathbf{u}_{h+1}$$

$\mathbf{u}_{h+1}$ is the $(h+1)$-th eigenvector of $\mathcal{C}$

$$\mathcal{C} = [\![\widetilde{\mathbf{C}} > 0]\!]^\top [\![\widetilde{\mathbf{C}} > 0]\!]$$

$\cong$ **Co-activity gradients** (diffusion-map embeddings). Popular low-dimensional representations of correlation networks in imaging neuroscience.

$$(\textbf{co-activity gradient})_h \equiv \boldsymbol{q}_{h+1} \oslash \boldsymbol{q}_1$$

$\boldsymbol{q}_{h+1}$ is the $(h+1)$-th right eigenvector of $\mathcal{R}$

$$\mathcal{R} = \text{diag}(\mathbf{R}\mathbf{1})^{-1}\mathbf{R}$$
$$\mathbf{R} = (\tilde{\boldsymbol{\eta}}\widetilde{\mathbf{C}}^\top \widetilde{\mathbf{C}} \tilde{\boldsymbol{\eta}}) \odot (\boldsymbol{\nu}\boldsymbol{\nu}^\top)$$

Correlation networks: 0.988 (0.958–0.993)

## Part 1. Unified or fused clustering and dimensionality reduction methods

| Standard methods | Unified methods | % Improvement of unified over standard methods on example networks. |
|---|---|---|
| **Lloyd** (for $k$-means clustering). **Louvain** (for modularity maximization). | $\Rightarrow$ **Loyvain**. ($\Rightarrow$ **co-Loyvain** for bipartite networks). | $k$-means clustering of correlation networks: 4.880 (0.507–6.675)% Spectral clustering of structural networks: 0.882 (0.154–2.275)% Spectral clustering of correlation networks: 0.911 (0.371–1.397)% |
| **Louvain** (for modularity maximization) **Continuous optimizer** (for m-umap optimization). | **m-umap optimizer** with modularity (binary m-umap) initialization and between-module force approximation. | Embeddings of vertex-level correlation networks: 7.560 (4.975–21.191)% |

**Part 2. Network centrality and dynamics**

| *Relatively more established, simple, or robust analysis* | *Relatively more novel, complex, or speculative analysis* | *Example correlations* |
|---|---|---|
| **Degree** (strength). Sum of connection weights from a node to all other network nodes. A basic measure of connectional magnitude. $$\mathbf{s} = \mathbf{W}\mathbf{1} = \sum_{i=1}^{n} \lambda_i \mathbf{v}_i \mathbf{v}_i^\top \mathbf{1}$$ | $\cong$ **Eigenvector centrality**\*. The contribution of a node to the dominant pattern of variation (leading eigenvector) of a network. $$(\text{eigenvector centrality}) \equiv \left(\lim_{\tau \to \infty} \mathbf{W}^\tau\right)\mathbf{1} \equiv \mathbf{v}_1$$ This is a well-known result (Chung et al. 2003; Borgatti 2005; Nadakuditi and Newman 2013). | Structural networks: 0.922 (0.921–0.924) |
| | $\cong$ **Diffusion efficiency**. The average inverse accessibility of a node from all other network nodes through random walks. $$(\text{diffusion efficiency})_i = \sum_{j=1}^{n} \left(\frac{\Sigma(w)}{s_j}(z_{jj} - z_{ij})\right)^{-1}$$ $$\mathbf{Z} = \mathbf{I} + \sum_{\tau=1}^{\infty}\left(\left(\text{diag}(\mathbf{s})^{-1}\mathbf{W}\right)^\tau - \frac{1}{\Sigma(w)}\mathbf{1}\mathbf{s}^\top\right)$$ | Structural networks: 0.961 (0.960–0.962) |
| **Second degree**. Sum of squared connection weights from a node to all other network nodes. A basic measure of connectional magnitude and dispersion. $$\mathbb{s}_i = \mathbf{w}_i^\top \mathbf{w}_i = \sum_{j=1}^{n} w_{ij}^2$$ | $\cong$ **Communicability centrality**. A weighted mean over all possible walks from a node to itself. $$(\text{communicability})_i = \left[\sum_{\tau=0}^{\infty} \frac{1}{\tau!}\mathbf{W}^\tau\right]_{ii}$$ | Structural networks: 0.999 (0.999–0.999) |
| | $\cong$ **Average controllability**. A theoretical propensity of a node to facilitate transitions to network activity states with little stimulation. $$(\text{average communicability})_i = \left[\sum_{\tau=0}^{\infty}\left(\frac{1}{\gamma}\mathbf{W}\right)^\tau \mathbf{B}\mathbf{B}^\top \left(\frac{1}{\gamma}\mathbf{W}^\top\right)^\tau\right]_{ii}$$ | Structural networks: 0.999 (0.999–0.999) |
| | $\equiv$ **Modal controllability**. A theoretical propensity of a node to facilitate transitions to difficult-to-reach network activity states. $$(\text{modal controllability})_i = \sum_{j=1}^{n} v_{ij}^2 (1 - \lambda_j^2)$$ | −1 |
| **Squared coefficient of variation**. The variance of nodal connection weights, divided by the square of the mean of these weights. A basic measure of relative connectional dispersion. $$(\text{coefficient of variation})_i^2 = \frac{\text{var}(\boldsymbol{w}_i)}{\text{mean}(\boldsymbol{w}_i)^2}$$ | $\cong$ **$k$-participation coefficient**. A measure of nodal connection diversity to distinct modules (normalized by module size). $$(k\text{-participation coefficient})_i = 1 - \sum_{h=1}^{k} \frac{1}{N_h}\left(\frac{s_{i(h)}}{s_i}\right)^2$$ | Co-neighbor structural networks: 0.927 (0.812–0.971) Co-neighbor correlation networks: 0.914 (0.822–0.951) |

## Part 3. Semi-analytical vignettes in imaging and network neuroscience

| *Relatively more established, simple, or robust analysis* | *Relatively more novel, complex, or speculative analysis* | *Example correlations* |
|---|---|---|
| **Residual degree**. (Negative) degree of correlation networks after global signal regression. $$\mathbf{d}' = \mathbf{C}'\mathbf{1}.$$ | $\Rightarrow$ **First co-activity gradient**. Popular representation of a canonical structural pattern in imaging neuroscience. $(\text{co-activity gradient})_1 \equiv \boldsymbol{q}_2 \oslash \boldsymbol{q}_1$, where $\boldsymbol{q}_h$ is the $h$-th right eigenvector of $\mathcal{R} = \text{diag}(\mathbf{R1})^{-1}\mathbf{R}$ and $\mathbf{R} = \left(\tilde{\boldsymbol{\eta}}\tilde{\mathbf{C}}^\top\tilde{\mathbf{C}}\tilde{\boldsymbol{\eta}}\right) \odot \left(\boldsymbol{vv}^\top\right)$. | Correlation networks: 0.922 (0.911–0.929) |
| **Shrunken proximity network**. Physical proximity after shrinkage, the weakening of dominant structural patterns in the network. $$\widehat{\mathbf{W}}_0 = \widehat{\mathbf{V}}\widehat{\boldsymbol{\Lambda}}'_*\widehat{\mathbf{V}}^\top$$ | $\Rightarrow$ **Structural connectivity network**. Approximation of the empirical network, possibly through a network-growth process. $$\widetilde{\mathbf{W}}_{\tau+1} = f\left(\boldsymbol{\Phi}^{\circ\alpha} \odot \boldsymbol{\mathcal{W}}_\tau^{\circ\beta} \odot [\![\widetilde{\mathbf{W}}_\tau = 0]\!]\right) + \widetilde{\mathbf{W}}_\tau$$ | Structural networks: 0.939 (0.936–0.940) |
| **Node-module correlation**. Average correlations of node time series with module centroid time series. $$(\text{node-module correlation})_{ih} = \bar{c}_{i(h)} = \frac{1}{N_h}d'_{i(h)}$$ | $\Rightarrow$ **Node-module dynamical affinity**. Fraction of times a node is placed in a module across individual time windows. $$(\text{node-model dynamical affinity})_{ih} = \frac{1}{T}\sum_{\tau=1}^{T}(\widetilde{m}_{ih})_\tau$$ | Correlation networks: 0.973 (0.948–0.980) |